\documentclass[twocolumn]{aastex7}
\usepackage{lineno}
\usepackage[caption=false]{subfig}
\usepackage{longtable}
\usepackage{multirow}
\usepackage{graphicx}
\usepackage{subcaption}
\usepackage{booktabs}
\usepackage{threeparttable}
\usepackage{amsmath,amssymb}
\usepackage[T1]{fontenc}
\usepackage{CJKutf8}
\usepackage{caption}
\usepackage{natbib}
\usepackage{siunitx}
\linenumbers
\graphicspath{{./}{figures/}}

\def\degr{^{\circ}}

\begin{document}
% \begin{CJK}{UTF8}{gbsn}

\title{Spectral Variations of $\gamma$-rays in Mrk 421}

\correspondingauthor{Jin Zhang}
\email{j.zhang@bit.edu.cn}

\author[0000-0002-4789-7703]{Rui-Qi Huang}
\affiliation{School of Physics, Beijing Institute of Technology, Beijing 100081, People's Republic of China; j.zhang@bit.edu.cn}
\email{j.zhang@bit.edu.cn}

\author{Xin-Ke Hu}
\affiliation{School of Physics, Beijing Institute of Technology, Beijing 100081, People's Republic of China; j.zhang@bit.edu.cn}
\email{j.zhang@bit.edu.cn}

\author[0009-0000-6577-1488]{Yu-Wei Yu}
\affiliation{School of Physics, Beijing Institute of Technology, Beijing 100081, People's Republic of China; j.zhang@bit.edu.cn}
\email{j.zhang@bit.edu.cn}

\author[0000-0003-2547-1469]{Ji-Shun Lian}
\affiliation{School of Physics, Beijing Institute of Technology, Beijing 100081, People's Republic of China; j.zhang@bit.edu.cn}
\email{j.zhang@bit.edu.cn}

\author[0000-0003-3554-2996]{Jin Zhang\dag}
\affiliation{School of Physics, Beijing Institute of Technology, Beijing 100081, People's Republic of China; j.zhang@bit.edu.cn}
\email{j.zhang@bit.edu.cn}

\begin{abstract}

We present a comprehensive analysis of the 17-year Fermi-LAT observational data of Mrk 421 to investigate the spectral variations in the $\gamma$-ray bands. The light curve of the source in the 0.1--1000 GeV band with a 14-day time bin exhibits significant variability at a confidence level exceeding 5$\sigma$, which is accompanied by spectral variation, displaying a {\it harder-when-brighter} behavior. Moreover, its flux variation can reach up to one order of magnitude within one day, with a daily flux up to $(1.19\pm0.84)\times10^{-8}~{\rm erg~cm^{-2}~s^{-1}}$ on MJD 56152. The 17-year integrated spectrum of Mrk 421 necessitates a complex model for explanation, whereas its time-resolved spectra over one-day or several-day time intervals can be well fitted by a power-law model. We propose that the complex spectral shape of the 17-year integrated spectrum stems from the superposition of different spectral shapes in different flux states. By generating the GeV spectra that are simultaneously observed with the archived TeV observations and constructing the combined GeV--TeV spectra, we find that some combined GeV--TeV spectral shapes clearly imply different radiation origins for the GeV and TeV emissions, challenging the one-zone leptonic model. It is found that the flux follows a lognormal distribution, while the photon spectral index distributions can be well fitted by either a lognormal or a Gaussian functions. The possible nature of the $\gamma$-ray variability in Mrk 421 is discussed. 

\end{abstract}


\section{Introduction}

Blazars represent a subclass of radio-loud active galactic nuclei (AGNs), distinguished by intense non-thermal emission from relativistic jets, which dominates across nearly their entire electromagnetic spectrum \citep[e.g.][]{Urry:1995mg}. Based on the optical emission line features, blazars are further classified into BL Lac objects (BL Lacs) and flat spectrum radio quasars (FSRQs). They are one of the main types of detected extragalactic $\gamma$-ray-emitting sources \citep{2023arXiv230712546B}. Especially in the very-high-energy (VHE) $\gamma$-ray band, the majority of identified extragalactic sources are BL Lacs to date\footnote{\url{https://www.tevcat.org/}}. The observed broadband energy spectral distributions (SEDs) of blazars commonly feature a bimodal structure. The low-energy component, peaking in the infrared-ultraviolet band (and sometimes extending into the X-ray band), is produced by the synchrotron radiation of relativistic electrons in jets. The high-energy component, spanning from the MeV to TeV $\gamma$-ray bands, is generally interpreted as arising from inverse-Compton (IC) scattering by the same electron population. This includes the synchrotron self-Compton (SSC) process, where synchrotron photons serve as seed photons \citep{1992ApJ...397L...5M,1996MNRAS.280...67G,1996A&AS..120C.503G,Massaro:2007ay,2009ApJ...704...38S,2012ApJ...752..157Z,2014MNRAS.439.2933Y}, and external Compton (EC) process \citep{1993ApJ...416..458D,1994ApJ...421..153S,2000ApJ...545..107B,2009MNRAS.397..985G,2014ApJ...788..104Z,2015ApJ...807...51Z}.

Blazars exhibit significant variability across multiple wavelengths on timescales ranging from minutes to years \citep{1997ARA&A..35..445U,2007ApJ...664L..71A,2009A&A...502..749A,2024PhRvD.109j3039B}. In the X-ray band, the variability is commonly characterized by correlations between spectral index and flux, as well as shifts in the peak energy of the synchrotron component toward higher energies during flaring states. For instance, the spectrum of Mrk 501 has been observed to harden during periods of increased brightness \citep{2006ApJ...646...61G,2025ApJ...980...88A}. Similarly, PKS 2155--304 demonstrates a hardening X-ray spectrum with increasing intensity \citep{1993ApJ...404..112S,2024ApJ...963L..41H}. Multi-epoch observations of several BL Lacs, including PKS 0548--322, H 1426+428, Mrk 501, 1ES 1959+650, and Mrk 421, conducted with BeppoSAX, XMM-Newton, and Swift, reveal a positive correlation between synchrotron peak energy and peak luminosity, as well as a negative correlation between synchrotron peak energy and spectral curvature \citep{2008A&A...478..395M}. An analogous spectral evolution characterized by a {\it harder-when-brighter} behavior has also been documented in the $\gamma$-ray regime in certain blazars (e.g., \citealt{2010ApJ...721.1425A,2020PASJ...72...44Z}). Furthermore, the seamless connection between the GeV and TeV $\gamma$-ray spectra of PKS 1510--089 suggests a common origin for both emission components \citep{2014A&A...569A..46A}.

Mrk 421 is a typical $\gamma$-ray-emitting BL Lac, located at a redshift of $z=0.031$ \citep{1975ApJ...198..261U}. Its GeV $\gamma$-ray emission has been detected by the EGRET telescope on the Compton Gamma Ray Observatory \citep{1999ApJS..123...79H}, and it was subsequently included in the catalog of sources detected during the first three months of the Fermi Large Area Telescope (Fermi-LAT) all-sky survey  \citep{2009ApJ...700..597A}. Mrk 421 is the first extragalactic source confirmed to emit TeV $\gamma$-rays, with its detection reported by the Whipple imaging atmospheric Cherenkov telescopes in 1992 \citep{1992Natur.358..477P}. The source exhibits violent variability at multi-frequencies, particularly in the X-ray and TeV bands \citep[e.g.,][]{1996Natur.383..319G,2005ApJ...630..130B,2025ApJ...980...88A}. Its X-ray spectrum is generally well described by a log-parabolic model and exhibits significant spectral evolution. Notable trends include a positive correlation between the SED peak flux and peak energy, and an anti-correlation between the curvature parameter and the peak energy, which are commonly interpreted as signatures of synchrotron radiation and statistical/stochastic acceleration of relativistic electrons \citep{2008A&A...478..395M,2009A&A...501..879T}. On the other hand, the {\it harder-when-brighter} spectral behavior frequently observed in Mrk 421 \citep[e.g.,][]{2015A&A...576A.126A,2022ApJ...938L...7D} and other BL Lacs is generally attributed to the injection of fresh high-energy electrons by a shock \citep{Kirk:1998kp,2022ApJ...938L...7D}. In the TeV band, significant variations in the spectral shape have also been detected, accompanied by a clear correlation between flux and spectral index \citep[e.g.,][]{MAGIC:2006cno,2002ApJ...575L...9K,2005A&A...437...95A}. To date, TeV emission from Mrk 421 has been detected by multiple facilities, such as Whipple, HEGRA, Tibet, MAGIC, HAWC and LHAASO \citep{2001ApJ...560L..45K,2001ICRC....7.2605K,2003ICRC....5.2595A,2007ApJ...663..125A,2010A&A...519A..32A,1999A&A...350..757A,2025ApJ...980...88A,2024ApJS..271...25C,2024ApJS..271...10W}, resulting in a substantial accumulation of observational data. Moreover, Fermi-LAT has been monitoring this source for over 17 years, making Mrk 421 an excellent target for comprehensive spectral variation investigation in the $\gamma$-ray band.

In this study, we systematically study the $\gamma$-ray emission properties and spectral evolution of Mrk 421 by using Fermi-LAT observations in conjunction with archival TeV data. The observations and data analysis are detailed in Section \ref{sec:obs&ana}. The results of the Fermi-LAT data analysis and the spectral variation in the GeV--TeV band are presented in Section \ref{sec:result}. The key findings are discussed in Section \ref{sec:dis}, and a summary is provided in Section \ref{sec:sum}.

\section{Observations and Data Analysis} \label{sec:obs&ana}
\subsection{Fermi-LAT}

It has been reported that 4FGL J1140.4+3812 is spatially coincident with Mrk 421 in the Fermi-LAT 14-year Source Catalog \citep[4FGL-DR4;][]{2023arXiv230712546B}. The Pass 8 data from Fermi-LAT observations spanning from 4 August 2008 to 26 March 2025 (MJD 54682--60760), extracted from a $15\degr$ region of interest (ROI) centered on the radio position of Mrk 421 (R.A. = $166\degr.114$, Decl. = $38\degr.2088$, J2000), are downloaded from the Fermi Science Support Center and analyzed in this work. The publicly available software \texttt{fermitools} (v.2.2.0) is used in our analysis \citep{2019ascl.soft05011F}. The $\gamma$-ray photons in the energy range of 0.1--1000 GeV are selected with a standard filter expression ``(DATA\_QUAL\textgreater0)\&\&(LAT\_CONFIG==1)''. Events with zenith angles exceeding 90$^\circ$ are excluded to minimize $\gamma$-ray contamination from the Earth’s limb. We use event class ``SOURCE'' (evclass=128) and event type ``FRONT+BACK'' (evtype=3) for a binned likelihood analysis. The P8R3\_SOURCE\_V3 set of instrument response functions is used. The background models include all $\gamma$-ray sources listed in the 4FGL-DR4 Catalog, along with the Galactic diffuse component (gll\_iem\_v07.fits) and the isotropic emission (iso\_P8R3\_SOURCE\_V3\_v1.txt). 

We use the maximum likelihood test statistic (TS), i.e., $\rm TS=2(log\mathcal{L}_1-log\mathcal{L}_0)$, where $\mathcal{L}_0$ is the likelihood of the background without the point source (null hypothesis) and $\mathcal{L}_1$ is the likelihood of the background including the point source, to estimate the significance of $\gamma$-ray signals. Generally, the $\gamma$-ray spectra of Fermi-LAT blazars are well described by a power-law (PL) model, given by  
\begin{equation}
\begin{aligned}
\frac{dN}{dE}=N_{0}(\frac{E}{E_{0}})^{-\Gamma_{\gamma}},
\end{aligned}
\end{equation}
where $N_{0}$ is the normalization parameter and $\Gamma_{\gamma}$ is the photon spectral index. However, for the BL Lac Mrk 421, the $\gamma$-ray spectrum in the 0.1--1000 GeV band requires a more complex functional form, as documented in the 4FGL-DR4 \citep{2023arXiv230712546B}. Specifically, the PLSuperExpCutoff4 model is used, defined as  
\begin{equation}
{\scriptsize \frac{dN}{dE}=\left\{\begin{aligned}
        & N_{0}(\frac{E}{E_{0}})^{{-\Gamma_{\gamma}}-\frac{\beta}{2}\ln\frac{E}{E_{0}}-\frac{{\beta}b}{6}\ln^{2}\frac{E}{E_{0}}-\frac{{\beta}b^{2}}{24}\ln^{3}\frac{E}{E_{0}}},~{\rm if}~|b\ln\frac{E}{E_{0}}|< 1e^{-2} \\
        & N_{0}(\frac{E}{E_{0}})^{-\Gamma_{\gamma}+\frac{\beta}{b}}\exp(\frac{\beta}{b^{2}}(1-(\frac{E}{E_{0}})^{b})),~{\rm otherwise},
       \end{aligned}
\ \right.}
\end{equation}
where $E_{0}=1328.17$ MeV denotes the energy scale parameter, $\beta$ is the curvature parameter, and $b$ is fixed at 0.657. The spectral parameters of all sources lying within $9\degr.5$ are left free, as are the normalization parameters of gll\_iem\_v07 and iso\_P8R3\_SOURCE\_V3\_v1, while the parameters of those sources lying beyond $9\degr.5$ are fixed to their 4FGL-DR4 values. 

We perform a maximum likelihood analysis to assess the relative statistical preference between the PLSuperExpCutoff4 model and the PL model. The analysis yields $\Delta$TS, which is defined as
\begin{equation}
\begin{aligned}
{\Delta {\rm TS}}=2(\log{\mathcal{L}_{\rm PLSC}}-\log{\mathcal{L}_{\rm PL}}),
\end{aligned}
\label{eq:TS}
\end{equation}
where $\mathcal{L}_{\rm PLSC}$ and $\mathcal{L}_{\rm PL}$ denote the likelihood values obtained from the PLSuperExpCutoff4 and PL models, respectively. The PL model is a special case of the PLSuperExpCutoff4 model when $\beta=0$. If $\Delta$TS $\ge$9, which corresponds to a 3$\sigma$ significance for the curvature \citep{Mattox:1996zz,2012ApJS..199...31N,2020ApJS..247...33A}, the PLSuperExpCutoff4 model is favored; otherwise, the PL model is preferred.

Based on the spectral fitting results, we generate the $\sim$17-year light curves using a time bin of two weeks (14 days). During the construction of the light curves, the parameters $N_{0}$, $\Gamma_{\gamma}$, and $d$, as well as the normalization parameters of gll\_iem\_v07 and iso\_P8R3\_SOURCE\_V3\_v1, are set free, while all other parameters are fixed to their 4FGL-DR4 values.

\subsection{Archival VHE Observations}\label{sec:TeV}

In this study, we consider only VHE spectra with simultaneous observations from Fermi-LAT; that is, the TeV observations of Mrk 421 that we selected were conducted following the commencement of Fermi-LAT operations. Additionally, only observations shorter than two weeks are included in our sample. The VHE data are compiled from published literature: observations on 17 February 2010 by VERITAS \citep{2011arXiv1109.6059G}; observations from 10 March 2010 to 22 March 2010 by MAGIC and VERITAS \citep{RosalesdeLeon:2021ape,Sahu:2021wue}; observations on 10 January 2013 by MAGIC and VERITAS \citep{2016ApJ...819..156B,Hu:2024xku}; observations on 28 December 2014 by TACTIC \citep{Singh:2018tzy}; observations from 10 January 2019 to 14 January 2019 by the FACT telescope \citep{Markowitz:2022jcp}; observations from 4 May 2022 to 6 May 2022 and 4 June 2022 to 6 June 2022 by MAGIC \citep{Abe:2023yoy}. A total of 18 TeV spectra for Mrk 421 are included in this work. 

\section{Results}\label{sec:result}
\subsection{Entire Observation Period of Fermi-LAT}

The 17-year integrated spectrum of Mrk 421 is well fitted by the PLSuperExpCutoff4 model, featuring a photon spectral index of $\Gamma_{\gamma}\sim1.744$ and a curvature parameter of $\beta\sim0.009$, as shown in Figure \ref{Spec_17y}. Based on the spectral fitting results, the derived long-term flux light curve in the 0.1--1000 GeV band and the corresponding spectral index curve, computed using a time bin of 14 days, are presented in Figure \ref{lc}, with an average flux of $F_{0.1-1000~{\rm GeV}}=(5.83\pm0.11)\times10^{-10}~{\rm erg~{cm}^{-2}~{s}^{-1}}$. The observed maximum and minimum fluxes are $(2.74\pm1.37)\times10^{-9}~{\rm erg~cm^{-2}~s^{-1}}$ and $(3.14\pm2.32)\times10^{-11}~{\rm erg~cm^{-2}~s^{-1}}$, respectively, spanning nearly two orders of magnitude. To quantify the variability, we assume a constant flux and calculate the $\chi^{2}$ and the corresponding null hypothesis probability. The result indicates that the light curve of Mrk 421 in the 0.1--1000 GeV band exhibits significant variability at a confidence level exceeding $5\sigma$. The spectral indices also show variation at a confidence level of $5.1\sigma$.

We further reanalyze the $\sim$ 17-year Fermi-LAT observational data in the 0.1--1 GeV, 1--100 GeV, and 100--1000 GeV bands, respectively. According to the criterion presented in Equation (\ref{eq:TS}), the PL model is preferred for the 0.1--1 GeV band, whereas the PLSuperExpCutoff4 model is favored for the 1--100 GeV band, with a $\Delta$TS value of 19.5. For the 100--1000 GeV band, the $\Delta$TS value of 3.0 does not indicate a significant preference for the PLSuperExpCutoff4 model. Nevertheless, we adopted the PLSuperExpCutoff4 model as well, based on clear visual evidence of spectral curvature in the high energy range. Additionally, the $\Delta$TS value in the full 0.1--1000 GeV band is 86.5, providing statistical preference support for the utilization of the PLSuperExpCutoff4 model. 

The long-term light curves in the 0.1--1 GeV, 1--100 GeV, and 100--1000 GeV bands, with a time bin of 14 days, are also presented in Figure \ref{lc}. The flux in both the 0.1--1 GeV and 1--100 GeV bands shows significant variability at a confidence level exceeding $5\sigma$, whereas no obvious variability is observed in the 100--1000 GeV band. The hardness ratio (HR) curve is also given in Figure \ref{lc}. Given the limited number of detection points in the light curve of the 100--1000 GeV band, we only take into account the HR between the 0.1--1 GeV and 1--100 GeV bands. Thus, the HR is defined as ${\rm HR}=\frac{F_{1-100~{\rm GeV}}-F_{0.1-1~{\rm GeV}}}{F_{1-100~{\rm GeV}}+F_{0.1-1~{\rm GeV}}}$, where $F_{1-100~{\rm GeV}}$ and $F_{0.1-1~{\rm GeV}}$ represent the fluxes in the 1--100 GeV and 0.1--1 GeV bands, respectively. As displayed in Figure \ref{lc}, the HR curve demonstrates substantial variability at a confidence level far exceeding 5$\sigma$, indicating notable spectral variations of the source.  

\subsection{Daily time-scale Variability in High-flux State}

To investigate the variability of Mrk 421 on a daily timescale, a time slice, which consists of 6 consecutive time bins with $F_{0.1-1000~{\rm GeV}}>10^{-9}~{\rm erg~{cm}^{-2}~{s}^{-1}}$, is selected from the 17-year light curve, as indicated by the shaded region in Figure \ref{lc}. We reanalyze the Fermi-LAT observational data within this time slice and obtain the light curves of the flux and photon spectral index using a time bin of 1 day, as presented in Figure \ref{lc_day}. The variability of Mrk 421 on a daily timescale during this time slice can be observed at a confidence level of $1.3\sigma$. However, the maximum flux variation is from $(9.69\pm2.93)\times10^{-10}~{\rm erg~cm^{-2}~s^{-1}}$ on MJD 56151 to $(1.19\pm0.84)\times10^{-8}~{\rm erg~cm^{-2}~s^{-1}}$ on MJD 56152, indicating that the flux variation of Mrk 421 can reach one order of magnitude within a day. Given the large errors of these data points in Figure \ref{lc_day}, further exploration of the shorter variability timescale is not conducted.

\subsection{Spectral Variation in the Fermi-LAT Band}

Both $\Gamma_{\gamma}$ and HR exhibit variations in their long-term curves, as displayed in Figure \ref{lc}, indicating that the Fermi-LAT spectra undergo significant changes. It can be noted that there are only a limited number of detection points in the long-term light curve in the 100-1000 GeV band. However, the PLSuperExpCutoff4 spectral shape of the 17-year integrated spectrum is significantly affected by the detection of high-energy photons. Therefore, the HR value is more appropriate to characterize the spectral slope than $\Gamma_{\gamma}$. In Figure \ref{Spec_17y}, we present HR as a function of the flux in the 0.1--100 GeV ($F_{0.1-100~{\rm GeV}}$) band. There appears to be a tendency of correlation between them. Given the considerable uncertainties related to the values of HR and $F_{0.1-100~{\rm GeV}}$, we employ a parametrized bootstrap method \citep{1979AnSta...7....1E} to estimate the correlation coefficient $r$ between these two variables. The bootstrap analysis yields $r = 0.42\pm0.04$, suggesting the \textit{harder-when-brighter} behavior of Mrk 421 in the 0.1--100 GeV band.

To further explore the spectral evolution properties in the Fermi-LAT energy band at different flux states, we first select the three highest and three lowest flux points from the long-term light curve shown in Figure \ref{lc}, and generate time-resolved spectra for the 6 time bins. Moreover, we analyze the Fermi-LAT data to obtain the simultaneous GeV spectra corresponding to the 18 TeV spectra. The results of the data analysis and the corresponding parameter values are presented in Table \ref{tab_LAT}. For some time-resolved spectra with a one-day timescale and the two time-resolved spectra in the low-flux state, we obtain low TS values. Recently,\citet{2025A&A...695A.217M} reported that they used averaged intervals exceeding 3 days centered at the MAGIC observation times for spectral analysis of the Fermi-LAT data, given Fermi-LAT's limited sensitivity to resolve the spectrum of Mrk 421 on a daily timescale.

The 24 time-resolved Fermi-LAT spectra are presented in Figure \ref{Spe_LAT_24}. All 18 time-resolved spectra with simultaneous TeV observations, together with the 3 time-resolved spectra in the low-flux state, can be well characterized by a simple PL model, which differs from the 17-year Fermi-LAT integrated spectrum. Visually, the 3 time-resolved spectra in the high-flux state exhibit a curved spectral shape. In particular, two of them have similar fitting parameters to that of the 17-year integrated spectrum, as listed in Table \ref{tab_LAT}. Therefore, we retain this fitting result for the three epochs, even though no significant statistical preference is obtained for the PLSuperExpCutoff4 model according to the criterion of Equation (3) and $\Delta$TS $\ge$9. Intriguingly, the spectrum from 2018 January 22 to 2018 February 5 exhibits a negative curvature parameter value, as illustrated in Figure \ref{Spe_LAT_24}. The low-energy end of the spectrum appears to be situated in the synchrotron-SSC radiation transition region for this high-synchrotron-peaked (HSP) BL Lac.

It can be observed that, in general, there are no detection points at the high-energy end of these time-resolved spectra, as shown in Figure \ref{Spe_LAT_24}. Therefore, we calculate the flux in the 0.1--100 GeV band for these time-resolved spectra, as given in Table \ref{tab_LAT}. In Figure \ref{Flux-Gamma}, we plot $\Gamma_{\gamma}$ as a function of $F_{0.1-100~{\rm GeV}}$ for the 24 time-resolved spectra. Additionally, we employ the bootstrap method to estimate the correlation coefficient $r$ between $\Gamma_{\gamma}$ and $F_{0.1-100~{\rm GeV}}$, yielding $r = -0.52\pm0.12$. Clearly, the source demonstrates a \textit{harder-when-brighter} behavior in the $\gamma$-ray band.

\subsection{Spectral Variation in the GeV--TeV Band}\label{sec:GeV-TeV}

To explore whether the GeV and TeV emissions of Mrk 421 share a common origin, we combine the 18 time-resolved Fermi-LAT spectra with their simultaneously observed TeV spectra, as presented in Figure \ref{Spe_GeV-TeV}. All VHE spectra have been corrected for the absorption effect of the extragalactic background light (EBL), with the intrinsic TeV emission reconstructed using the EBL model in \citet{Finke:2022uvv}. Given that the X-ray spectra of Mrk 421 generally exhibit a log-parabolic curvature (e.g., \citealt{2007A&A...466..521T, 2009A&A...501..879T}), similar to numerous other HSP BL Lacs (e.g., \citealt{2008A&A...478..395M}), this implies a log-parabolic distribution of electron energy. Under the one-zone leptonic model, the SSC bump is also likely to be characterized by a log-parabolic shape (e.g., \citealt{2006A&A...448..861M}). Therefore, we use a log-parabola (LP) model to fit the 18 combined GeV--TeV spectra. The LP model is defined as follows \citep{2004A&A...413..489M}:
\begin{equation}
\begin{aligned}
\frac{dN}{dE}=N_{0}\left(\frac{E}{E_{\rm b}}\right)^{-(\alpha+\beta_1\ln\left(\frac{E}{E_{\rm b}}\right))},
\end{aligned}
\end{equation}
where $\alpha$ is the spectral index, $\beta_1$ is the spectral curvature parameter, and $E_{\rm b}$ is a scale parameter fixed at 1.5 GeV.

The fitting lines using the LP model for these combined spectra are also shown in Figure \ref{Spe_GeV-TeV}. Here, we just provide a visually fitting result. Most GeV spectra are smoothly connected to the TeV spectra, and the combined GeV--TeV spectra can be fitted or roughly fitted by the LP model. Nevertheless, for some spectra, such as those from 2010 March 10, 2010 March 11, and 2010 March 17, a LP function is inadequate to fit these GeV--TeV spectra, indicating that the GeV and TeV emissions are not produced by the same IC process and likely have different origins. These findings pose a challenge to the typical one-zone leptonic model in explaining the broadband SEDs of BL Lacs.

Additionally, we roughly fit the 18 TeV spectra using a simple PL function and estimate the flux within the 0.1--10 TeV band ($F_{0.1-10~{\rm TeV}}$) based on the fitting results. The relationship between $F_{0.1-100~{\rm GeV}}$ and $F_{0.1-10~{\rm TeV}}$ for the 18 GeV--TeV spectra is displayed in Figure \ref{Flux-Gamma}. Pearson correlation analysis yields a coefficient of $r=0.572$ and a chance probability of $p=0.013$, indicating that the TeV emission of Mrk 421 does not exhibit a strong correlation with the flux state in the GeV band. 

\section{Discussion}\label{sec:dis}

Previous studies have shown that most of the HSP BL Lacs exhibit a stable, or even low, flux in the GeV band and no significant variability is observed \citep{2011ApJ...743..171A, 2024ApJS..270...22W, 1962SvA.....6..317K, 2026arXiv260115142L}. However, Mrk 421 is an exception. Its GeV $\gamma$-ray emission exhibits significant variability on a timescale of 14 days, accompanied by obvious spectral variations. Moreover, the significant flux variations of Mrk 421 in the GeV band can occur on a daily timescale. Recently, \citet{2026arXiv260115142L} analyzed the 16-year Fermi-LAT observational data for 25 extreme HSP BL Lacs in the 0.1--1000 GeV and found that, with the exception of Mrk 421, the 16-year integrated spectra of the remaining 24 extreme HSP BL Lacs can be well described by PL or LP models. It can be observed that the significant spectral variation of Mrk 421 is also presented in these time-resolved spectra, as illustrated in Figure \ref{Spe_LAT_24}. Nevertheless, these time-resolved spectra are generally fitted by a PL model, distinct from the 17-year integrated spectrum. Although the flux exhibits significant variability in the 0.1--1 GeV and 1--100 GeV bands, no significant variability is found in the 100--1000 GeV band, and no detection point is obtained for most of the time bins, as shown in Figure \ref{lc}. These results suggest that the complex spectral shape of the 17-year integrated spectrum of Mrk 421 in the 0.1--1000 GeV band is due to the superposition of spectra in different flux states.  

As shown in Figure \ref{lc}, the variability of Mrk 421 is accompanied by the spectral variation. Nevertheless, unlike NGC 1275, no long-term constant photon indices are observed. Therefore, its variability cannot be ascribed to the variations of the jet Lorentz factor and/or the alterations of the viewing angle, as reported in NGC 1275 \citep{2018ApJ...860...74T}. In the GeV band, Mrk 421 exhibits a significant \textit{harder-when-brighter} behavior. This \textit{harder-when-brighter} behavior of Mrk 421 has been extensively investigated and documented in the X-ray band (e.g., \citealt{2008A&A...478..395M, 2009A&A...501..879T, 2022ApJ...938L...7D}), which is typically interpreted as the injection of high-energy electrons into the emission region \citep{1998A&A...333..452K,2022ApJ...938L...7D,2024ApJ...965...58Z}. Under the one-zone leptonic model, the X-rays of Mrk 421 are generated by high-energy electrons through the synchrotron process, while the GeV emission is attributed to the IC scattering of low-energy photons by low-energy electrons (see Figure 2 in \citealt{1998ApJ...509..608T}). The recent X-ray polarization measurement by IXPE supports that the X-ray emission of these HSP BL Lacs, including Mrk 421, is produced by the synchrotron radiation of relativistic electrons \citep{2022ApJ...938L...7D, 2024A&A...684A.127A,2024A&A...681A..12K}. Therefore, the X-ray and the GeV $\gamma$-ray emissions of Mrk 421 may be contributed by electrons of different energies through different radiation processes. This is also consistent with the weak correlation between the X-ray flaring activity and the GeV $\gamma$-ray flux variability \citep{2024ApJS..275...23K}. 

\begin{figure*}
    \centering
    \begin{minipage}{0.45\textwidth}
        \centering
        \includegraphics[angle=0, width=\textwidth]{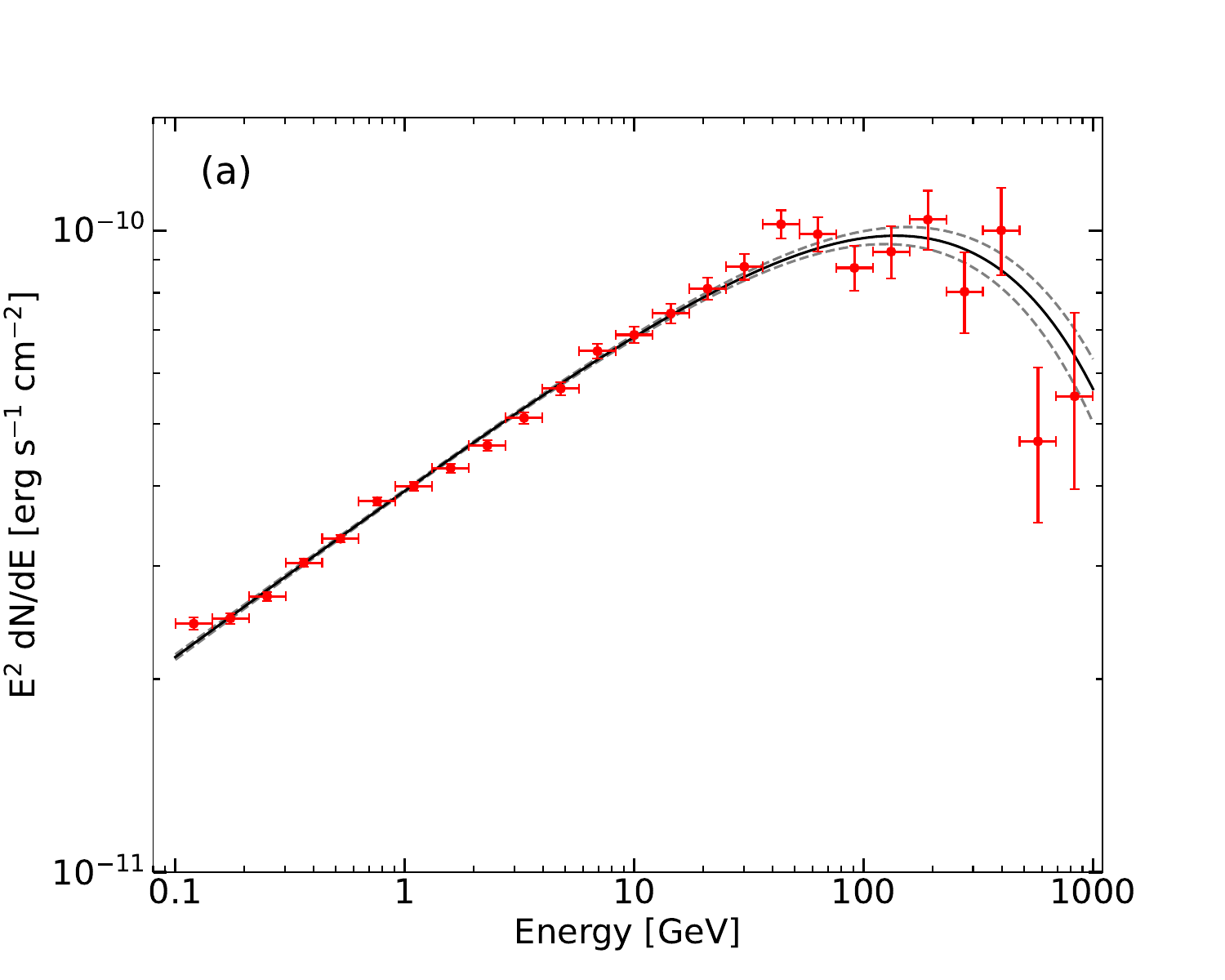} 
    \end{minipage}
    \begin{minipage}{0.48\textwidth}
        \centering
        \includegraphics[angle=0, width=\textwidth]{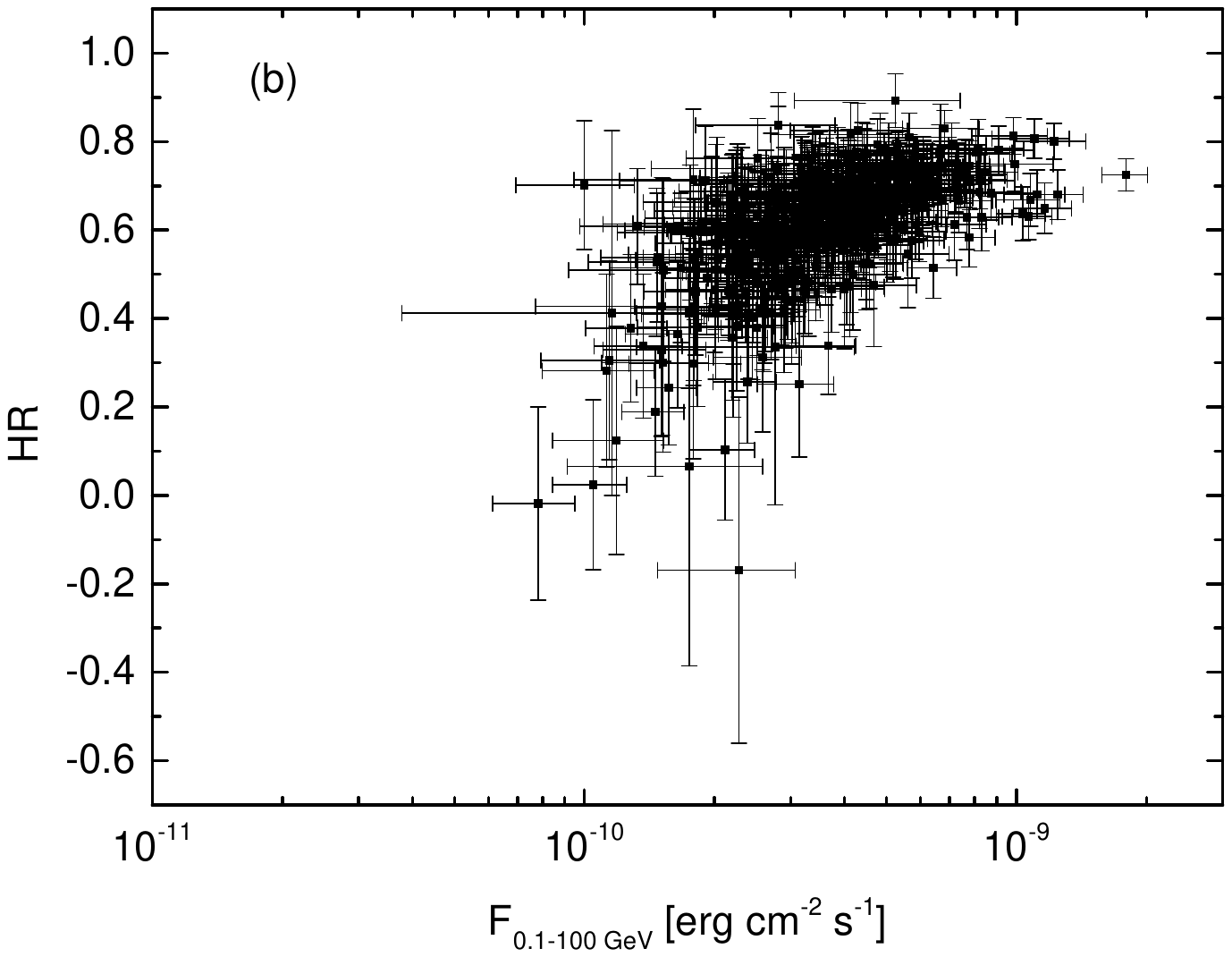} 
    \end{minipage}
    \begin{minipage}{0.45\textwidth}
        \centering
        \includegraphics[angle=0, width=\textwidth]{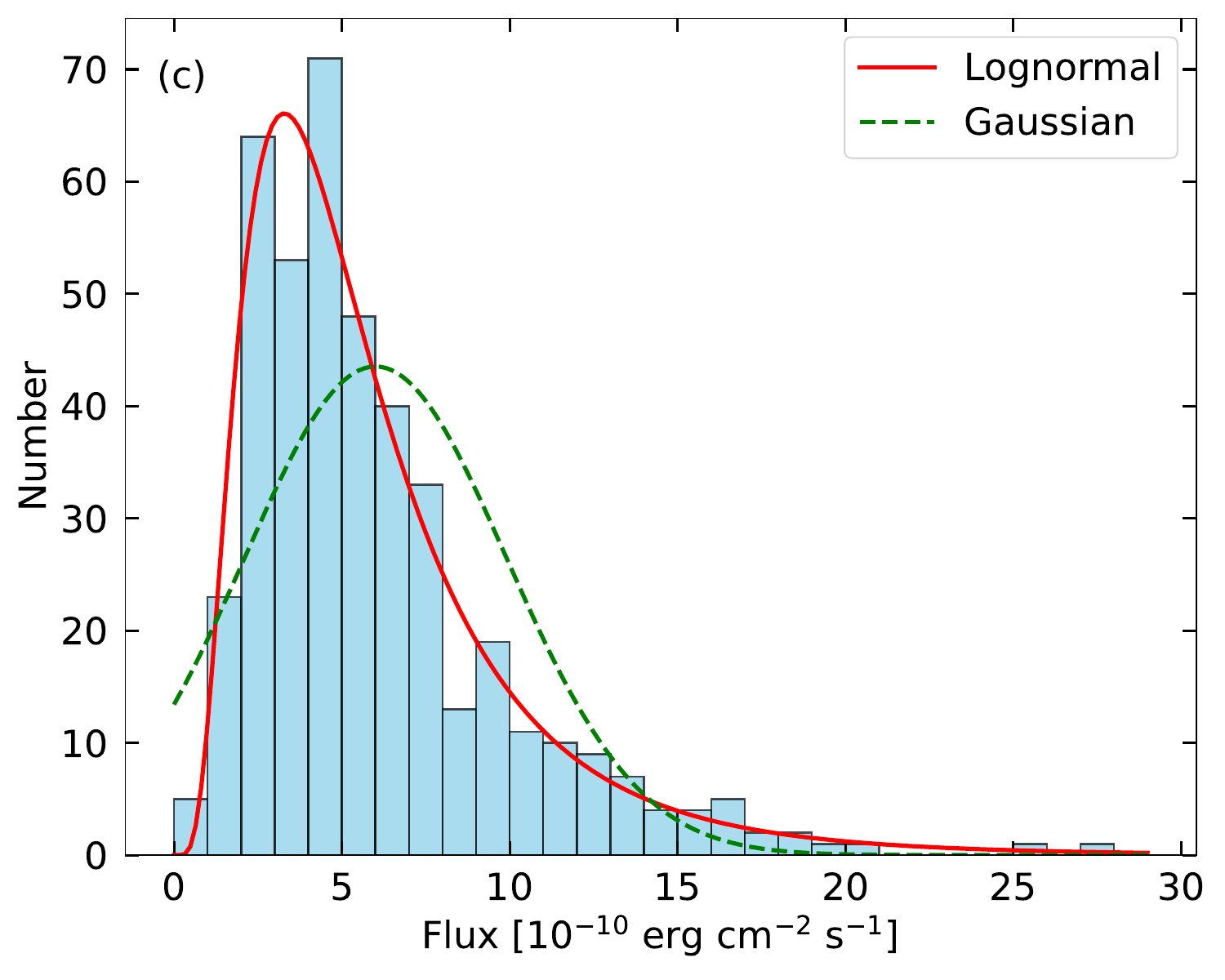} 
    \end{minipage}
    \begin{minipage}{0.45\textwidth}
        \centering
        \includegraphics[angle=0, width=\textwidth]{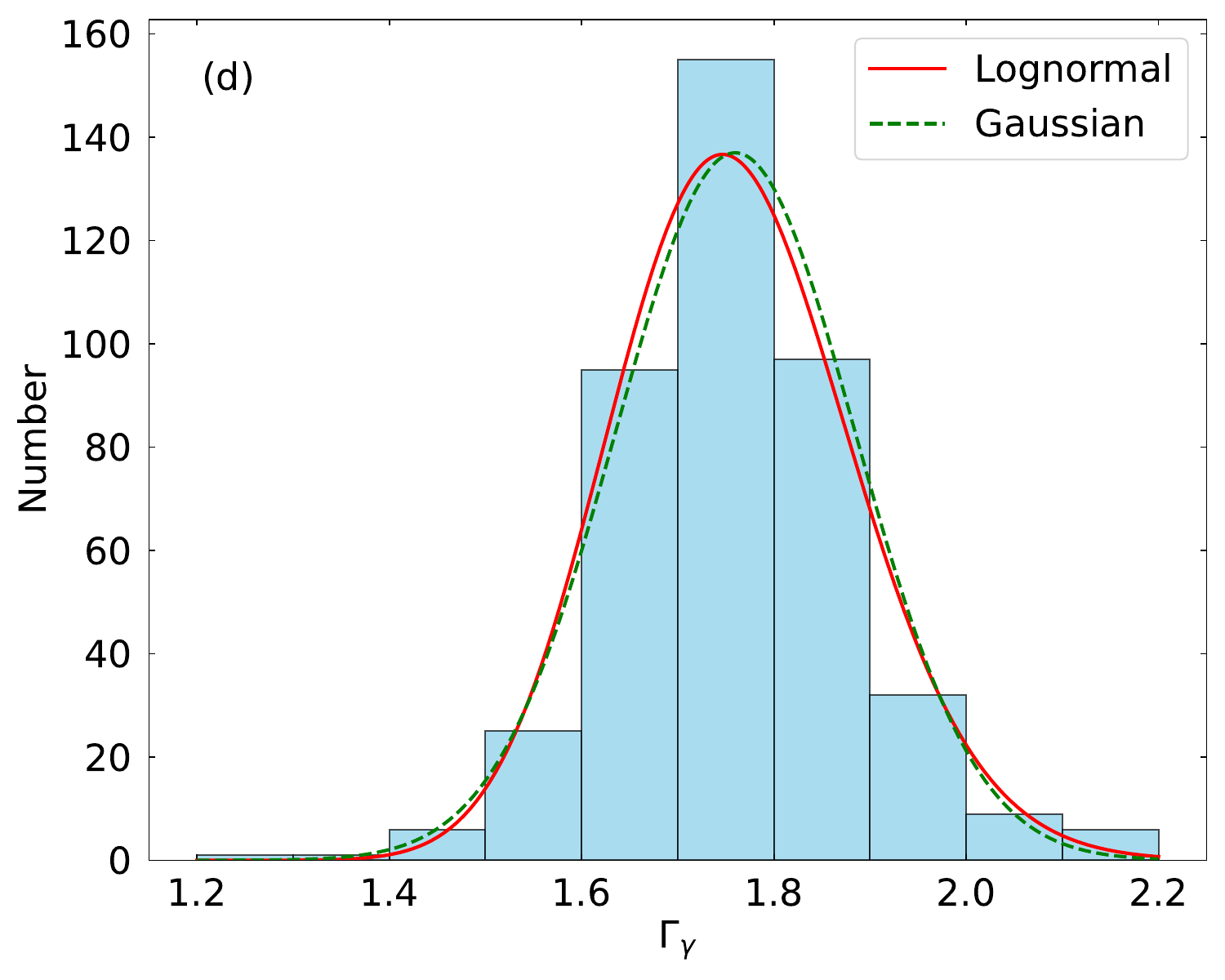} 
    \end{minipage}
    \caption{Panel (a): the 17-year integrated spectrum of Mrk 421 obtained by Fermi-LAT observations in the 0.1--1000 GeV band. The black solid line and gray dashed lines represent the spectral fitting result and the corresponding $1\sigma$ uncertainties, respectively. Panel (b): HR as a function of $F_{0.1-100~{\rm GeV}}$, where ${\rm HR}=\frac{F_{1-100~{\rm GeV}}-F_{0.1-1~{\rm GeV}}}{F_{1-100~{\rm GeV}}+F_{0.1-1~{\rm GeV}}}$ and $F_{0.1-100~{\rm GeV}}=F_{0.1-1~{\rm GeV}}+F_{1-100~{\rm GeV}}$. The values of $F_{0.1-1~{\rm GeV}}$ and $F_{1-100~{\rm GeV}}$ are derived using a 14-day time bin, consistent with the data in panel (b) and panel (c) of Figure \ref{lc}. Panels (c) and (d): the histogram distributions of the flux and $\Gamma_{\gamma}$ with the lognormal (red solid lines) and Gaussian (green dashed lines) fits, respectively. The values of the flux and $\Gamma_{\gamma}$ are derived from the 17-year Fermi-LAT observational data within the 0.1--1000 GeV band, using a 14-day time bin, consistent with the data in panel (a) and panel (e) of Figure \ref{lc}.}
    \label{Spec_17y}
\end{figure*}

\begin{figure*}
    \centering
    \includegraphics[angle=0, scale=0.75]{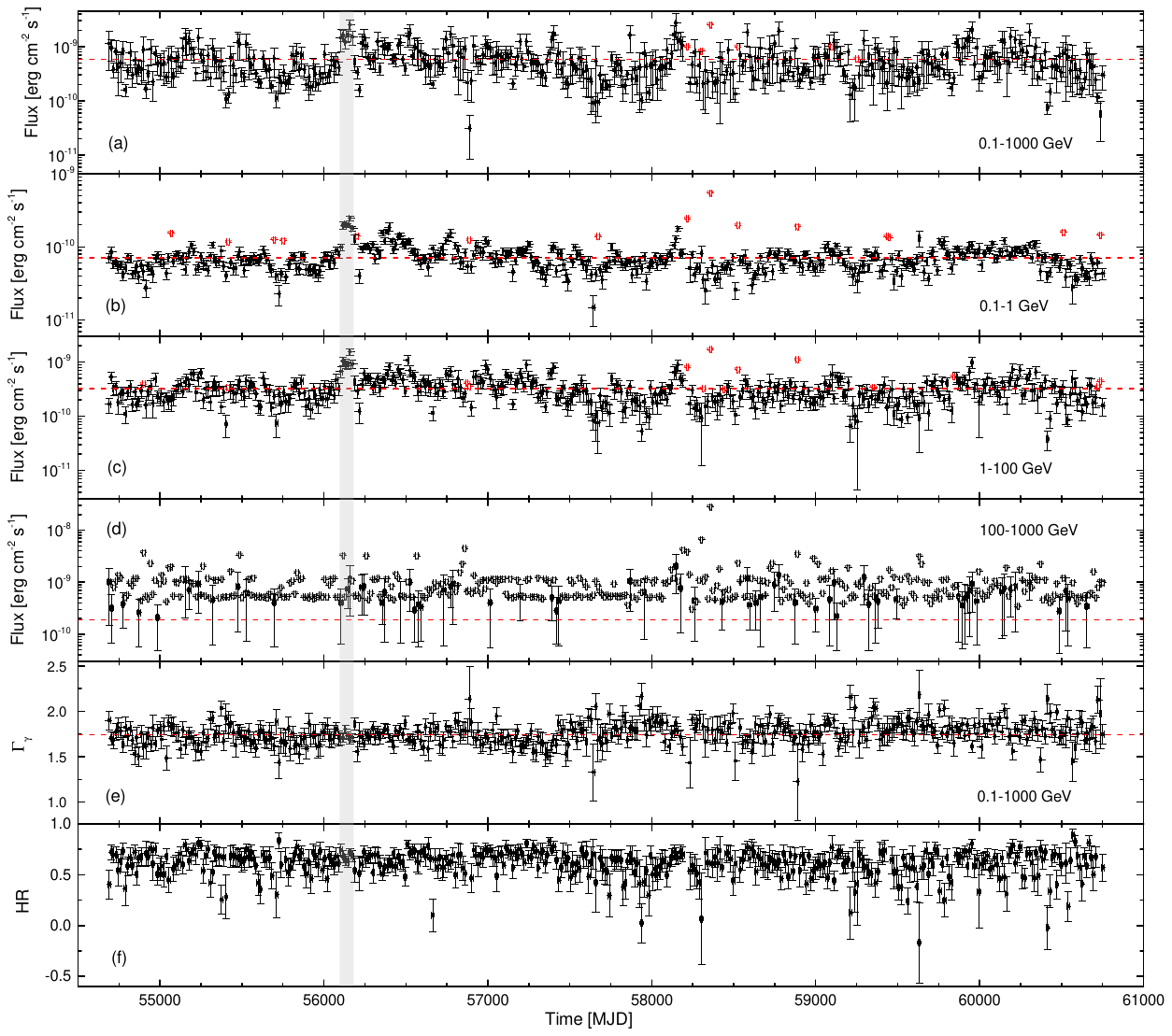}
    \caption{The 17-year long-term light curves, derived with a 14-day time bin, in the 0.1--1000 GeV band (panel (a)), 0.1--1 GeV band (panel (b)), 1--100 GeV band (panel (c)), 100--1000 GeV band (panel (d)), as well as the curves of $\Gamma_{\gamma}$ (panel (e)) and HR (panel (f)). The red horizontal dashed lines represent the average values. If TS $<$ 9, an upper limit of flux (red or black open triangles) is provided for that time bin. The gray shaded area indicates the time interval during which the light curve on a daily timescale is derived, as depicted in Figure \ref{lc_day}.}
    \label{lc}
\end{figure*}

\begin{figure*}
    \centering
    \includegraphics[angle=0, width=1\textwidth]{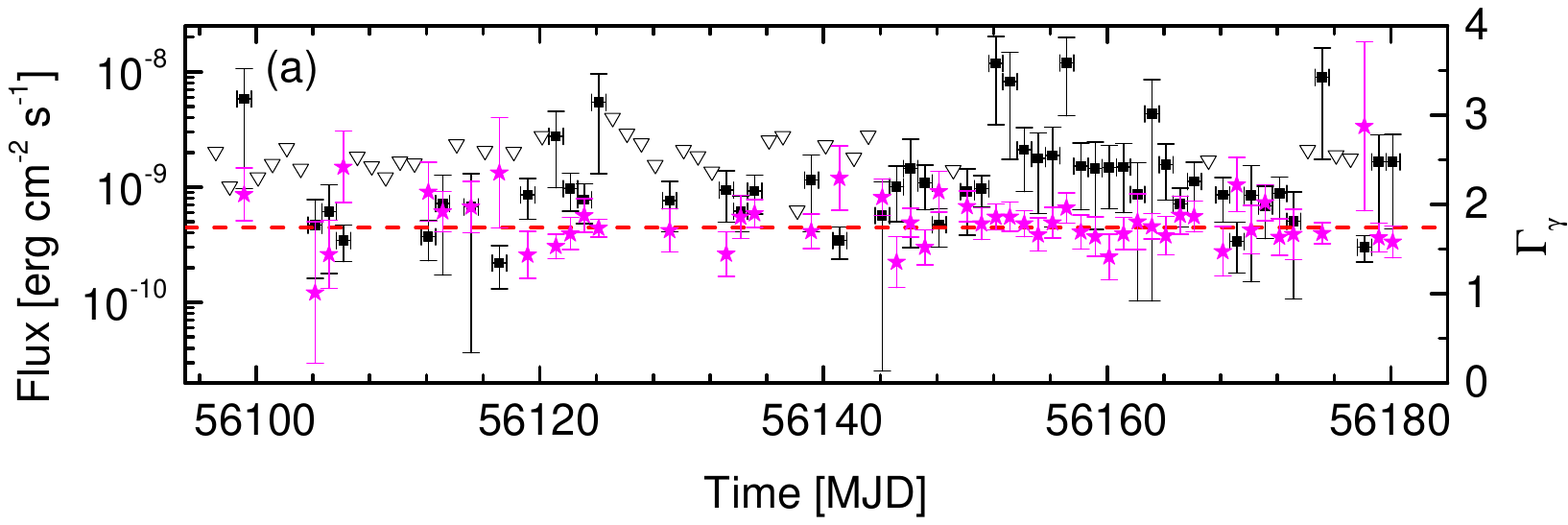}
    \includegraphics[angle=0, width=0.4\textwidth]{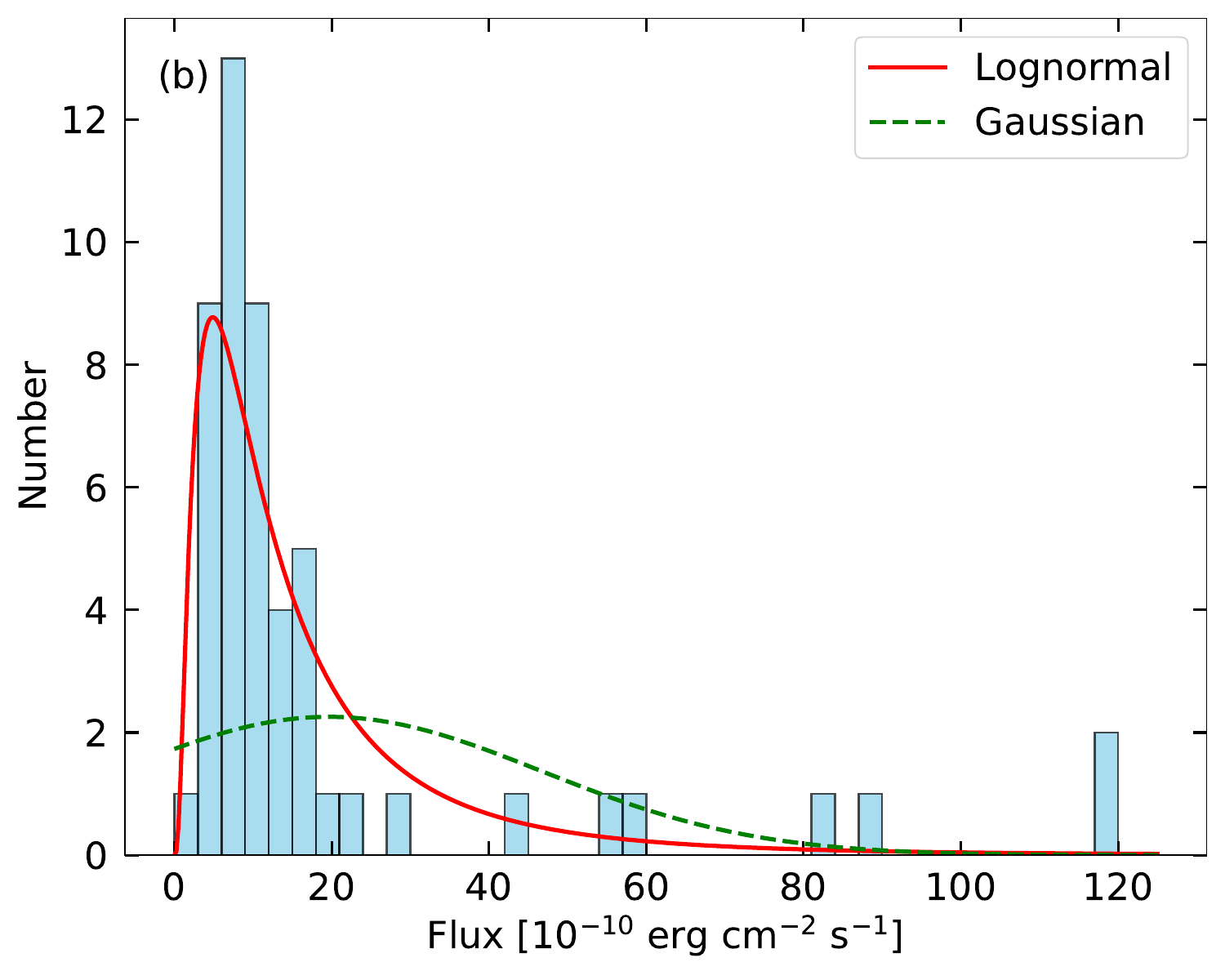} 
    \includegraphics[angle=0, width=0.4\textwidth]{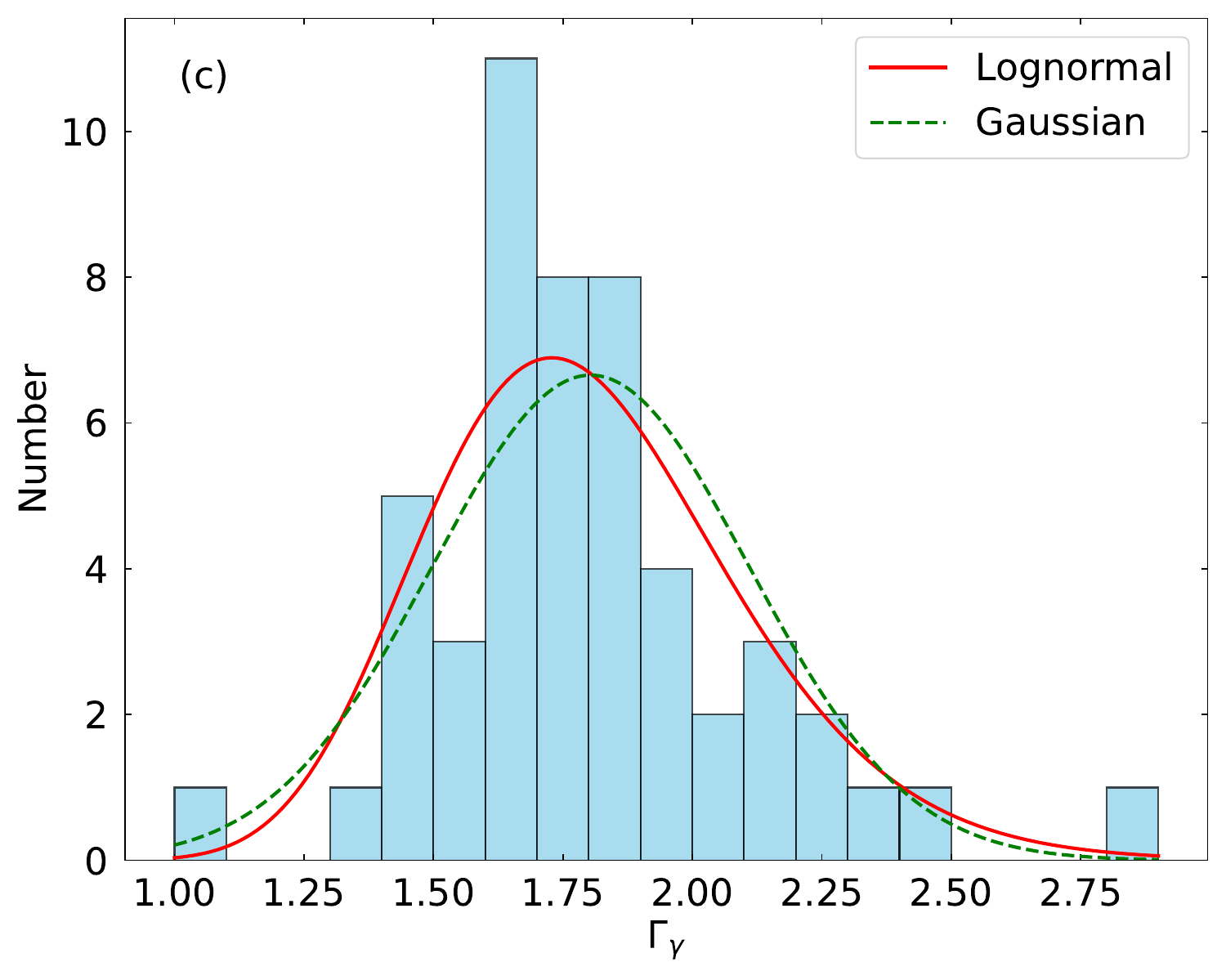} 
    \caption{Panel (a): the light curve (black solid squares) derived with a 1-day time bin in the 0.1--1000 GeV band, along with the curves of $\Gamma_{\gamma}$ (magenta solid stars). The red horizontal dashed line represents the weighted average flux. If TS $<$ 9, an upper limit of flux (black open triangles) is provided for that time bin. Panels (b) and (c): the histogram distributions of the flux and $\Gamma_{\gamma}$ with the lognormal (red solid lines) and Gaussian (green dashed lines) fits, respectively. The values of flux and $\Gamma_{\gamma}$ are consistent with the data presented in panel (a).}
    \label{lc_day}
\end{figure*}

Additionally, the X-ray spectra of Mrk 421 generally display a LP shape. The peak energy is strongly correlated with the peak flux and the spectral curvature parameter, and these relations are consistent with the statistical or stochastic acceleration of relativistic electrons (e.g., \citealt{2007A&A...466..521T, 2009A&A...501..879T}). Although a \textit{harder-when-brighter} tendency is observed for Mrk 421 in the GeV band, a large scatter in the correlation between flux and HR is presented, as illustrated in Figure \ref{Spec_17y}. This loose correlation may be attributed to the fact that the evolutions of the flares differ from one another due to different variability factors, such as different high-energy particle injection rates, particle acceleration rates, and acceleration mechanisms.

As displayed in Figures \ref{Spec_17y} and \ref{lc_day}, the flux demonstrates a lognormal distribution, whereas both the lognormal and Gaussian functions yield comparably good fits to the photon spectral index distributions. A lognormal distribution of Fermi-LAT flux has been observed in numerous bright blazars (e.g., \citealt{2018RAA....18..141S, 2020ApJ...891..120B}). The lognormal flux distribution was first observed in the X-ray emission of the black hole binary Cygnus X-1 \citep{2001MNRAS.323L..26U}, and this characteristic is attributed to the multiplicative processes from the accretion disc \citep{1997MNRAS.292..679L, 2005MNRAS.359..345U, 2010LNP...794..203M}. Although it is well known that the non-thermal emission from the jet dominates the electromagnetic spectrum of blazars, the coupling between the jet and the accretion disk \citep{1999ApJ...525..909A, 2002MNRAS.332..165M, 2004ApJ...615L...9W, 2004ApJ...613..716C, 2014ApJ...783...51C, 2014ARA&A..52..529Y} may cause the fluctuations from the disk to propagate through the relativistic jets and affect the jet processes. Therefore, the flux distribution in blazars could be an imprint of the fluctuations in the disk and an indication of a disk--jet connection (e.g., \citealt{2020ApJ...891..120B}). However, this model cannot account for the variability of blazars on a short timescale in $\gamma$-rays \citep{2018Galax...6..135R, 2018MNRAS.480L.116S}. \citet{2018MNRAS.480L.116S} reported that linear Gaussian variations of the intrinsic particle acceleration or escape timescales can produce a lognormal flux distribution, and a lognormal flux distribution with a Gaussian spectral index distribution could be attributed to the fluctuations in the
acceleration rate of particles (see also \citealt{Kapanadze:2025rer}). The distributions in Figures \ref{Spec_17y} and \ref{lc_day} seem to be consistent with this explanation. Nevertheless, the $\gamma$-ray variability in blazars could stem from a combination of numerous complex factors, in both the disk and the jet (e.g., \citealt{2020ApJ...891..120B}). 

The broadband SEDs of BL Lacs are generally explained by the one-zone leptonic model \citep{1992ApJ...397L...5M, 2010MNRAS.402..497G, 2010MNRAS.401.1570T, 2012ApJ...752..157Z, 2014MNRAS.439.2933Y}, namely, their GeV--TeV spectra, including those of Mrk 421, are produced by the SSC process of relativistic electrons within jets \citep{2010MNRAS.401.1570T, 2012ApJ...752..157Z, 2014MNRAS.439.2933Y}. However, some of the derived 18 simultaneously observed GeV--TeV spectra, as depicted in Figure \ref{Spe_GeV-TeV}, indicate that a single SSC component is insufficient to characterize their spectral shapes. \citet{2005ApJ...630..130B} have reported that more complex multi-zone models are often required to accurately reproduce the observed broadband emission when the source exhibit high-flux activity in the X-ray and/or TeV $\gamma$-ray bands. Recently, \citet{2025A&A...695A.217M} used a two-zone model, in which a compact zone overlaps an extended zone, to reproduce the observed broadband SEDs of Mrk 421. Specifically, the GeV and TeV spectra are respectively dominated by two different SSC components originating from the two zones. In addition, the hybrid leptonic-hadronic jet model has also been proposed to account for the TeV excess of the SED in Mrk 421 \citep{2022ApJ...925L..19C}. Moreover, taking into account the contribution of the proton radiation process can also account for the possible neutrino emission from Mrk 421 \citep{2015MNRAS.448.2412P, 2021MNRAS.501.2198R}. Given the significant flux and spectral variations in the 17-year observational light curves, we propose that each flux variability event in Mrk 421 may involve different physical processes, such as particle injection or acceleration, dissipation and propagation within the emission region, and radiation mechanisms.

\section{Summary}\label{sec:sum}

Mrk 421 is a typical $\gamma$-ray emitting BL Lac, and it has amassed over 17 years of Fermi-LAT observational data, along with a substantial amount of observational data in the VHE band. In this paper, we conducted a comprehensive analysis of the 17-year Fermi-LAT observational data of Mrk 421, in conjunction with some archived TeV data, to explore its spectral variation characteristics in the $\gamma$-ray band and the potential physical causes. The main findings are summarized as follows.

\begin{itemize}

  \item The $\gamma$-ray emission of Mrk 421 in the GeV band exhibits significant variability in the two-week binned light curve, with a maximum-to-minimum flux ratio of nearly two orders of magnitude. By conducting an analysis of the daily-timescale light curve during a time interval when the source was in a high-flux state, we found that the flux can vary by up to one order of magnitude within a single day.

  \item The 17-year integrated spectrum of Mrk 421 requires a complex model for explanation, whereas the PL model suffices to describe its time-resolved spectra over one-day or several-day time intervals. 

  \item A significant spectral variation is observed for Mrk 421, with a \textit{harder-when-brighter} trend in the GeV band. It seems that the complex spectral shape of the 17-year integrated spectrum stems from the superposition of different spectral shapes in different flux states.  

  \item The fluxes in both 14-day and 1-day time bins follow a lognormal distribution. Meanwhile, the photon spectral index distributions can be well fitted by either a lognormal or a Gaussian function. These phenomena may be associated with fluctuations in the particle acceleration rate, or with a combination of numerous complex physical processes. 

  \item The majority of the simultaneously observed GeV--TeV spectra constructed in this work for the source can be fitted by a LP model, indicating that they might originate from the same SSC process. However, some combined GeV--TeV spectral shapes clearly suggest different radiation origins for the GeV and TeV emission. 

\end{itemize}

%***********************************************************

\begin{acknowledgments}

This work is supported by the National Natural Science Foundation of China (grants 12022305 and 11973050).

\end{acknowledgments}

\clearpage

\bibliography{ref}{}
\bibliographystyle{aasjournalv7}

\clearpage
%%%%%%%%%%%%%%%%%%%%%%%%%%%%%%%%%%%%%%%%%%%%%%%%%%%%%%%%%%%%%%

\begin{figure*}
    \centering
    \begin{minipage}{0.3\textwidth}
        \centering
        \includegraphics[angle=0, width=\textwidth]{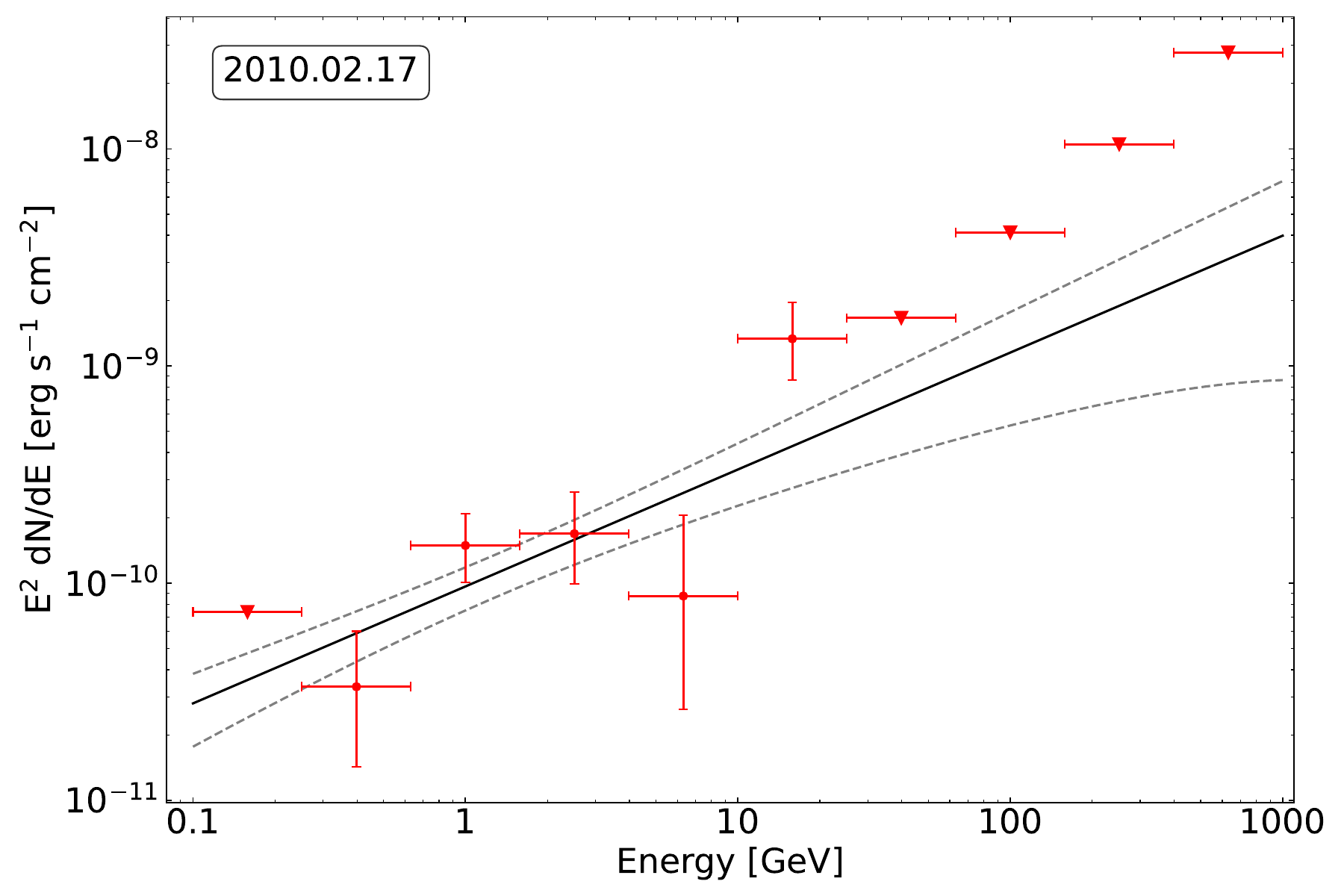} 
            % \makebox{(a)}
    \end{minipage} 
    \begin{minipage}{0.3\textwidth}
        \centering
        \includegraphics[angle=0, width=\textwidth]{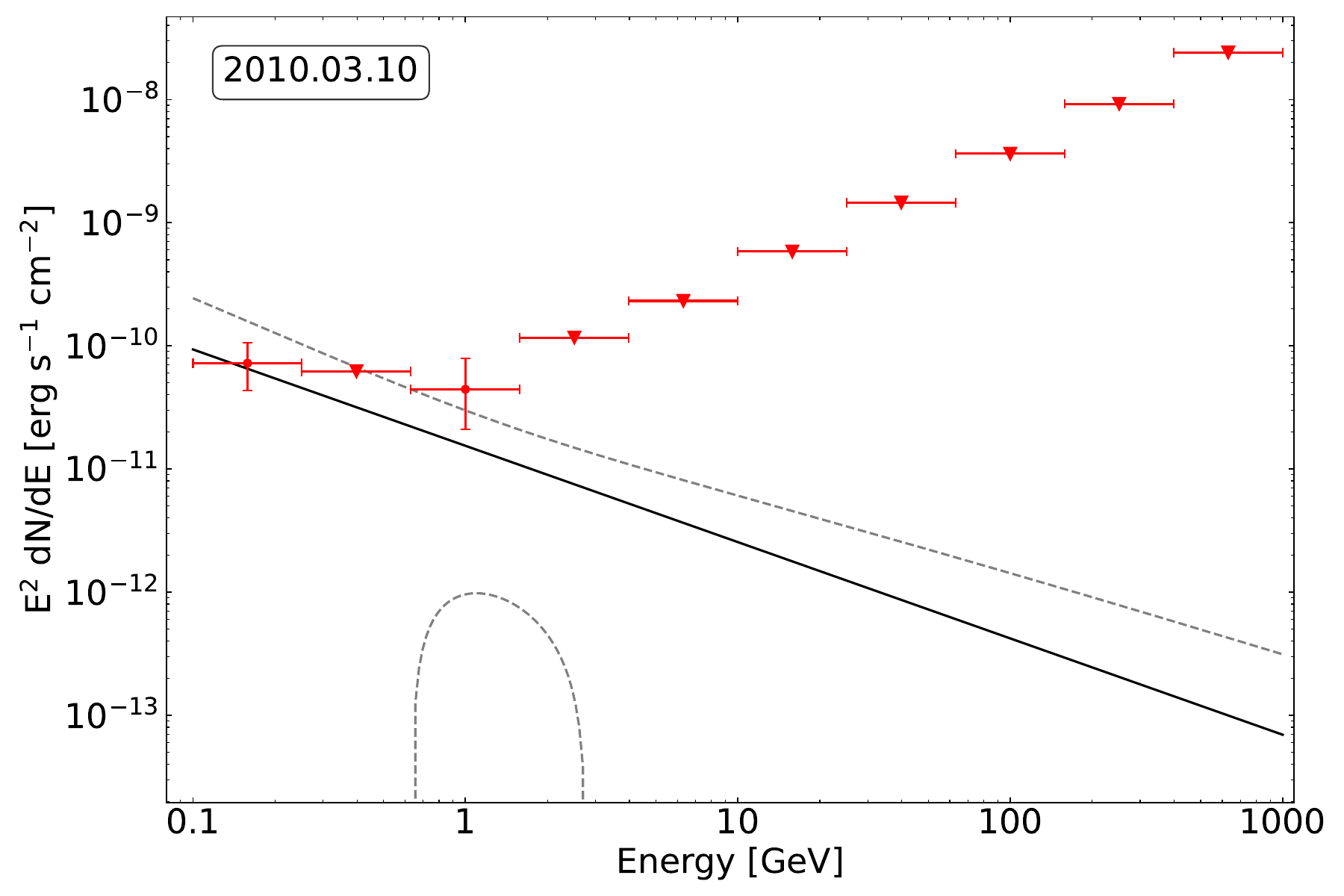} 
    \end{minipage} 
    \begin{minipage}{0.3\textwidth}
        \centering
        \includegraphics[angle=0, width=\textwidth]{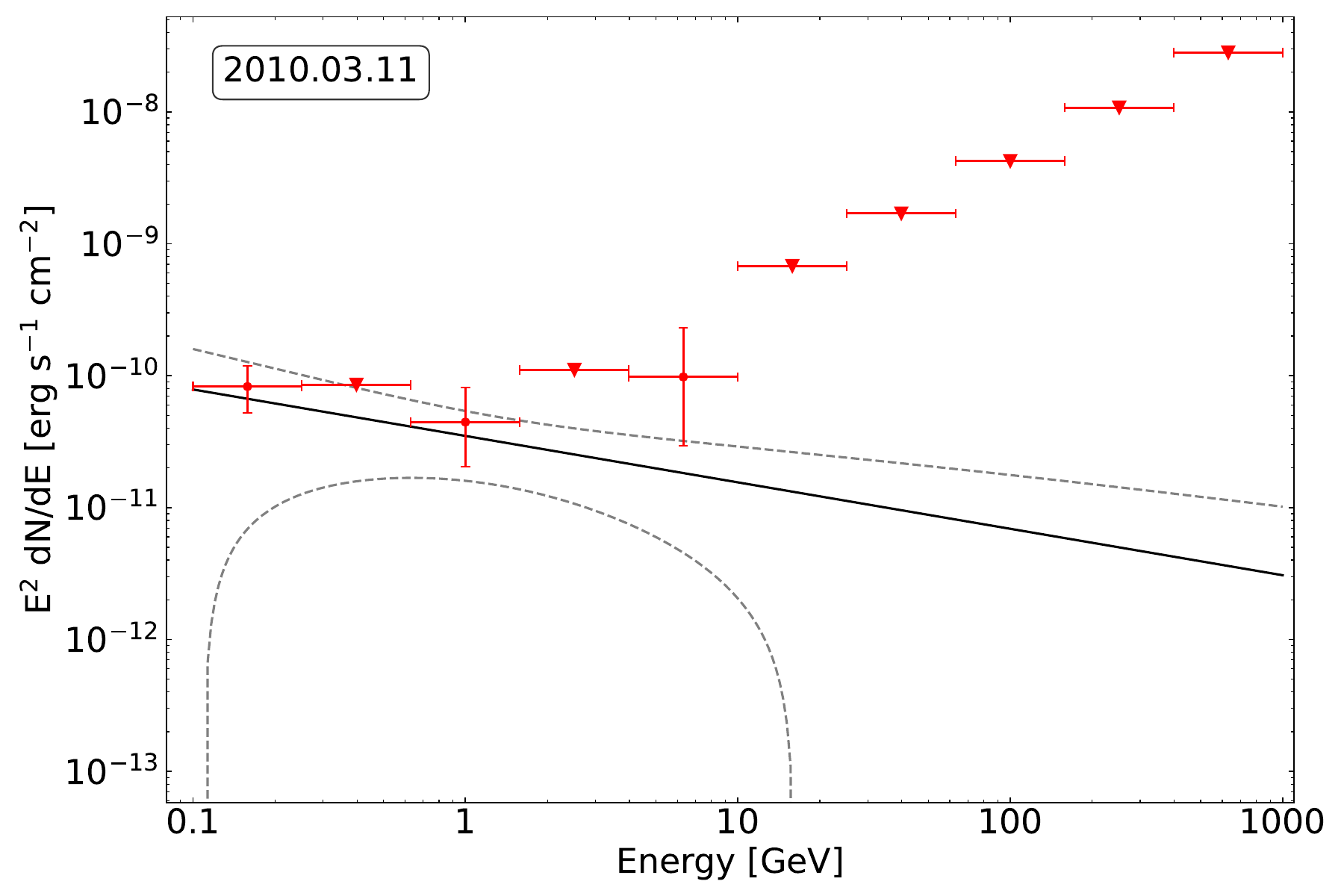} 
    \end{minipage}
    \begin{minipage}{0.3\textwidth}
        \centering
        \includegraphics[angle=0, width=\textwidth]{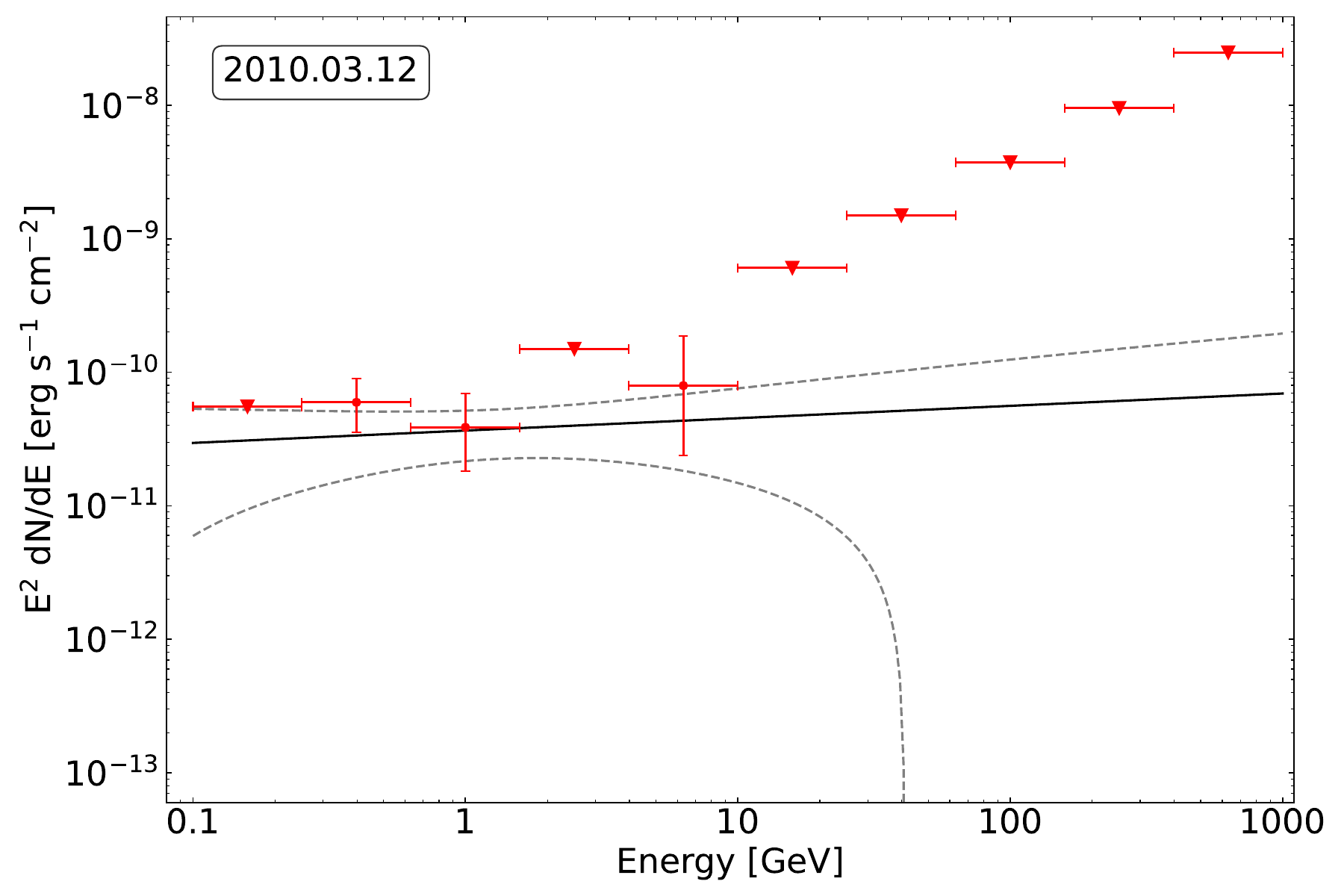} 
    \end{minipage}
    \begin{minipage}{0.3\textwidth}
        \centering
        \includegraphics[angle=0, width=\textwidth]{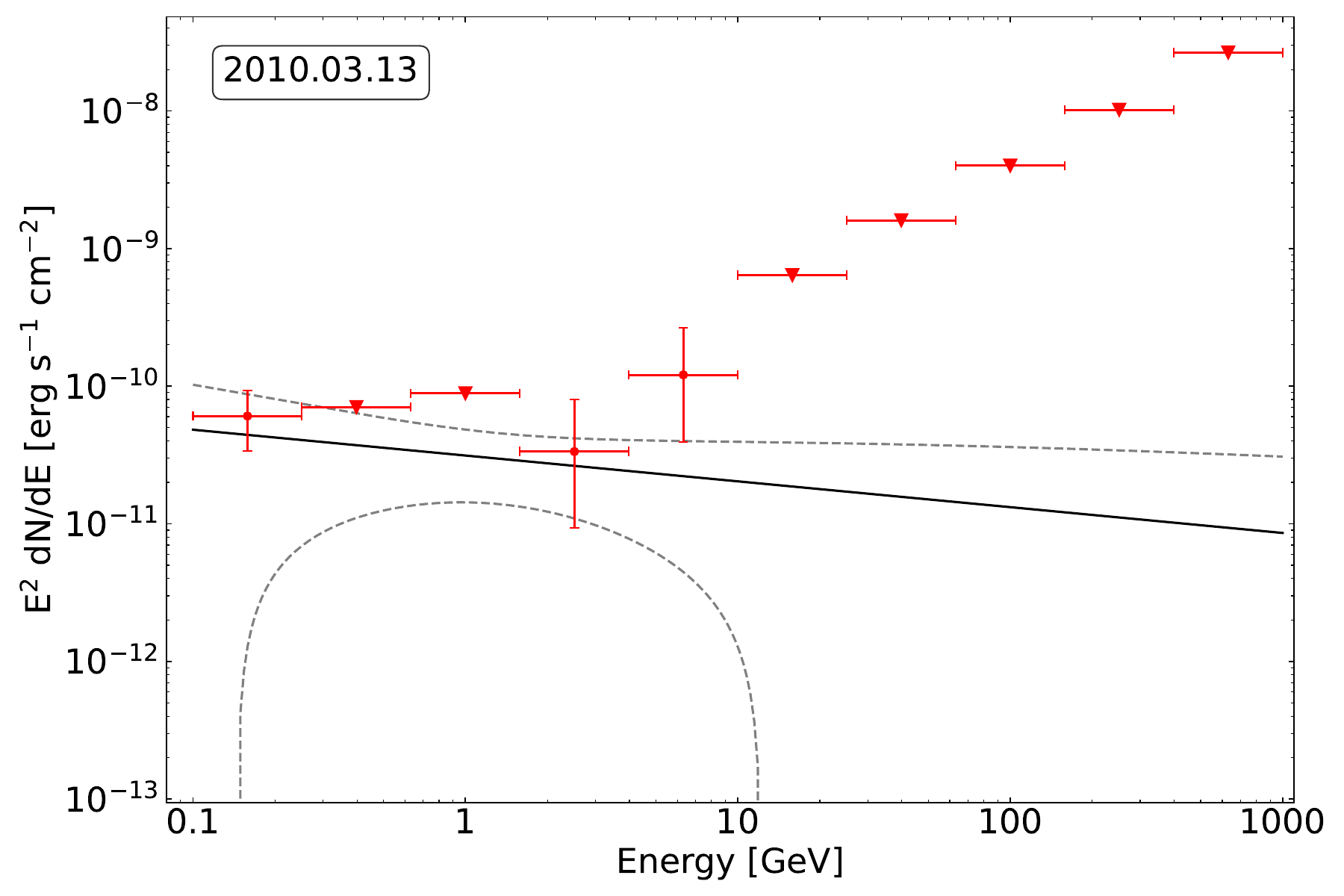} 
    \end{minipage}
    \begin{minipage}{0.3\textwidth}
        \centering
        \includegraphics[angle=0, width=\textwidth]{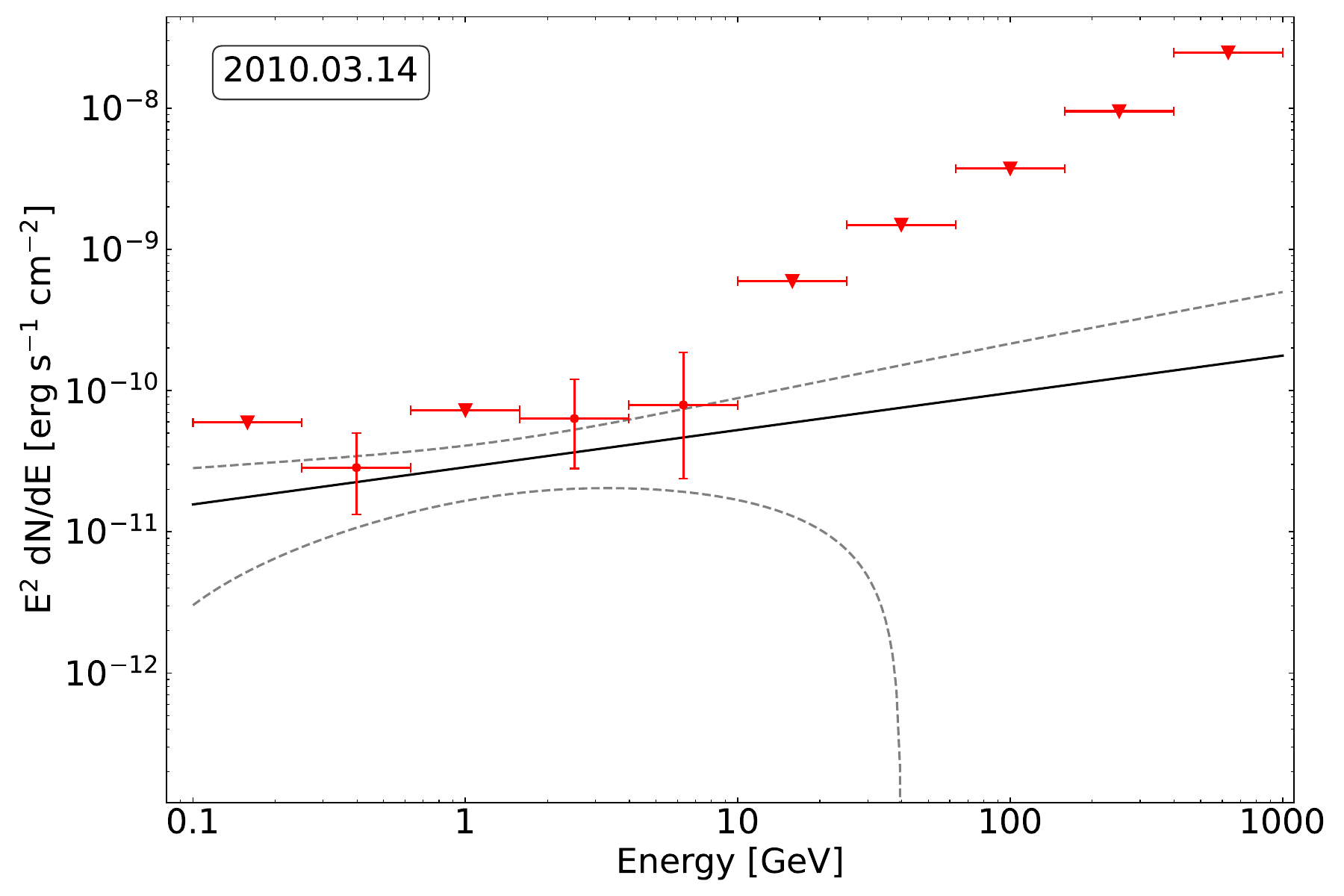} 
    \end{minipage}
    \begin{minipage}{0.3\textwidth}
        \centering
        \includegraphics[angle=0, width=\textwidth]{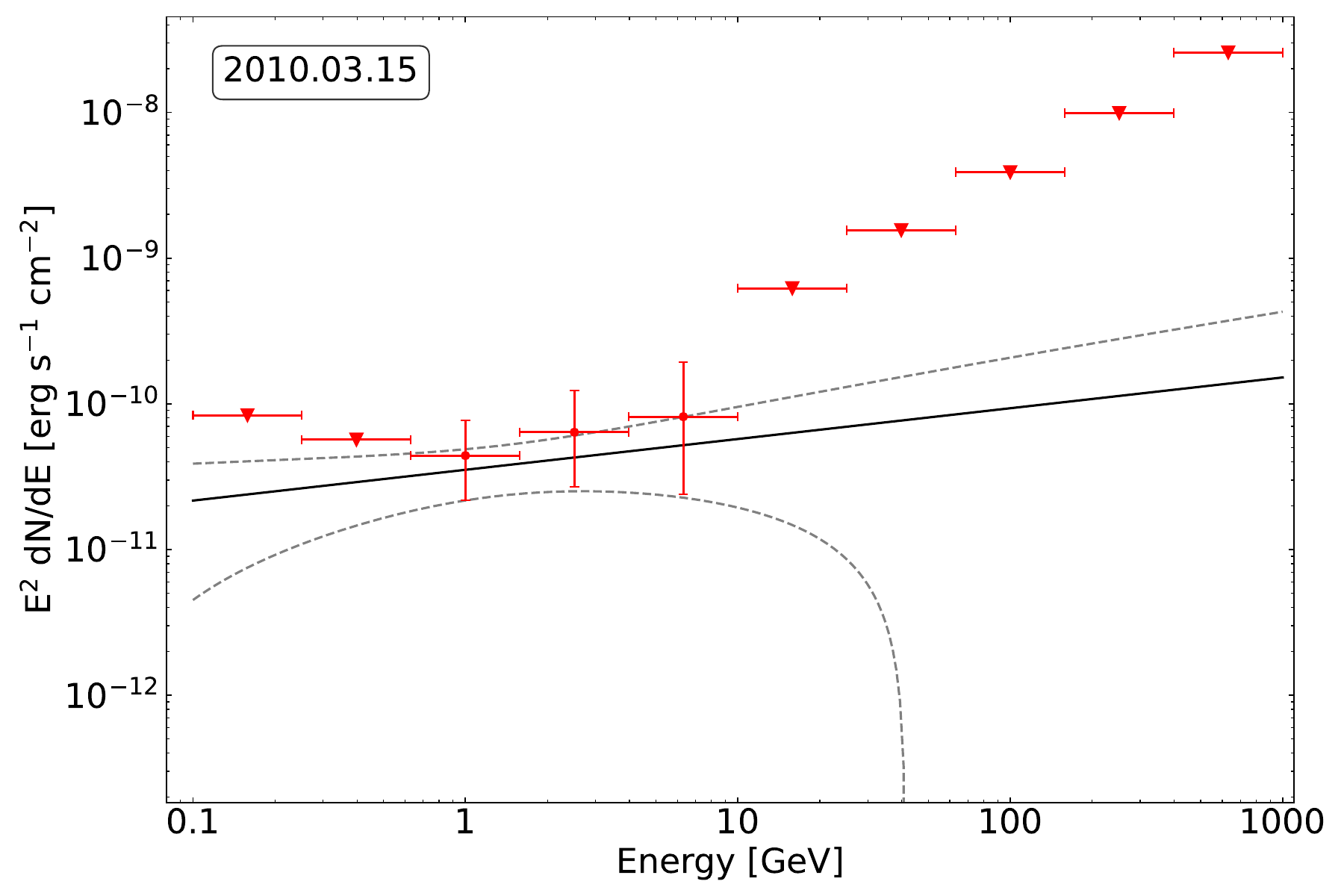} 
    \end{minipage}
    \begin{minipage}{0.3\textwidth}
        \centering
        \includegraphics[angle=0, width=\textwidth]{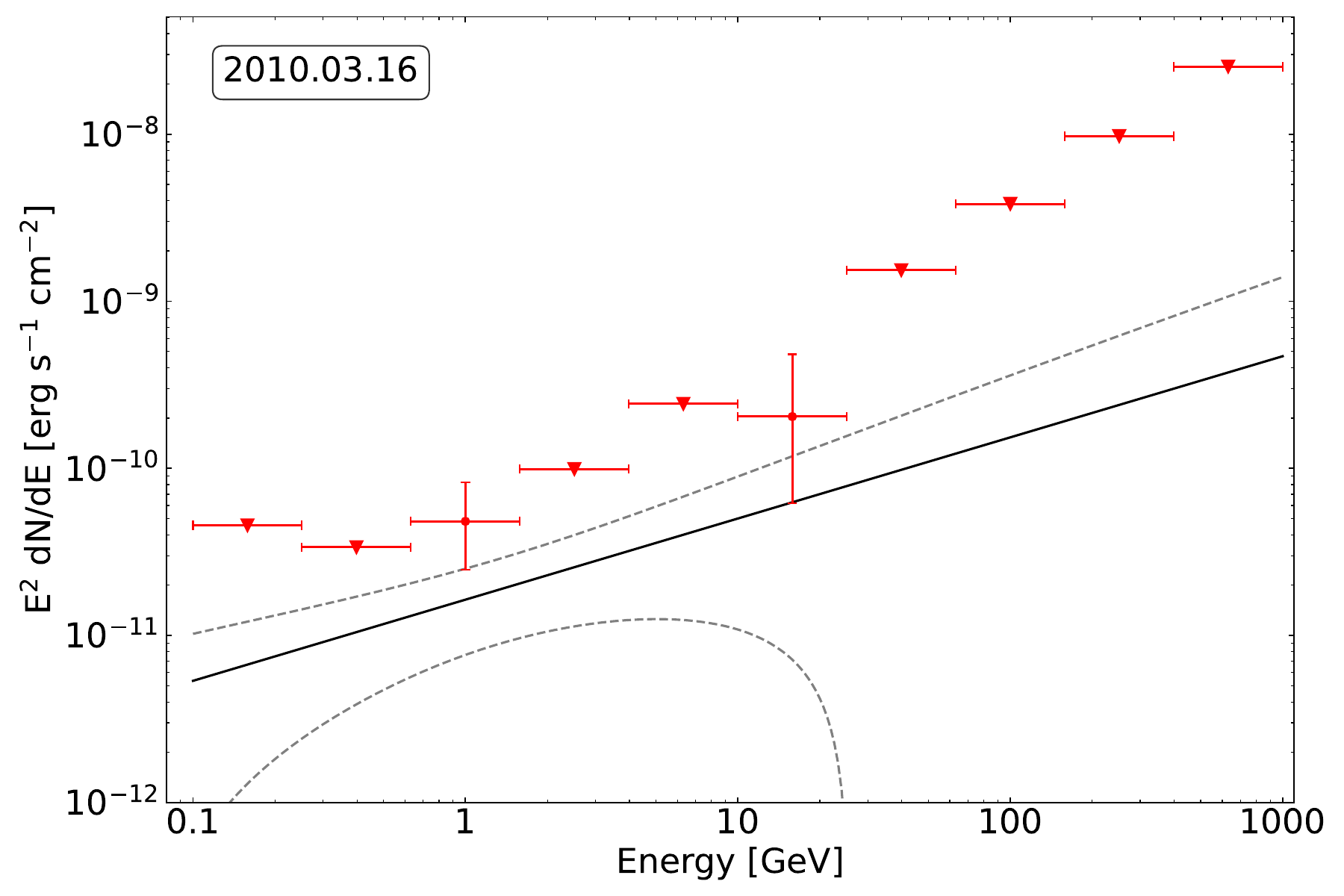} 
    \end{minipage}
    \begin{minipage}{0.3\textwidth}
        \centering
        \includegraphics[angle=0, width=\textwidth]{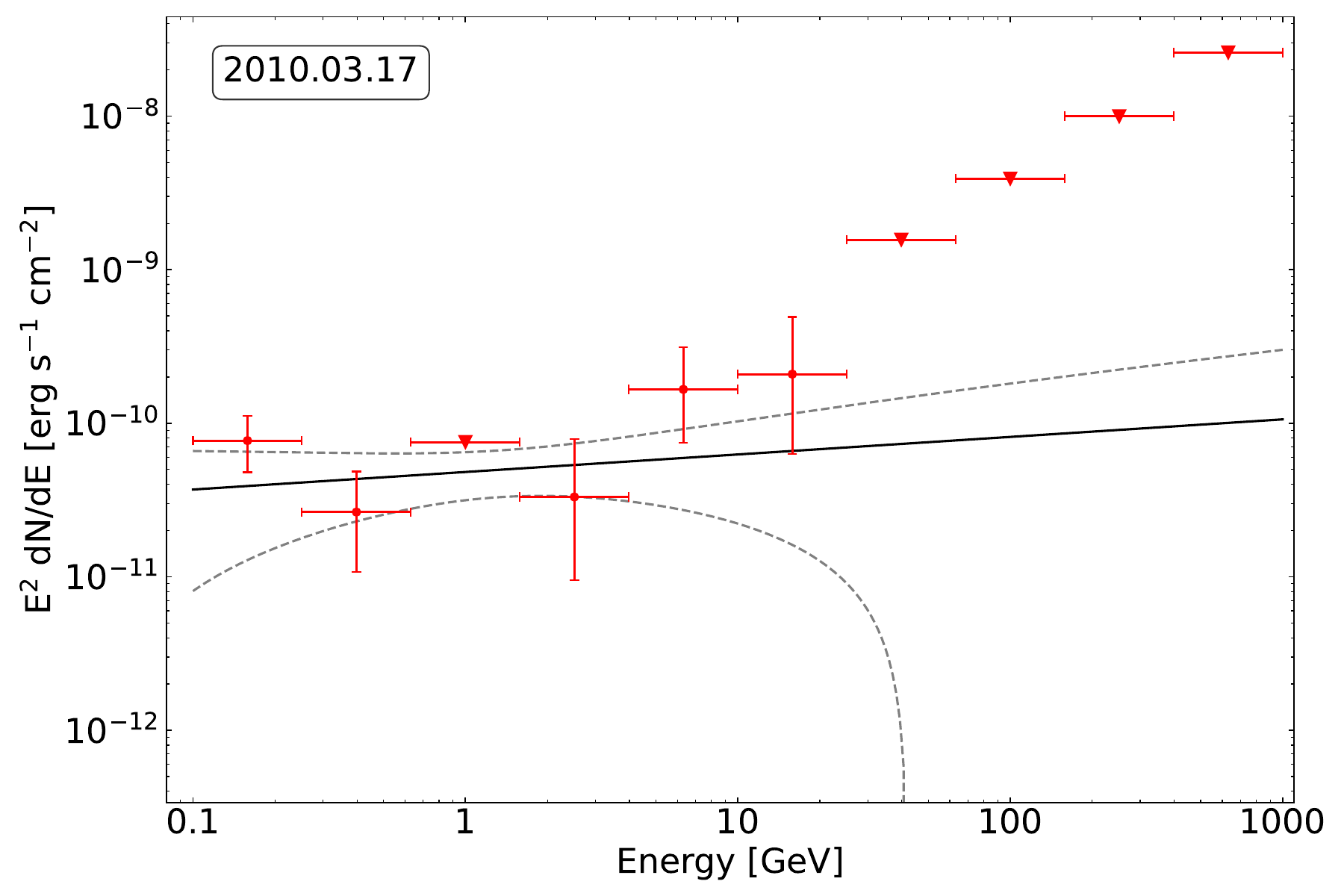} 
    \end{minipage}
    \begin{minipage}{0.3\textwidth}
        \centering
        \includegraphics[angle=0, width=\textwidth]{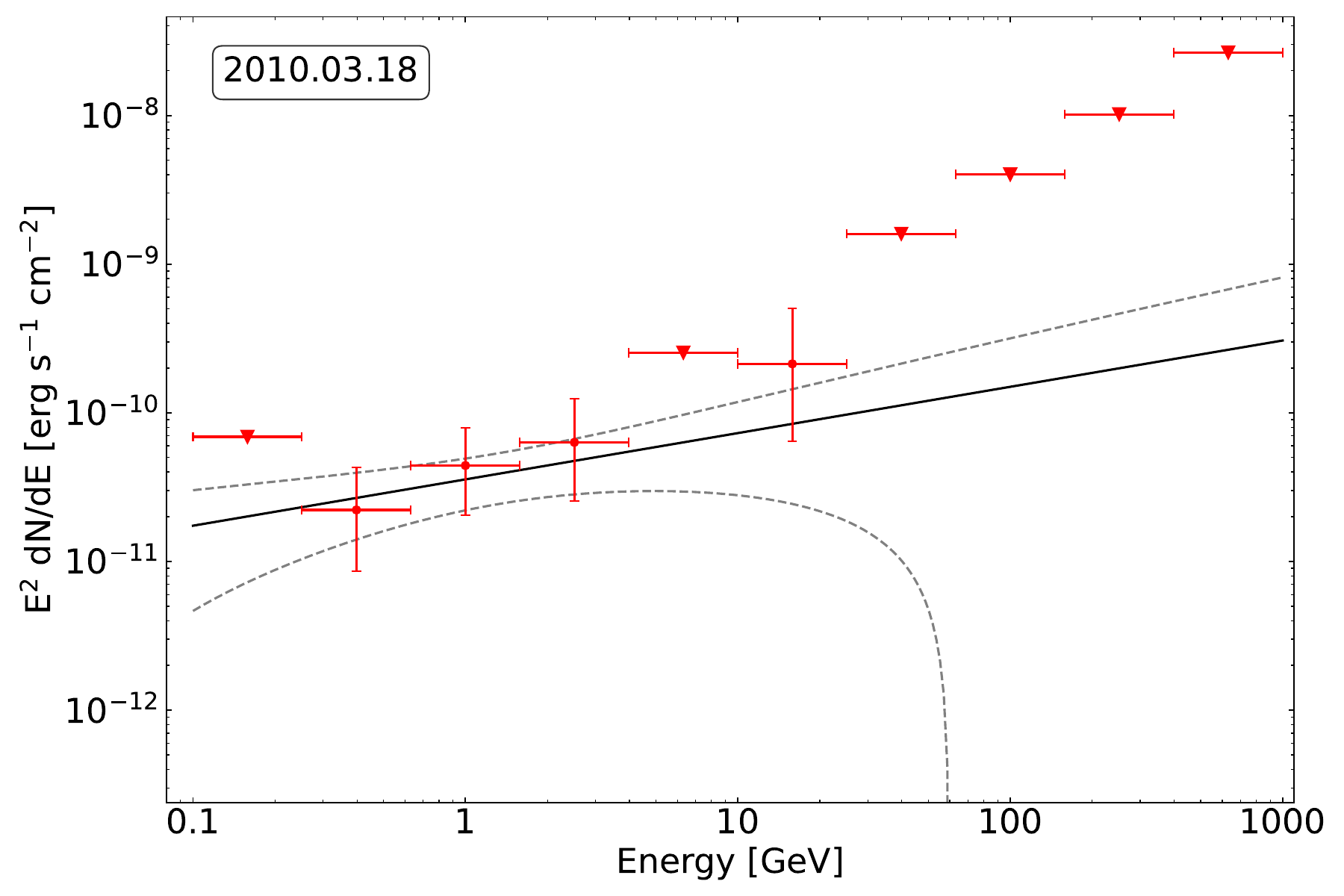} 
    \end{minipage}
    \begin{minipage}{0.3\textwidth}
        \centering
        \includegraphics[angle=0, width=\textwidth]{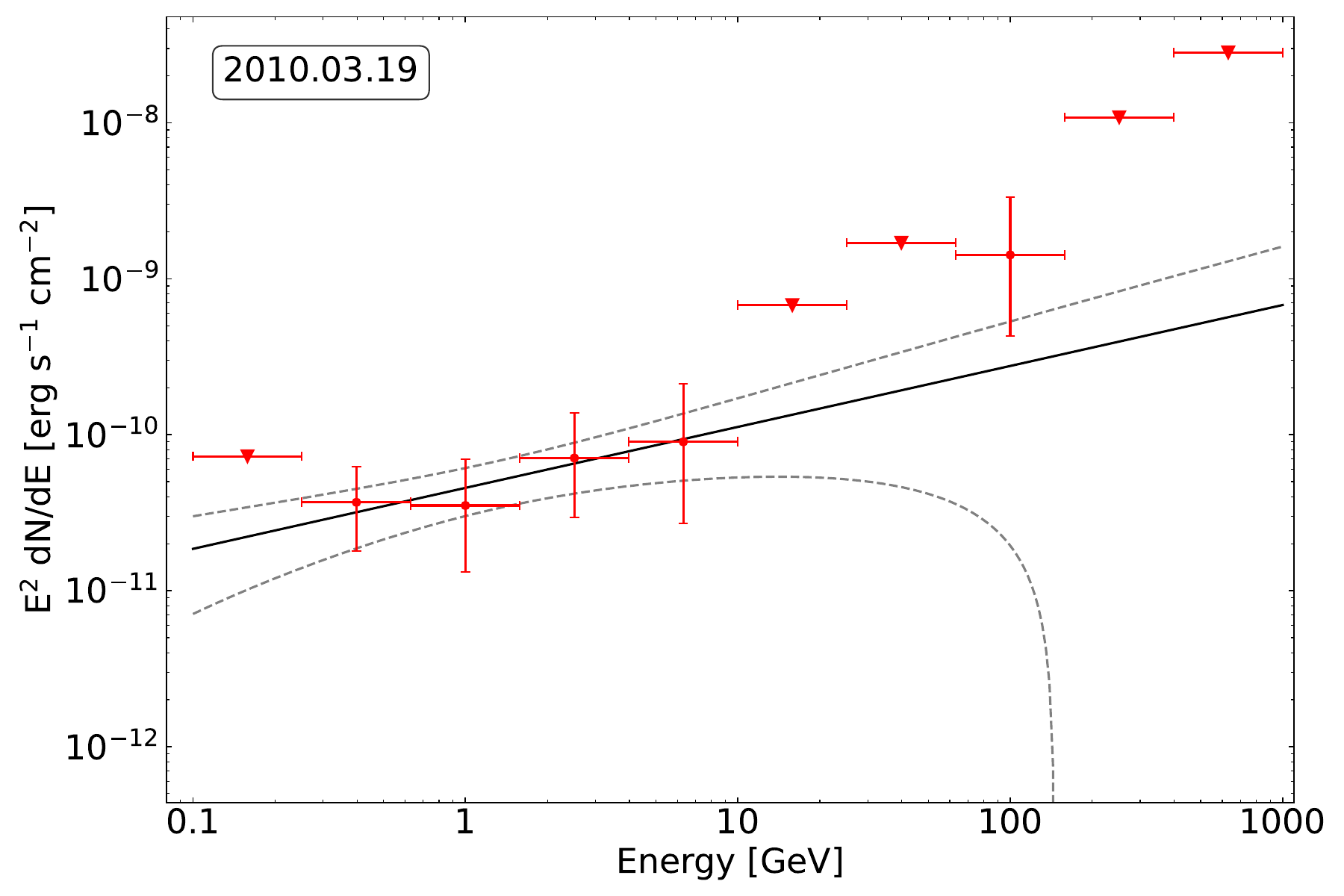} 
    \end{minipage}
    \begin{minipage}{0.3\textwidth}
        \centering
        \includegraphics[angle=0, width=\textwidth]{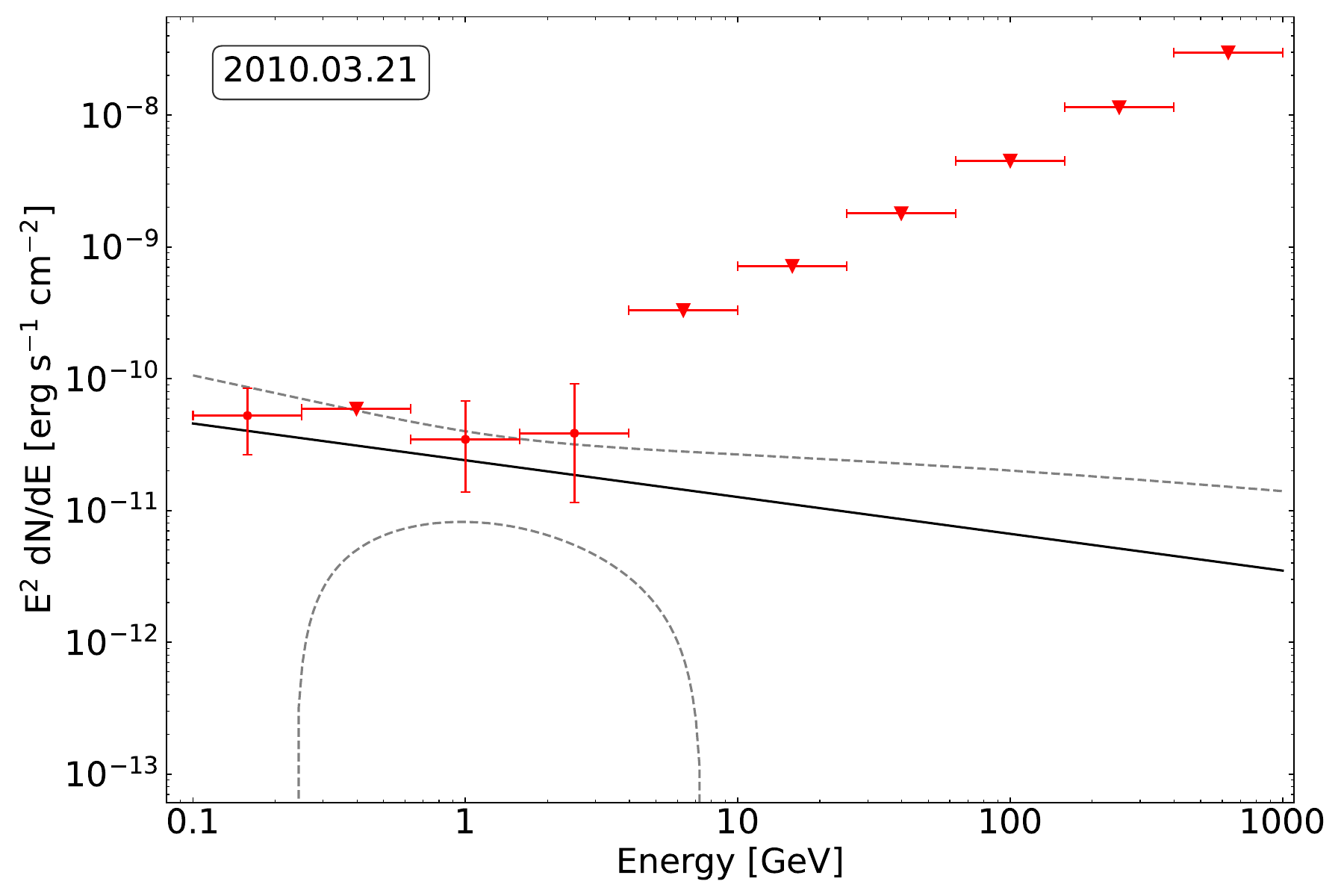} 
    \end{minipage}
    \begin{minipage}{0.3\textwidth}
        \centering
        \includegraphics[angle=0, width=\textwidth]{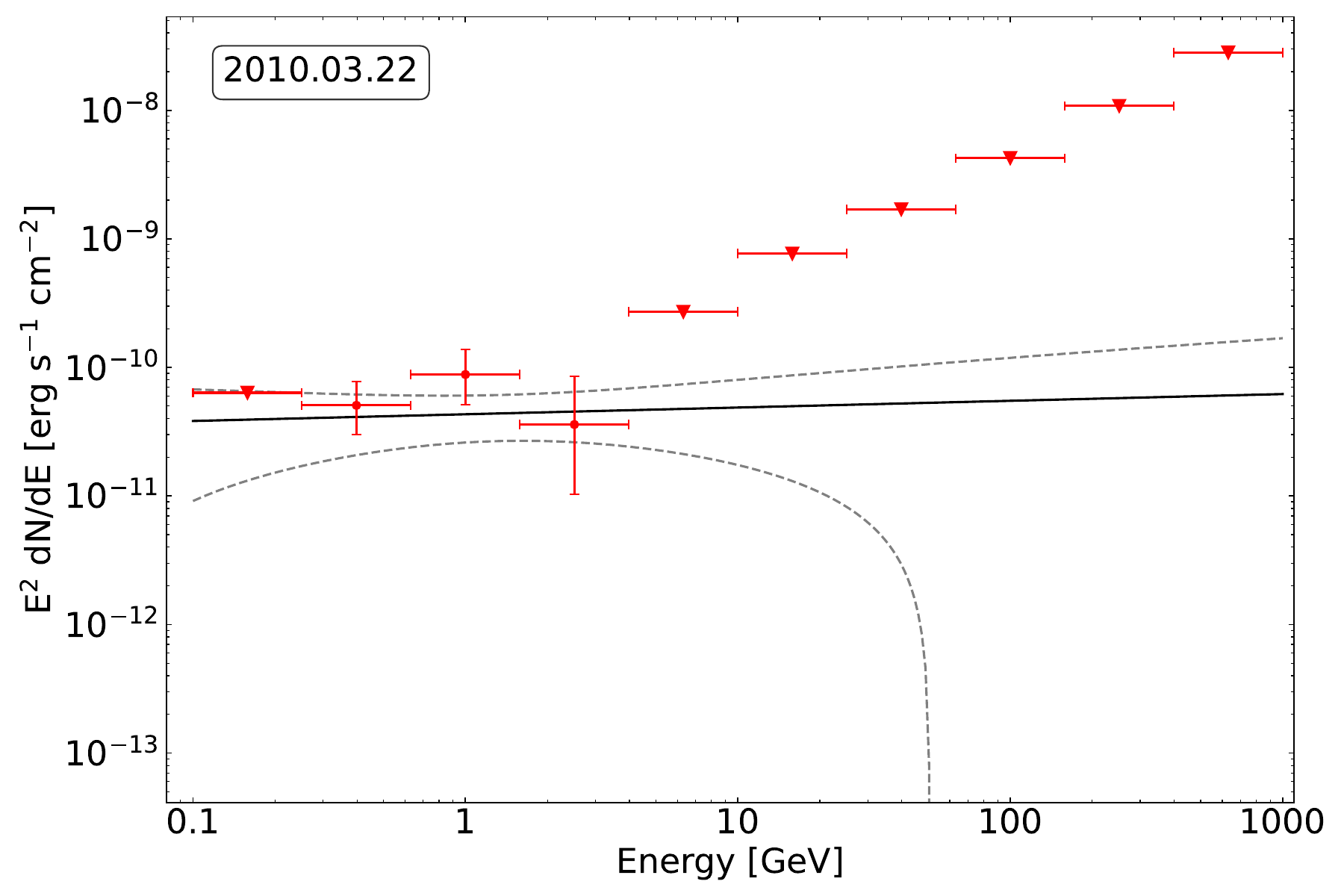} 
    \end{minipage}
    \begin{minipage}{0.3\textwidth}
        \centering
        \includegraphics[angle=0, width=\textwidth]{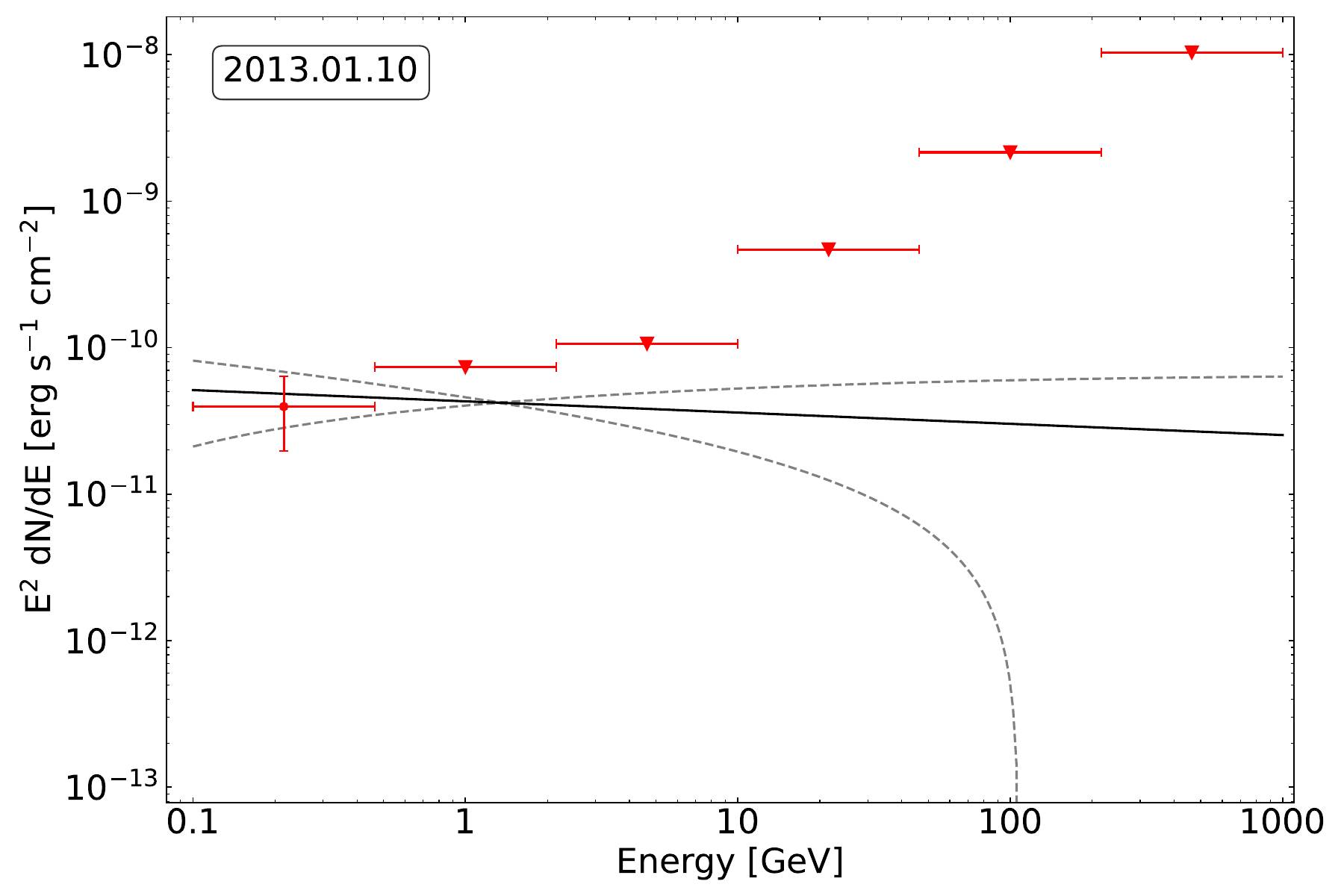} 
    \end{minipage}
    \begin{minipage}{0.3\textwidth}
        \centering
        \includegraphics[angle=0, width=\textwidth]{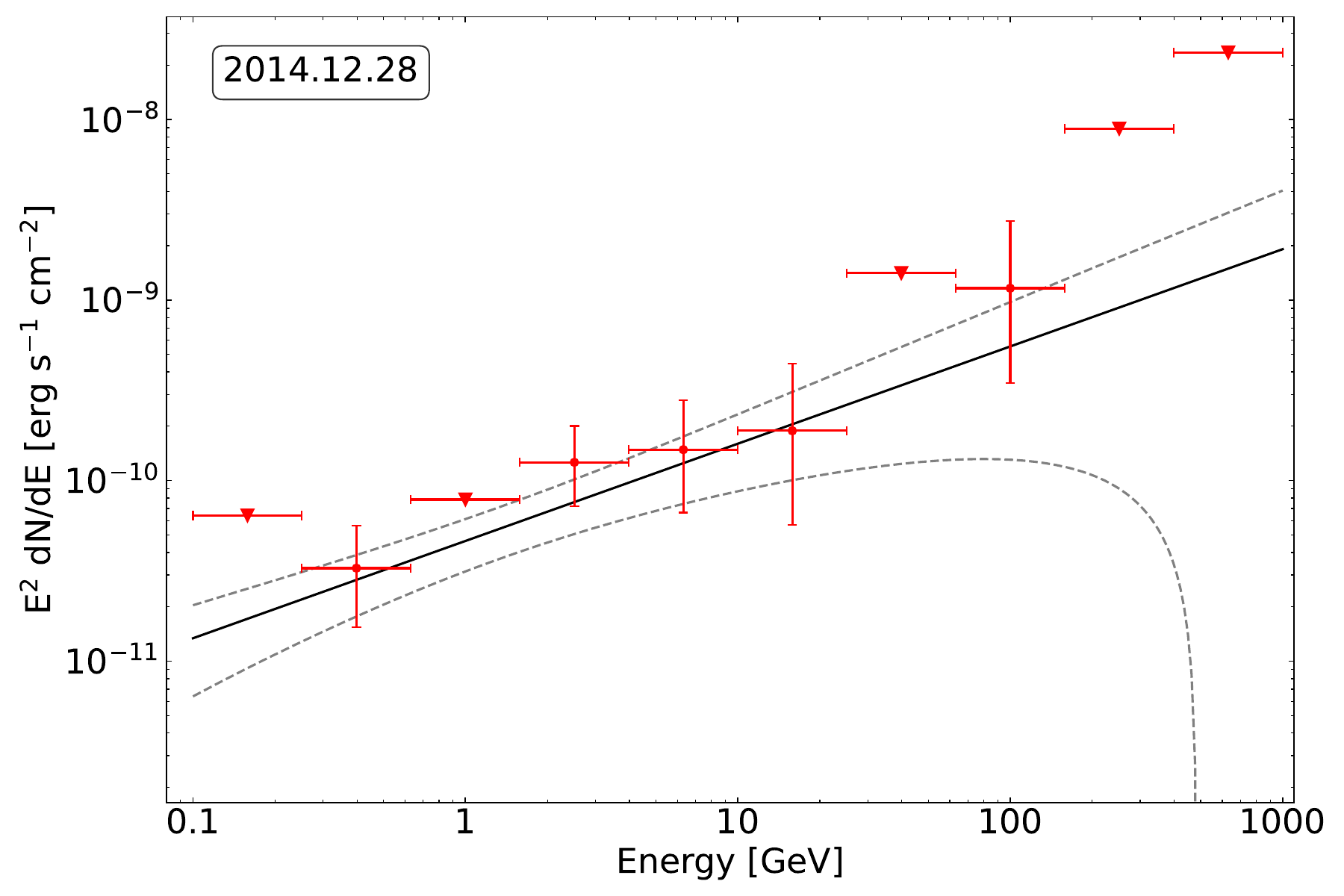} 
    \end{minipage}
        \caption{The 24 time-resolved spectra of Mrk 421 obtained through Fermi-LAT observations in the 0.1--1000 GeV band. The first 18 spectra were simultaneously obtained with the archived TeV observations, and they are consistent with the data presented in Figure \ref{Spe_GeV-TeV}. The three panels in the second row from the bottom correspond to the three lowest-flux points in panel (a) of Figure \ref{lc}, while the three panels in the first row from the bottom correspond to the three highest-flux points in panel (a) of Figure \ref{lc}. If TS $<$ 9, an upper limit (inverted triangles) is given for that energy bin. The black solid lines and gray dashed lines represent the spectral fitting results and the corresponding $1\sigma$ uncertainties, respectively. }
    \label{Spe_LAT_24}
\end{figure*}

\begin{figure*}
    \ContinuedFloat 
    \centering
    \begin{minipage}{0.3\textwidth}
        \centering
        \includegraphics[angle=0, width=\textwidth]{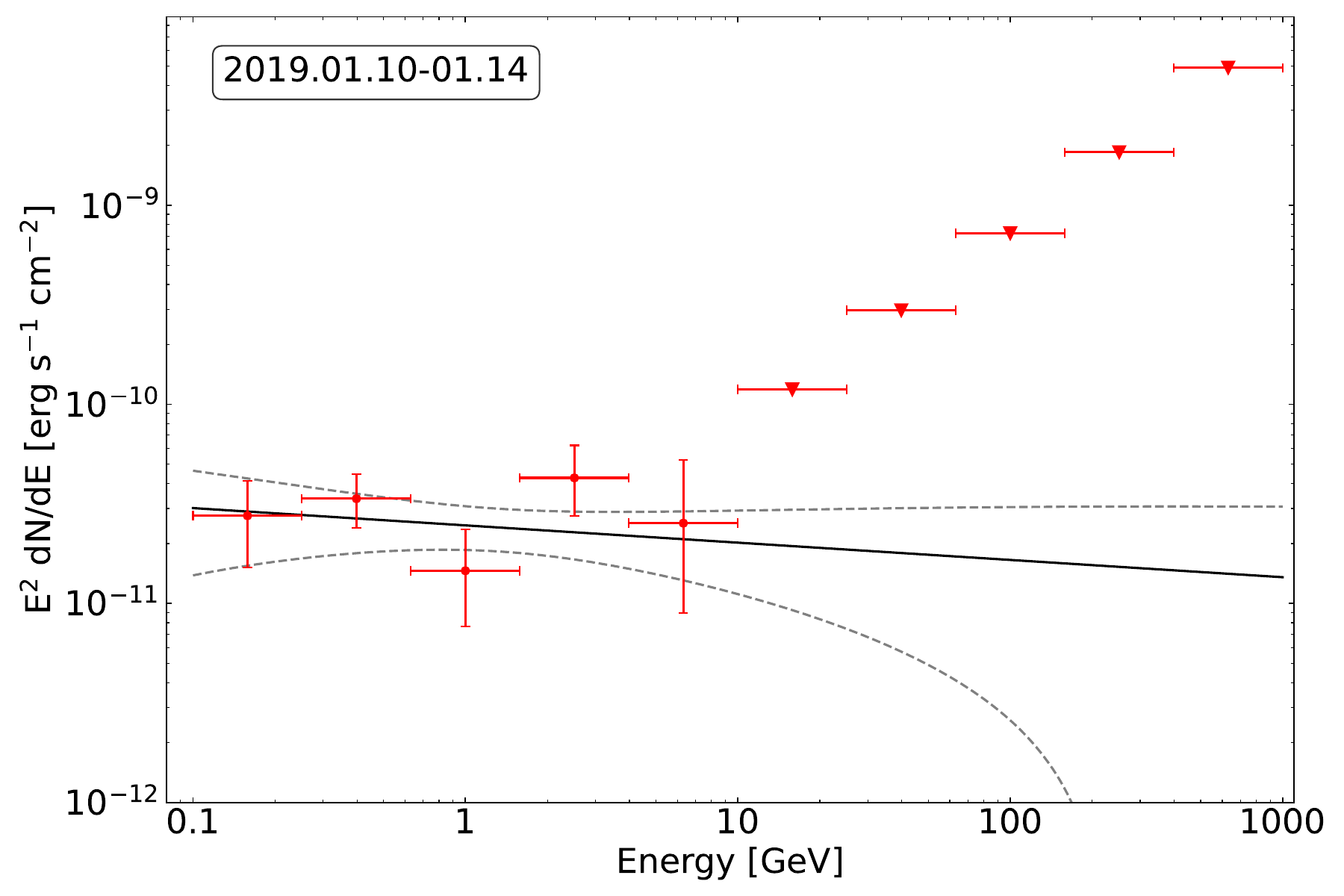} 
    \end{minipage}
    \begin{minipage}{0.3\textwidth}
        \centering
        \includegraphics[angle=0, width=\textwidth]{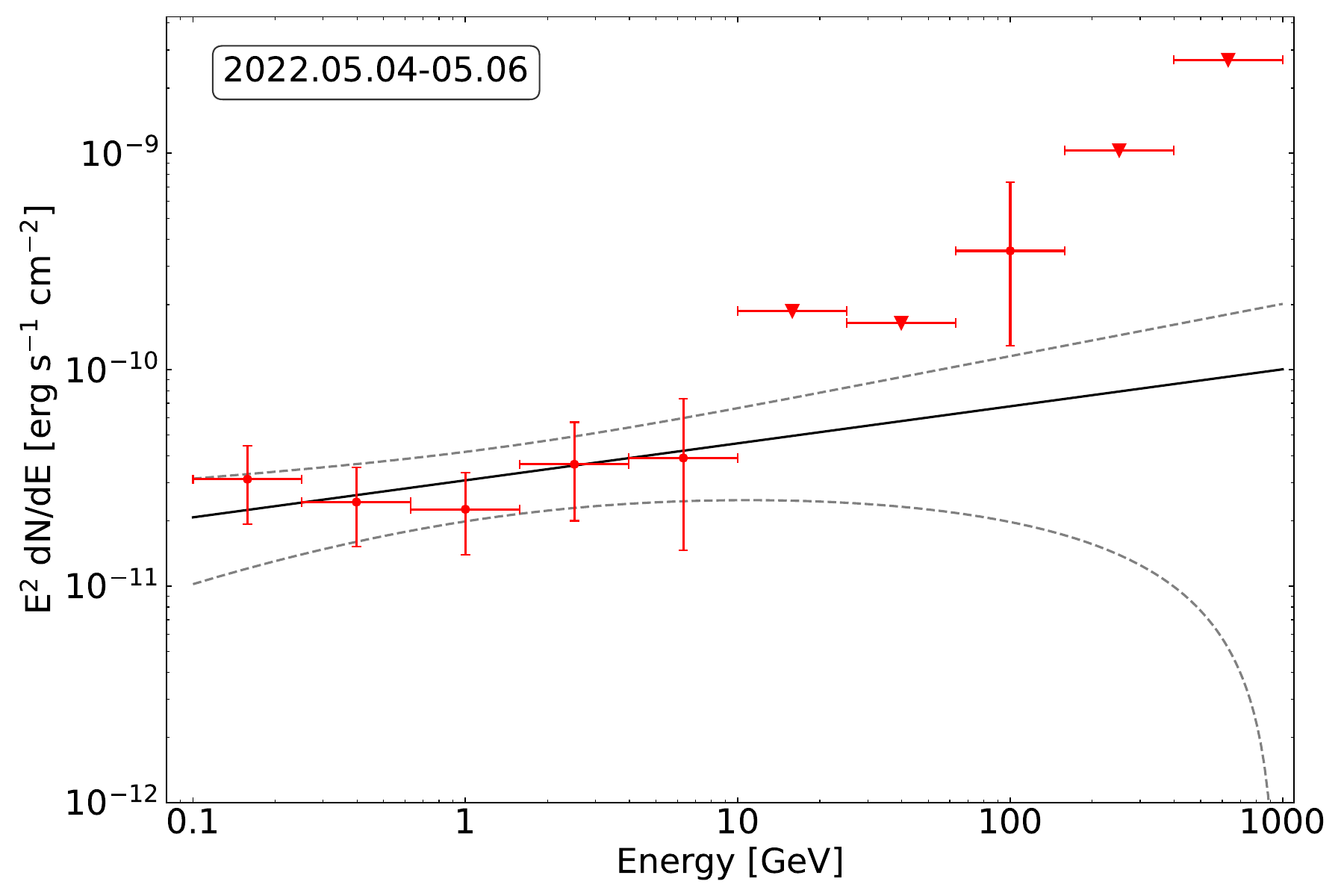} 
    \end{minipage}
    \begin{minipage}{0.3\textwidth}
        \centering
        \includegraphics[angle=0, width=\textwidth]{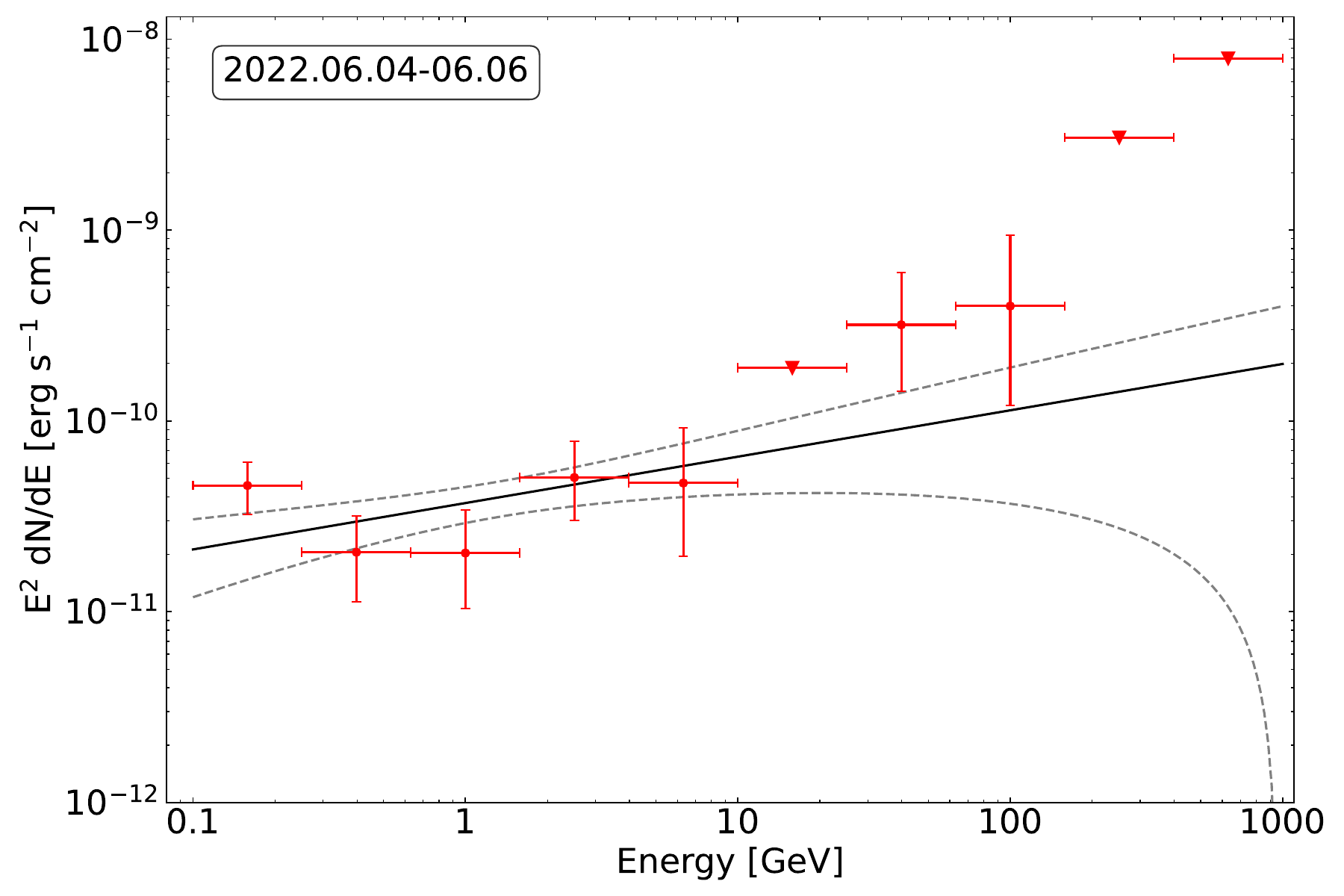} 
    \end{minipage}
    \begin{minipage}{0.3\textwidth}
        \centering
        \includegraphics[angle=0, width=\textwidth]{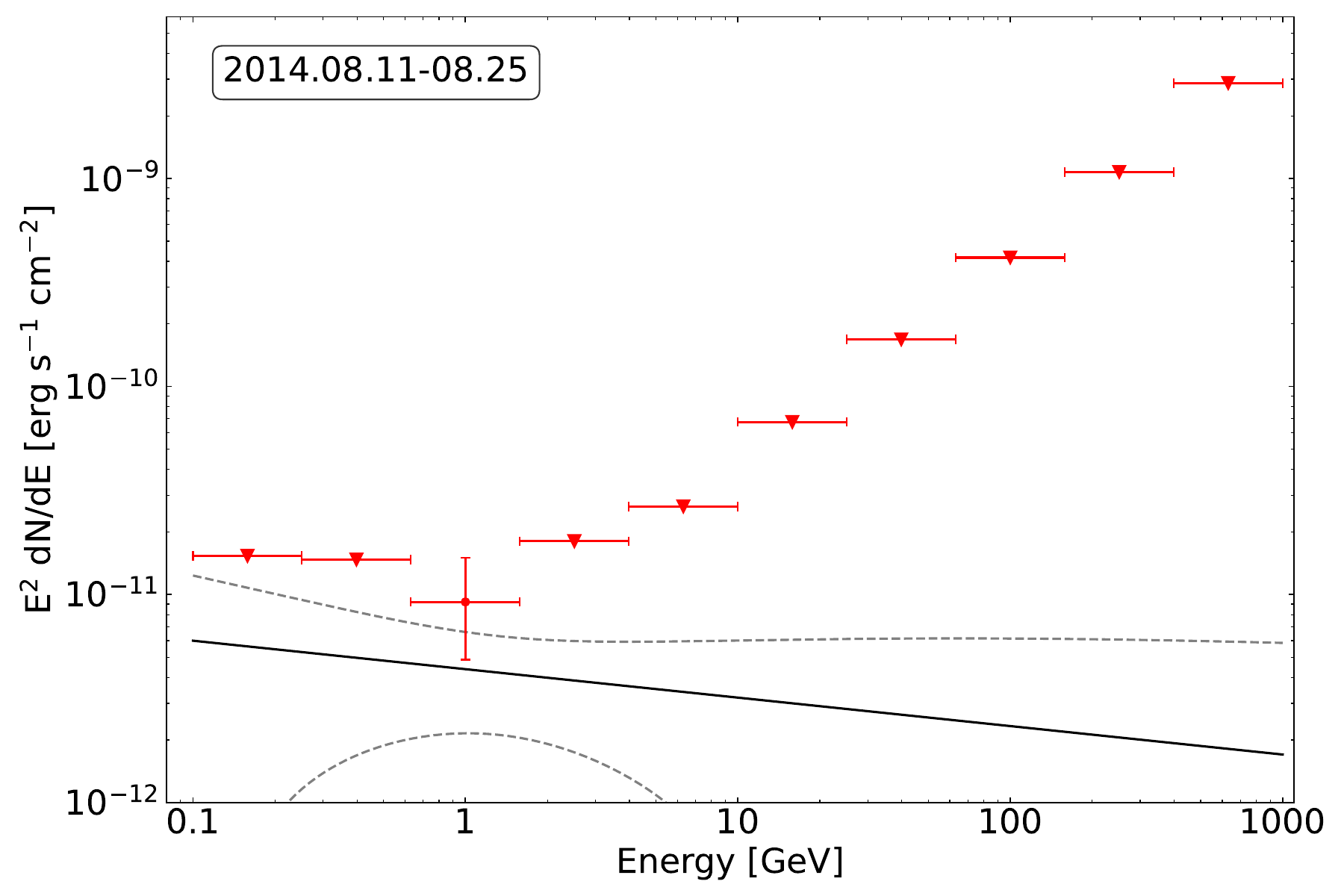} 
    \end{minipage}
    \begin{minipage}{0.3\textwidth}
        \centering
        \includegraphics[angle=0, width=\textwidth]{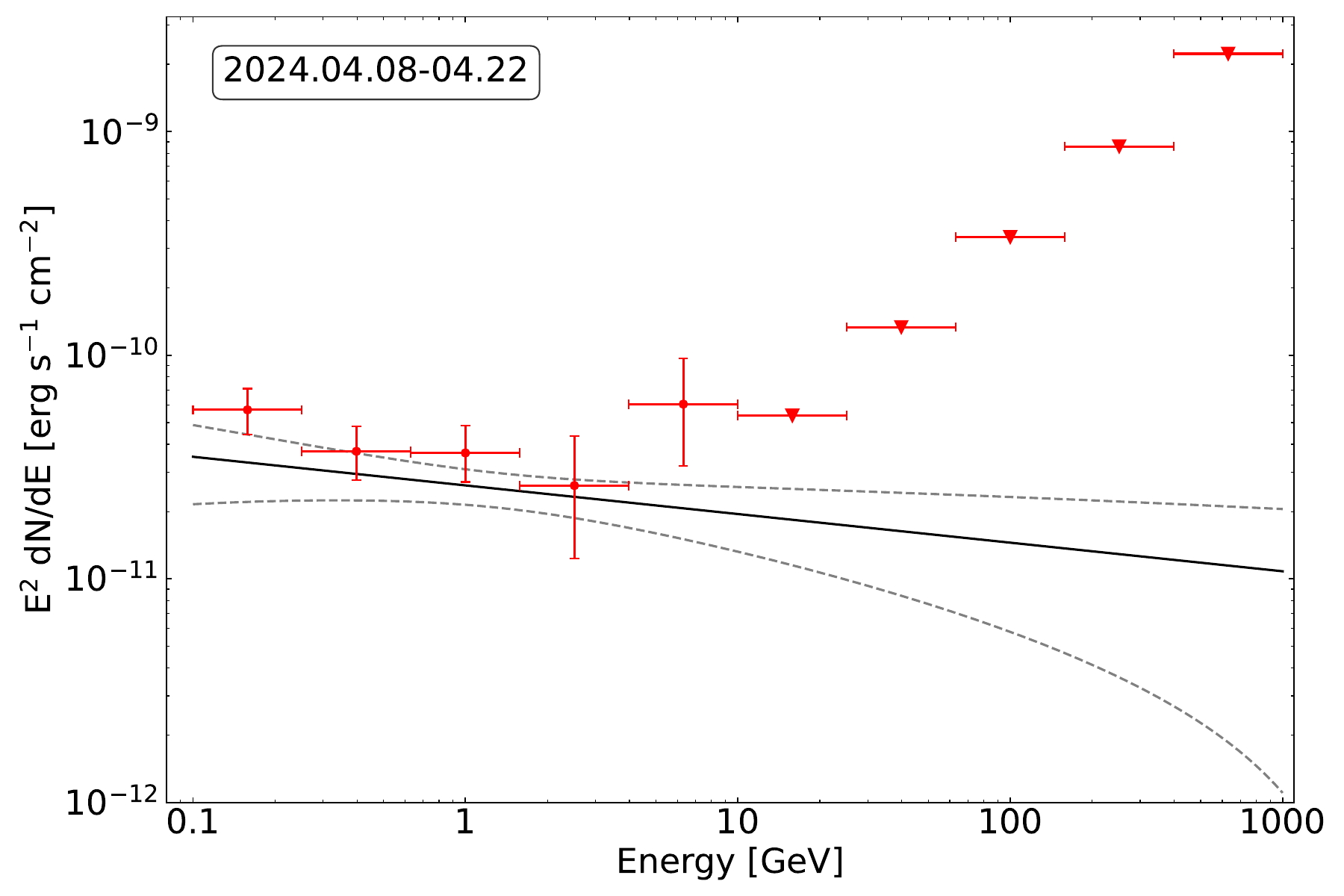} 
    \end{minipage}
    \begin{minipage}{0.3\textwidth}
        \centering
        \includegraphics[angle=0, width=\textwidth]{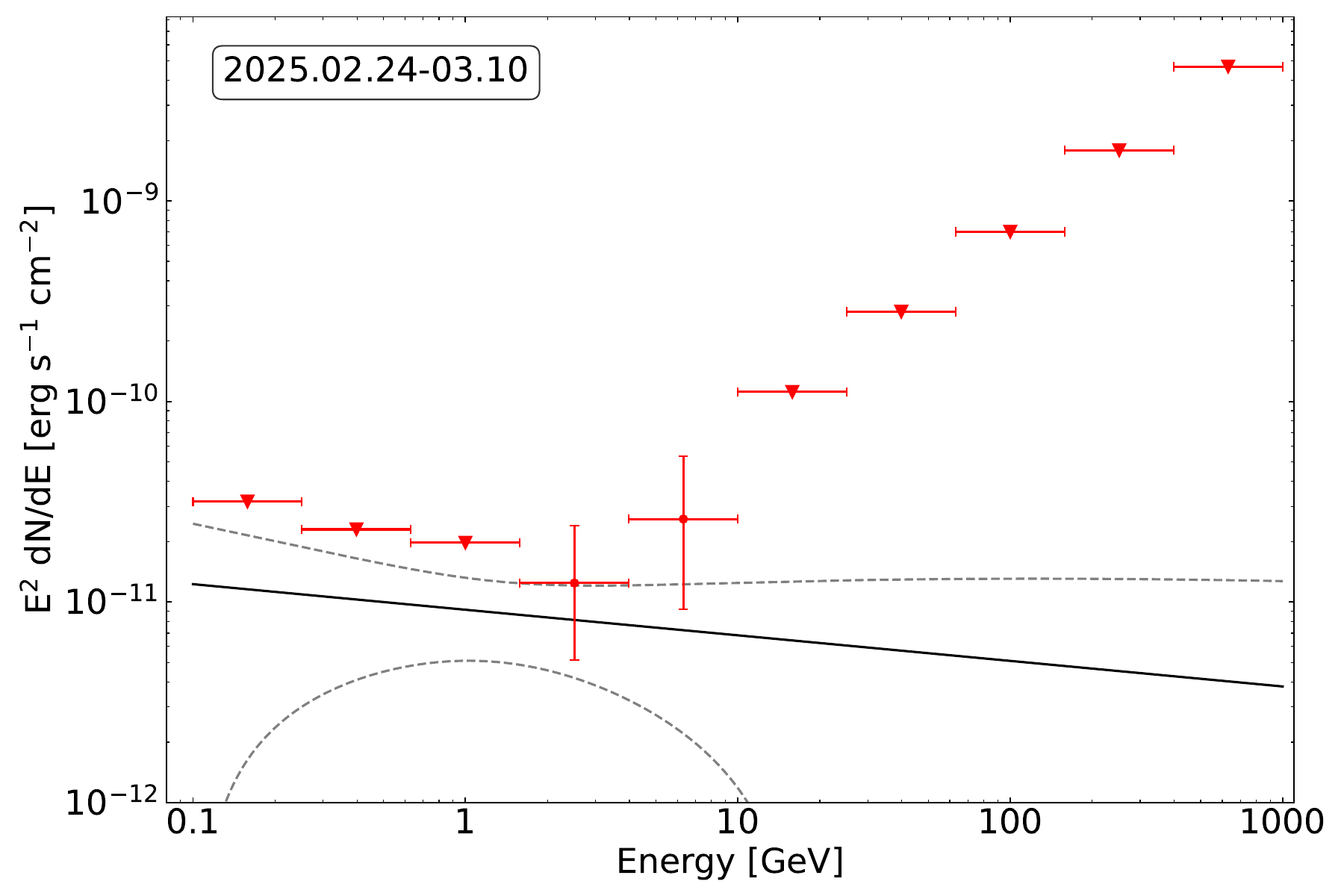} 
    \end{minipage}
    \begin{minipage}{0.3\textwidth}
        \centering
        \includegraphics[angle=0, width=\textwidth]{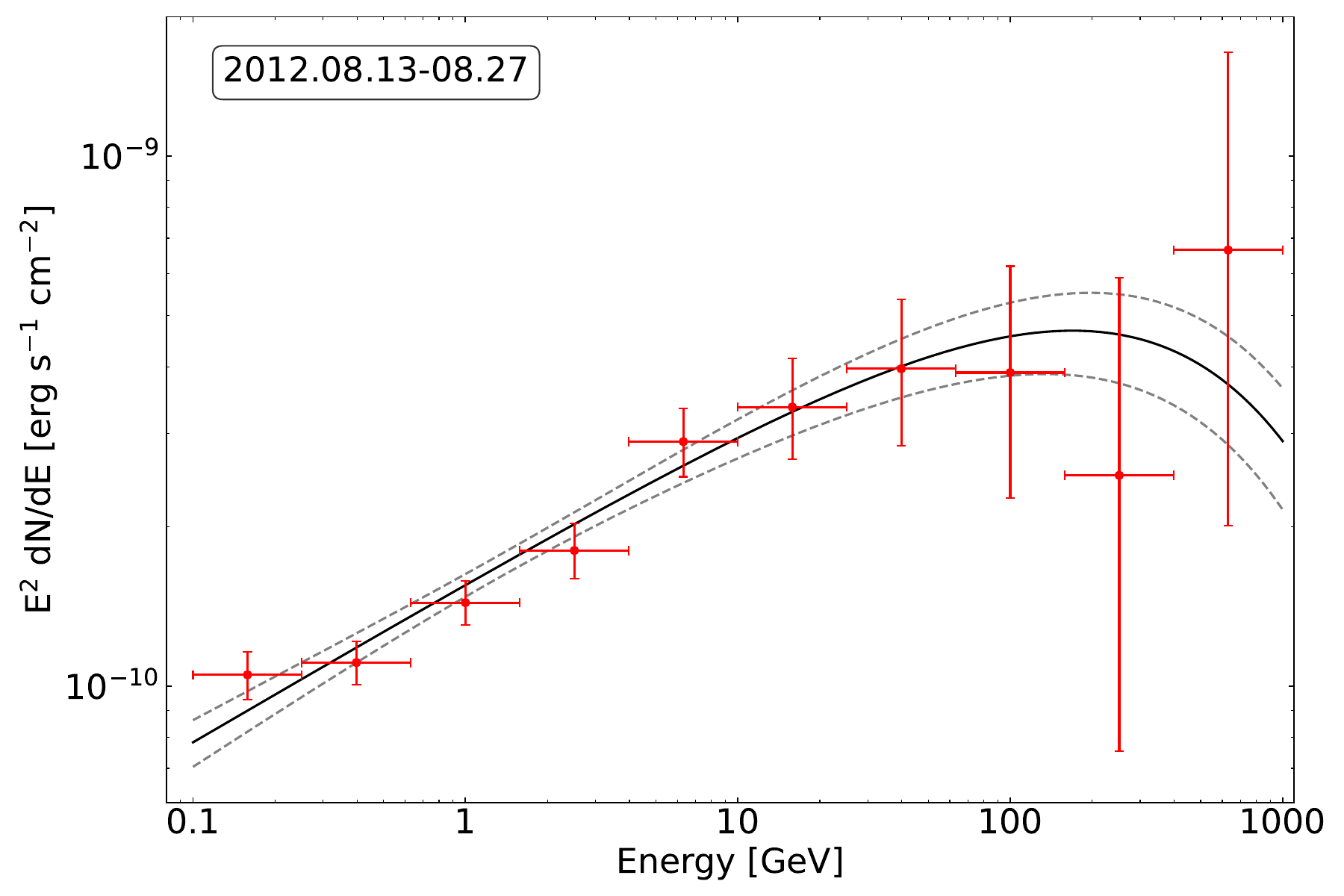} 
    \end{minipage}
    \begin{minipage}{0.3\textwidth}
        \centering
        \includegraphics[angle=0, width=\textwidth]{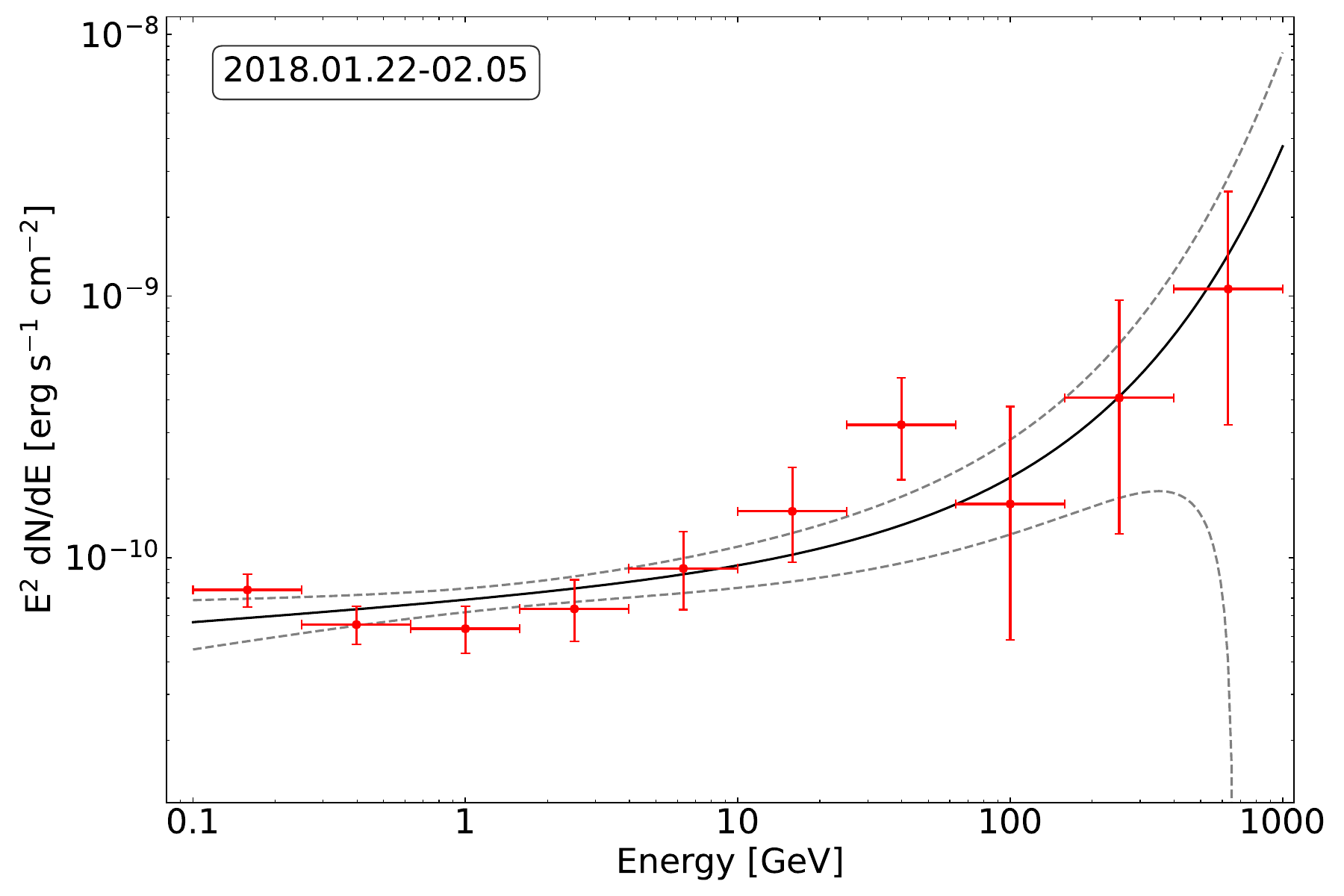} 
    \end{minipage}
    \begin{minipage}{0.3\textwidth}
        \centering
        \includegraphics[angle=0, width=\textwidth]{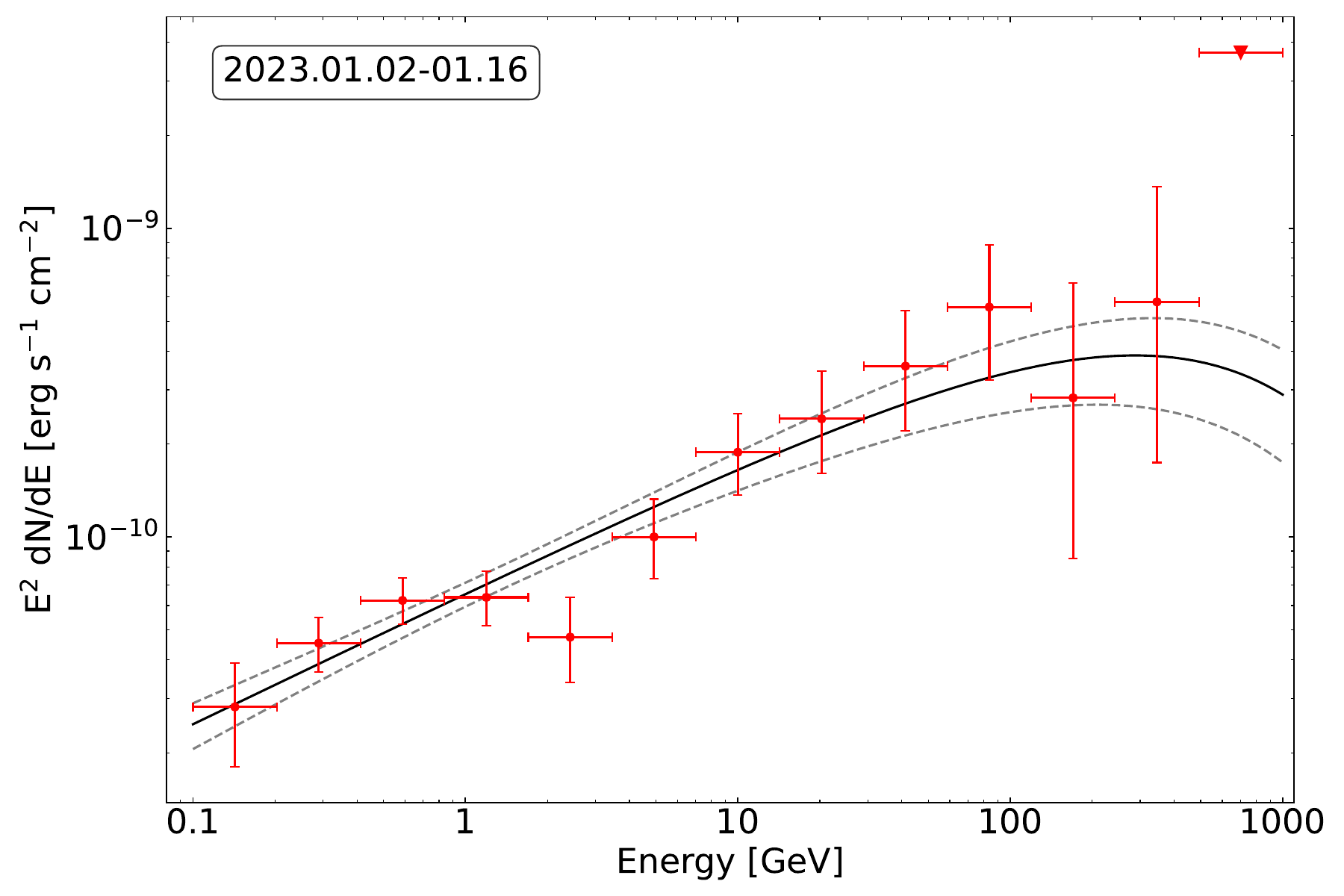} 
    \end{minipage}
    \caption{(Continued.)}
\end{figure*}

\begin{figure*}
    \centering
    \includegraphics[angle=0, width=0.5\textwidth]{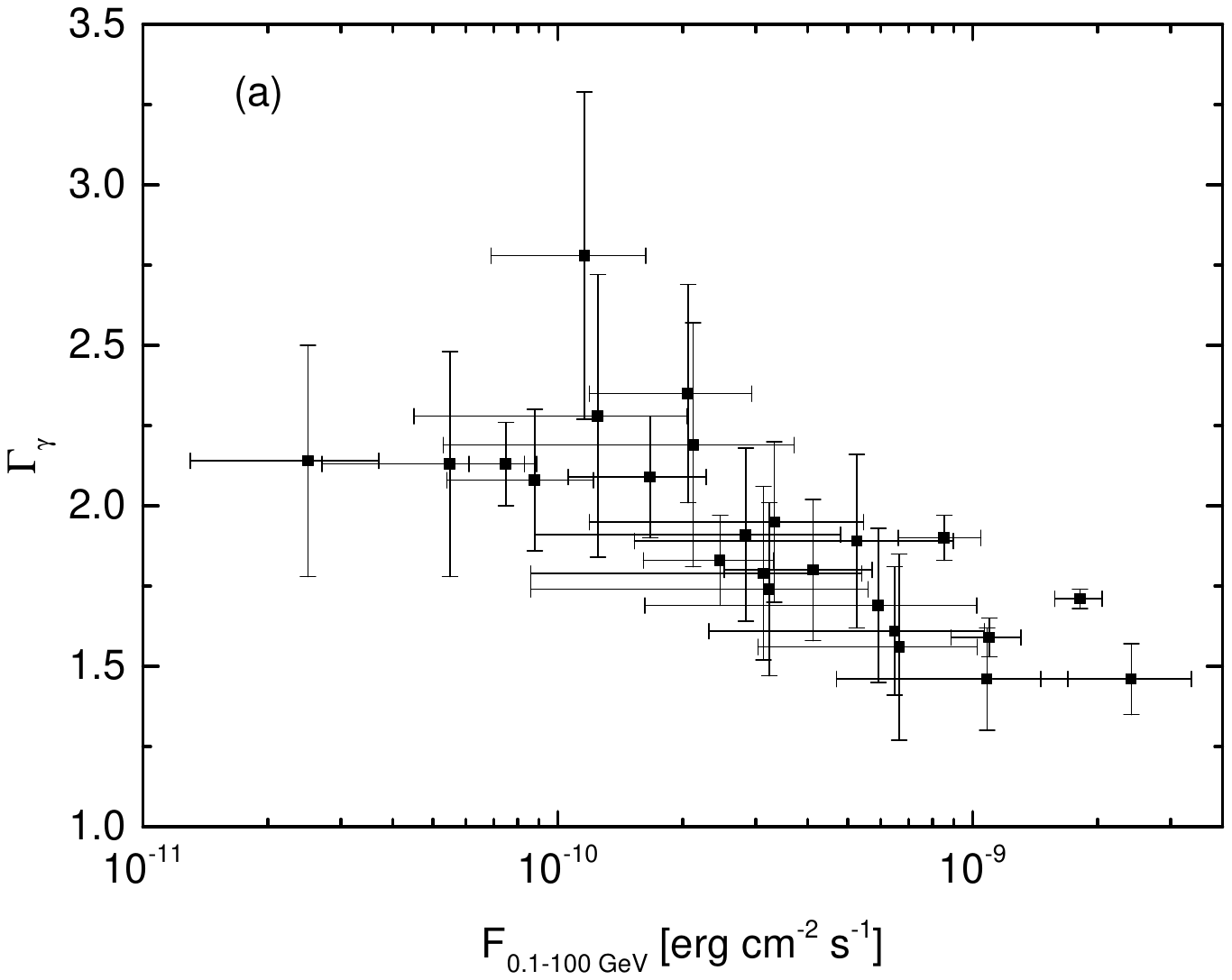}\hspace{-1.4cm}
    \includegraphics[angle=0, width=0.5\textwidth]{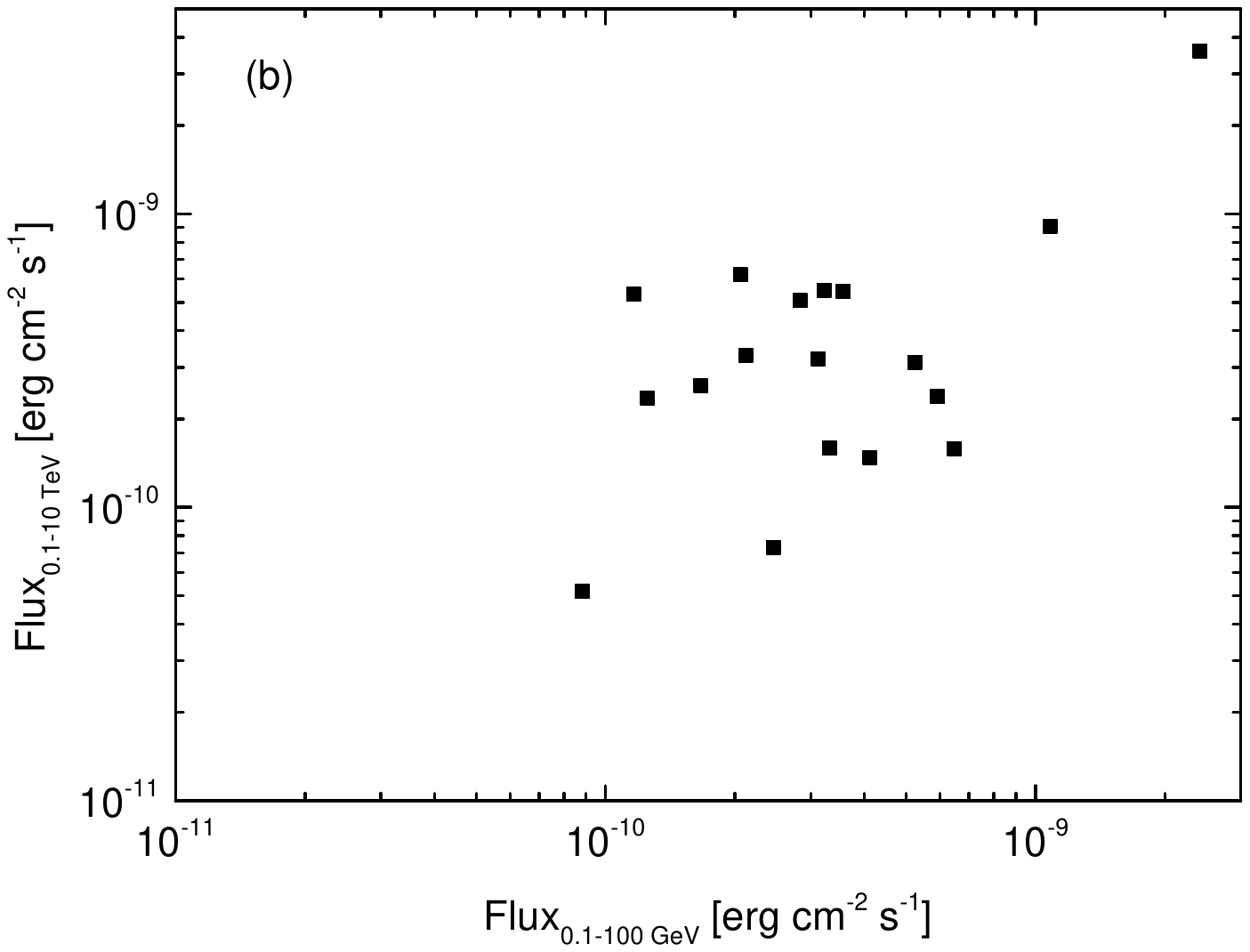} 
    \caption{Panel (a): $\Gamma_{\gamma}$ as a function of the flux in the 0.1--100 GeV band for the 24 time-resolved spectra, as given in Table \ref{tab_LAT}. Panel(b): $F_{0.1-10~{\rm TeV}}$ vs. $F_{0.1-100~{\rm GeV}}$ for the 18 GeV--TeV spectra shown in Figure \ref{Spe_GeV-TeV}. The values of $F_{0.1-100~{\rm GeV}}$ are taken from Table \ref{tab_LAT}, while the $F_{0.1-10~{\rm TeV}}$ values are derived through a rough fitting of the 18 TeV spectra using a simple PL function; see Section \ref{sec:GeV-TeV} for more details.}
    \label{Flux-Gamma}
\end{figure*}

\begin{figure*}
    \centering
    \begin{minipage}{0.3\textwidth}
        \centering
        \includegraphics[angle=0, width=\textwidth]{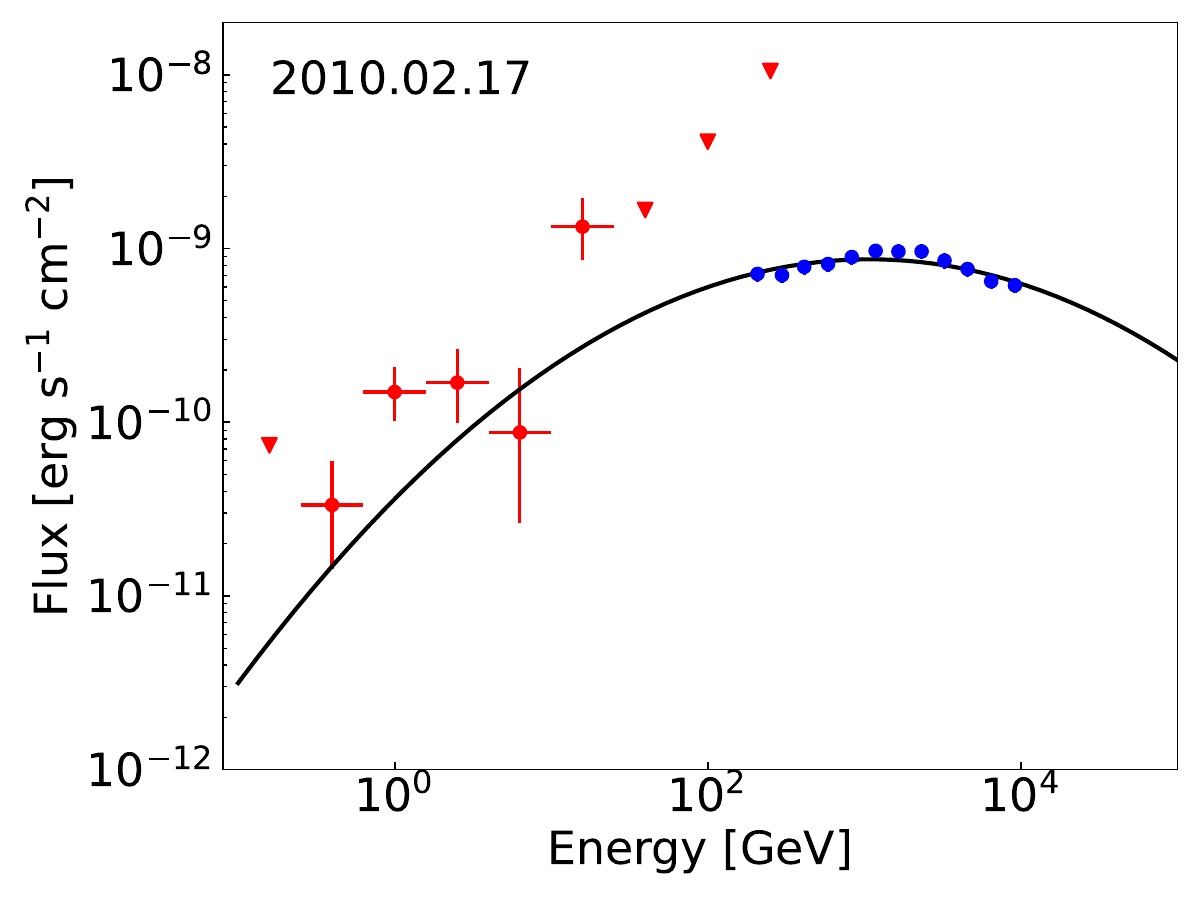} 
            % \makebox{(a)}
    \end{minipage} 
    \begin{minipage}{0.3\textwidth}
        \centering
        \includegraphics[angle=0, width=\textwidth]{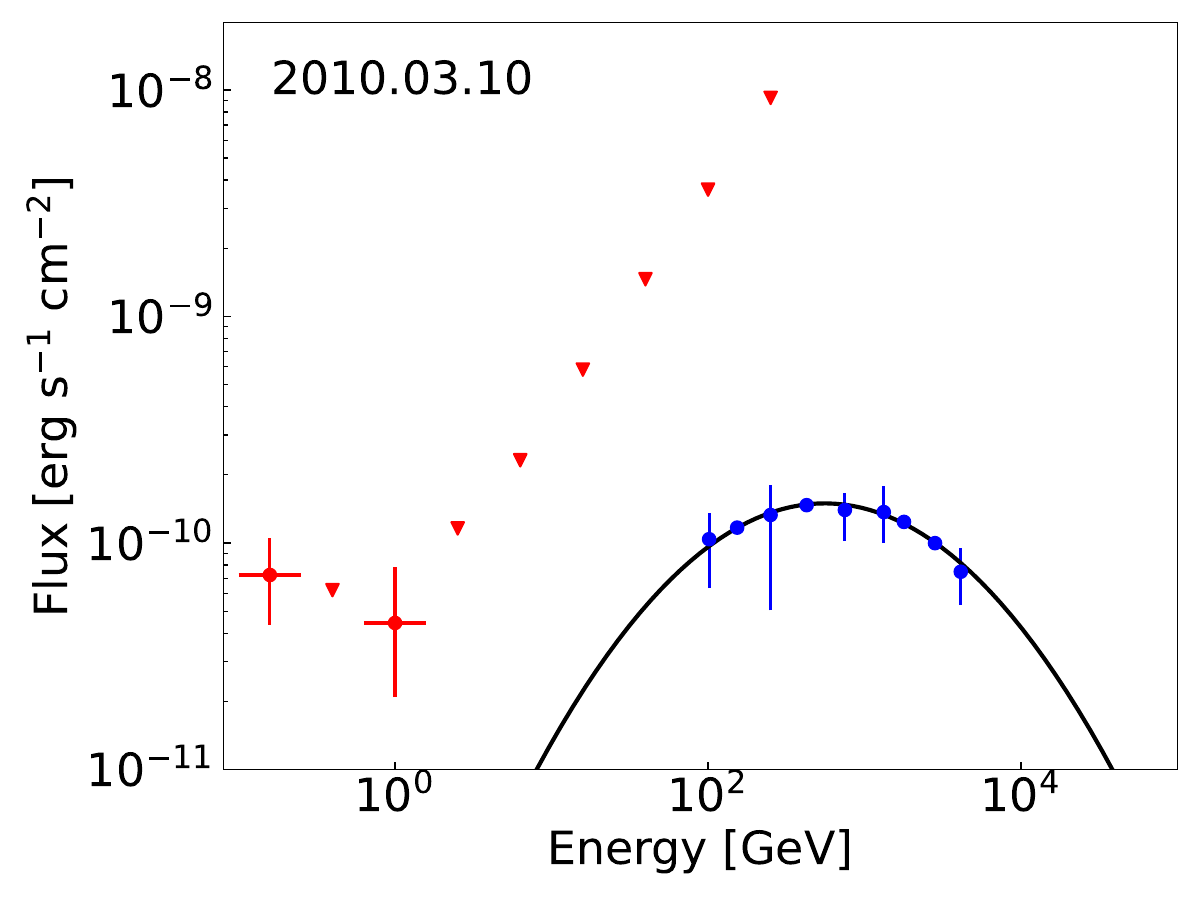} 
    \end{minipage} 
    \begin{minipage}{0.3\textwidth}
        \centering
        \includegraphics[angle=0, width=\textwidth]{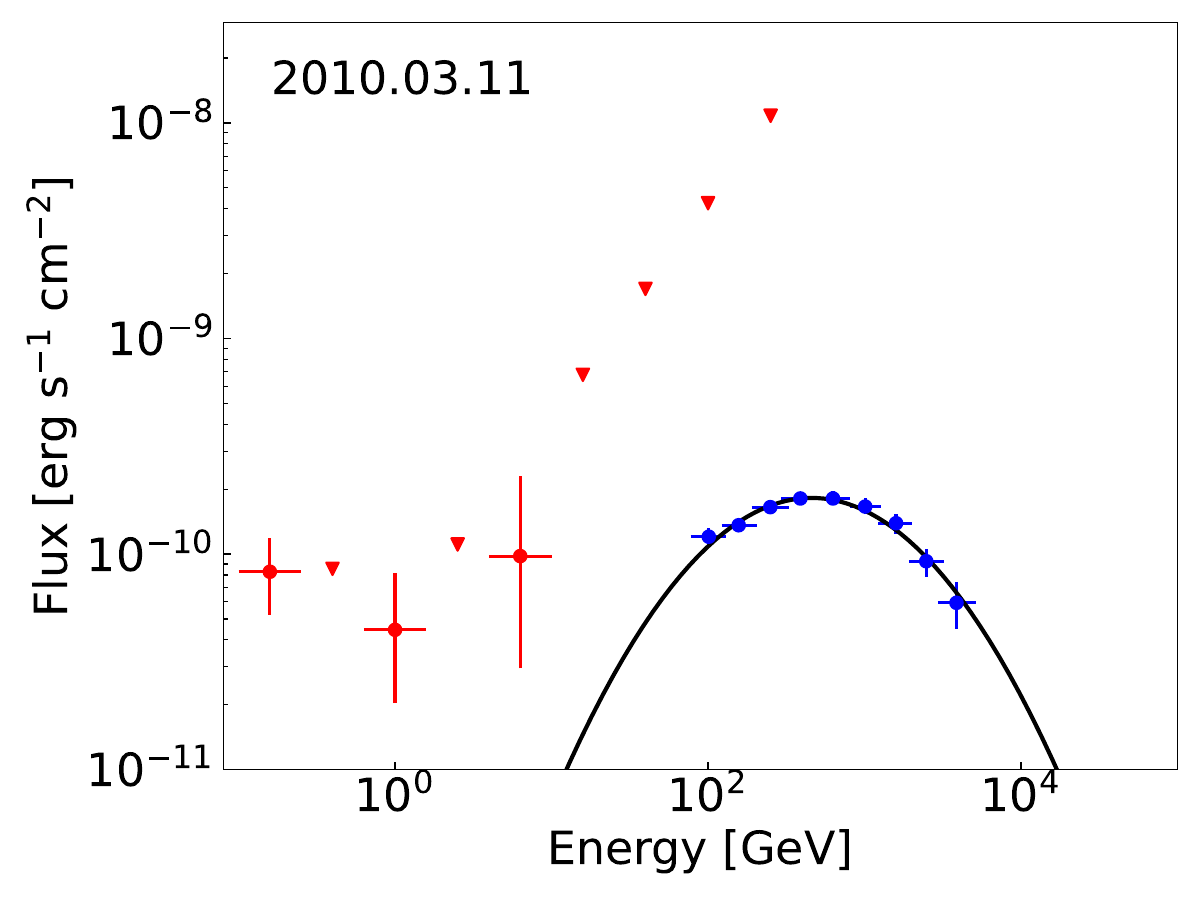} 
    \end{minipage}
    \begin{minipage}{0.3\textwidth}
        \centering
        \includegraphics[angle=0, width=\textwidth]{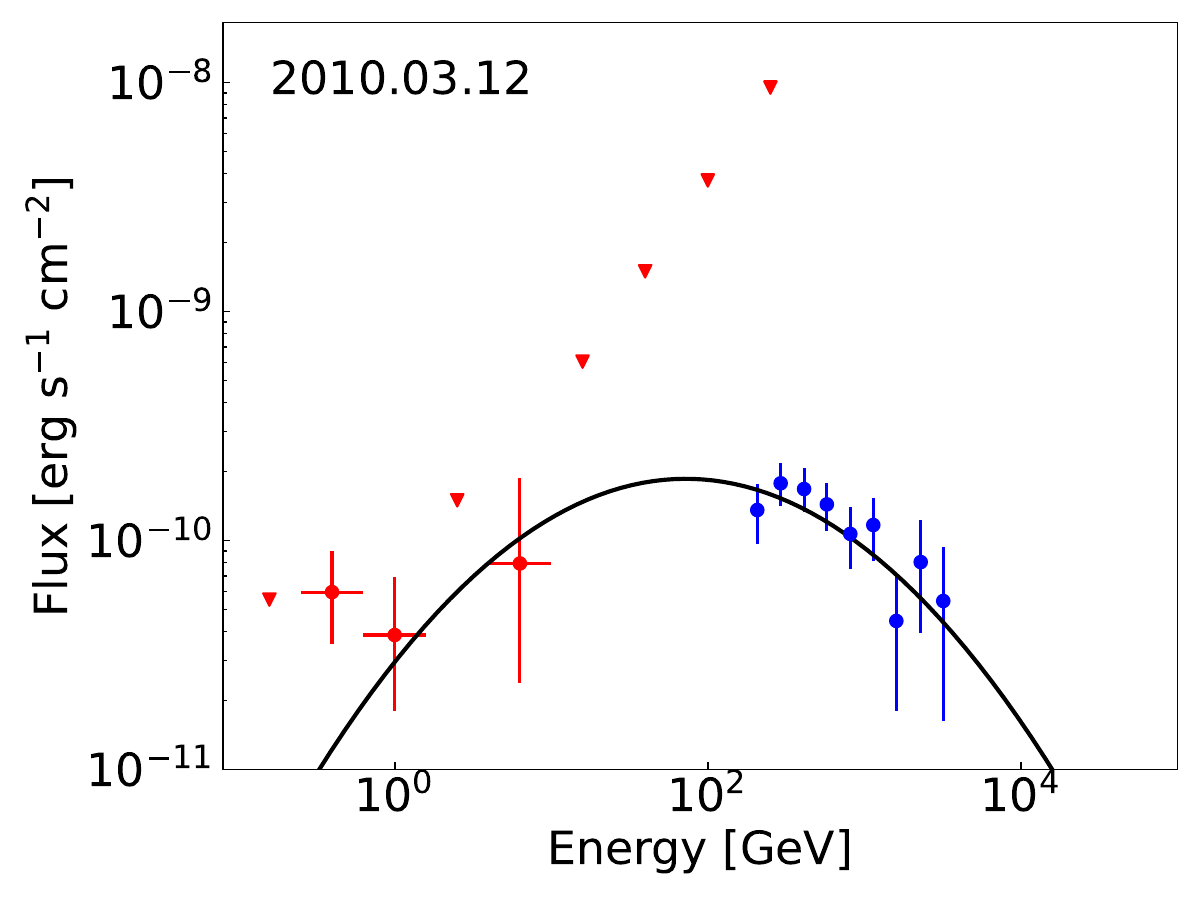} 
    \end{minipage}
    \begin{minipage}{0.3\textwidth}
        \centering
        \includegraphics[angle=0, width=\textwidth]{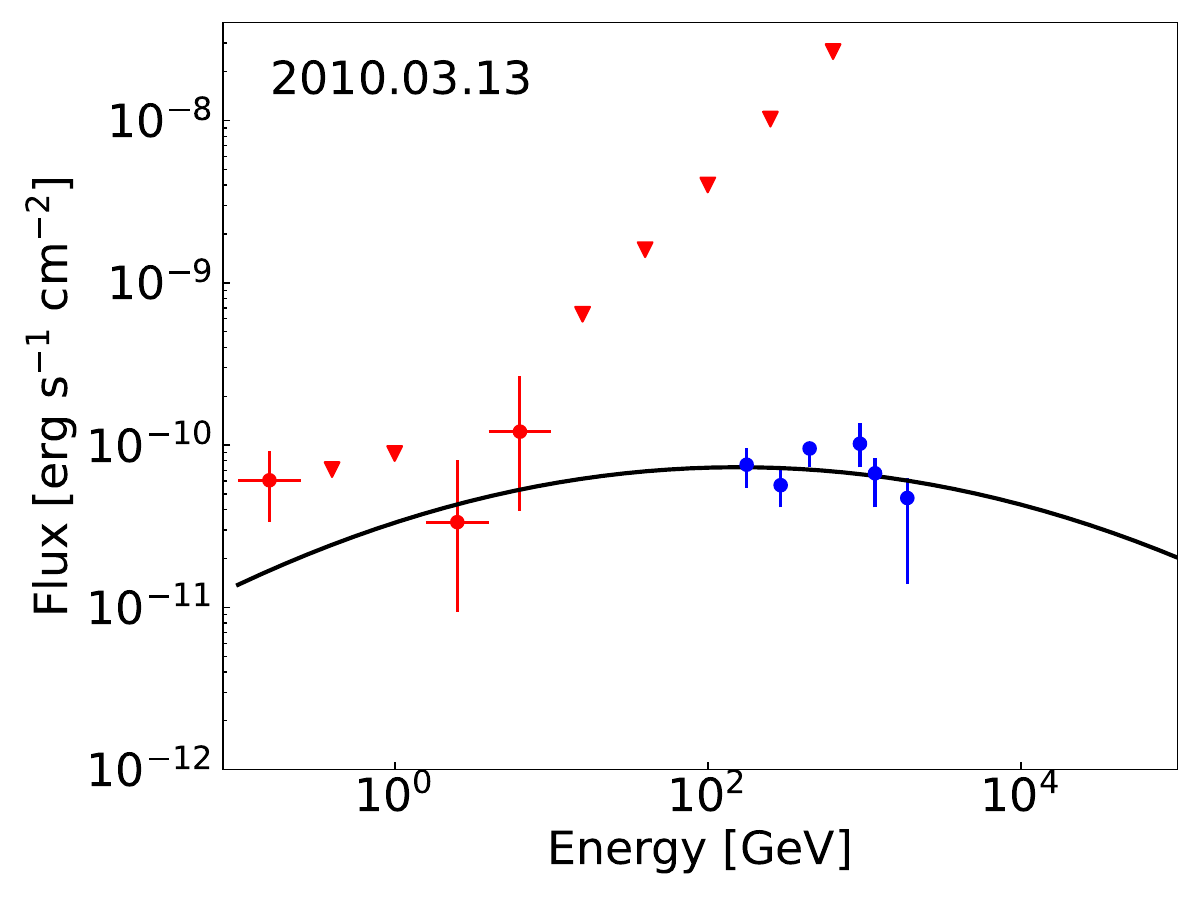} 
    \end{minipage}
    \begin{minipage}{0.3\textwidth}
        \centering
        \includegraphics[angle=0, width=\textwidth]{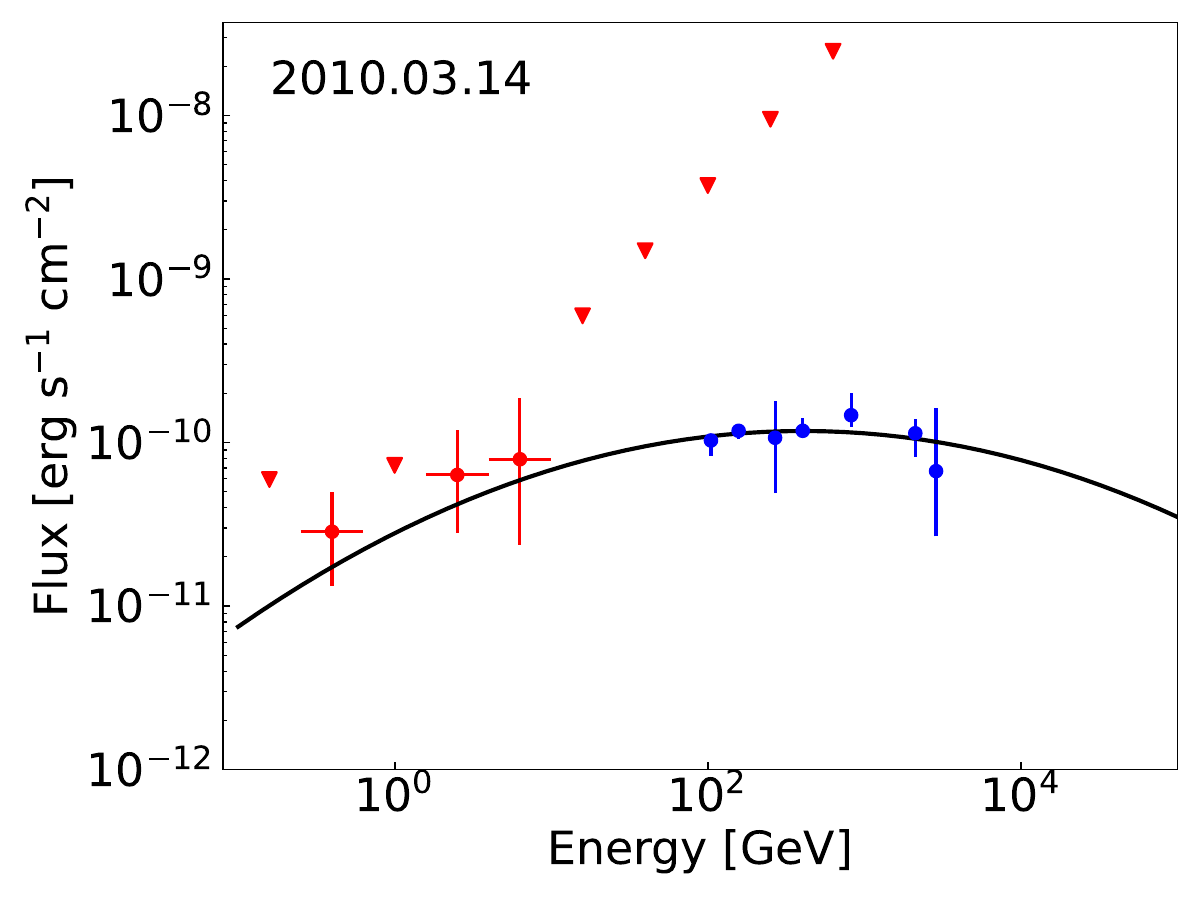} 
    \end{minipage}
    \begin{minipage}{0.3\textwidth}
        \centering
        \includegraphics[angle=0, width=\textwidth]{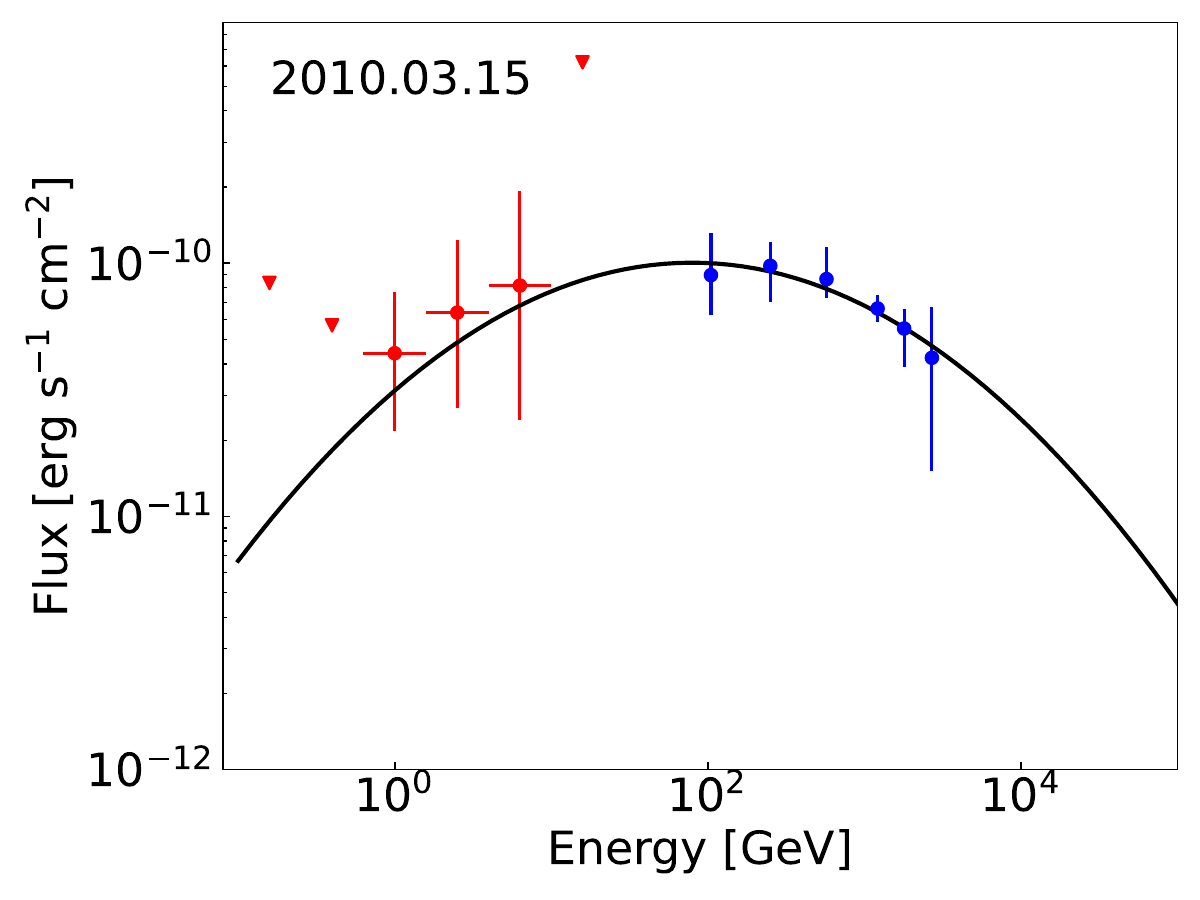} 
    \end{minipage}
    \begin{minipage}{0.3\textwidth}
        \centering
        \includegraphics[angle=0, width=\textwidth]{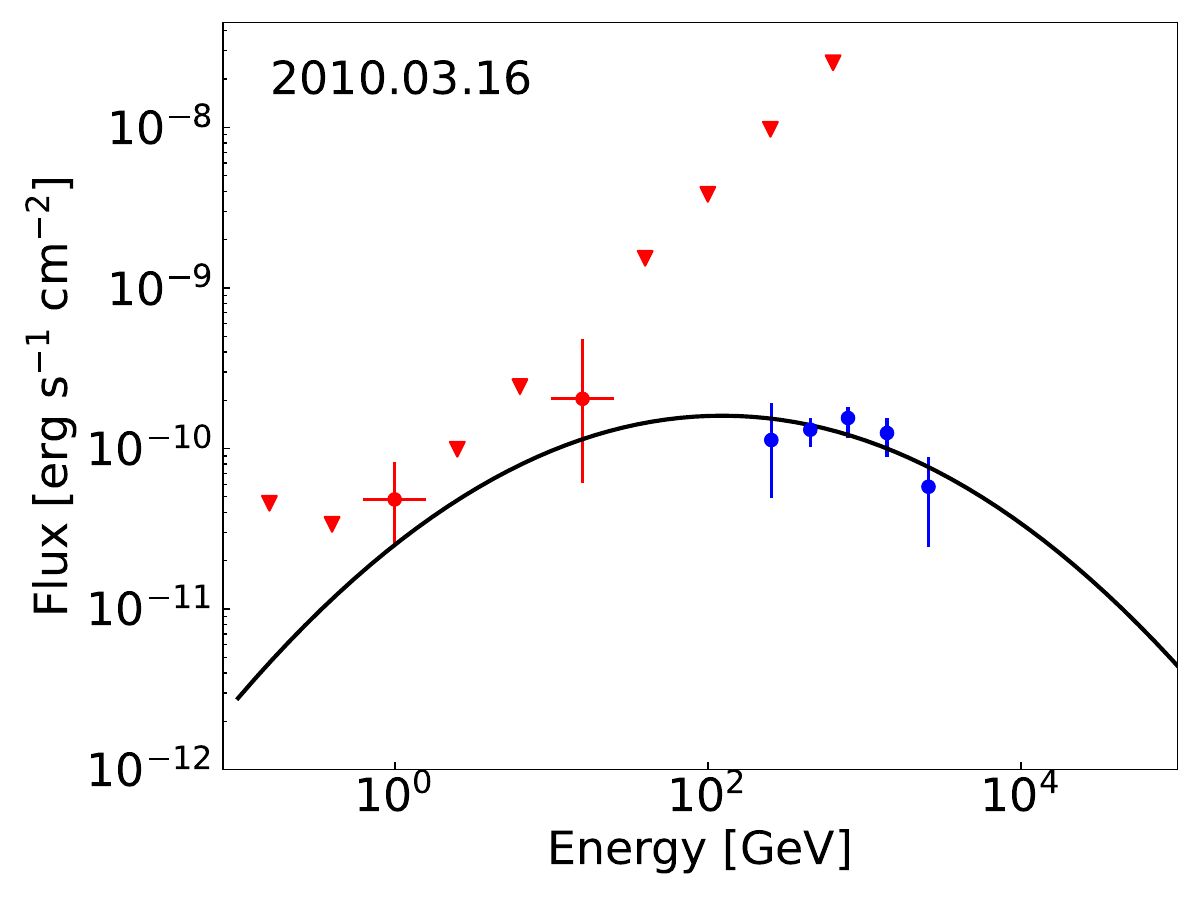} 
    \end{minipage}
    \begin{minipage}{0.3\textwidth}
        \centering
        \includegraphics[angle=0, width=\textwidth]{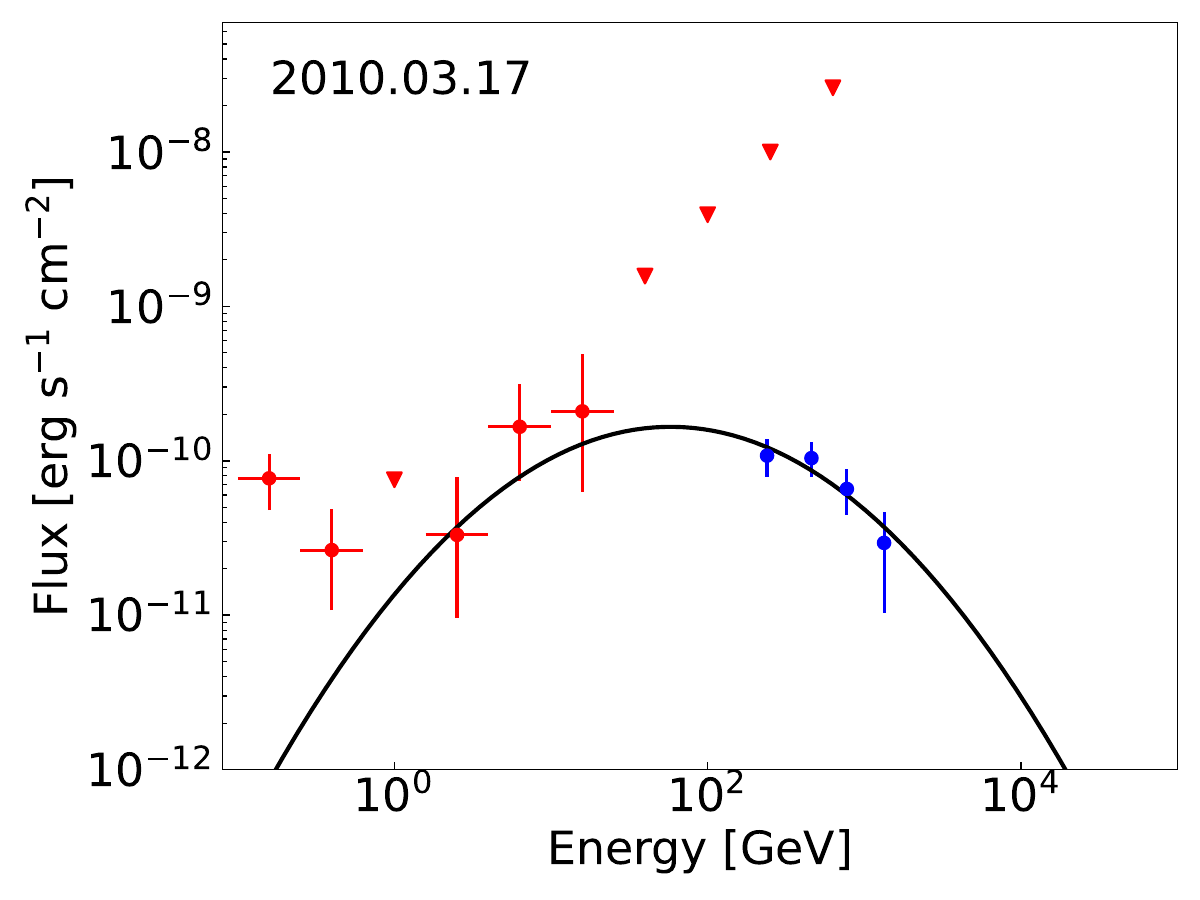} 
    \end{minipage}
    \begin{minipage}{0.3\textwidth}
        \centering
        \includegraphics[angle=0, width=\textwidth]{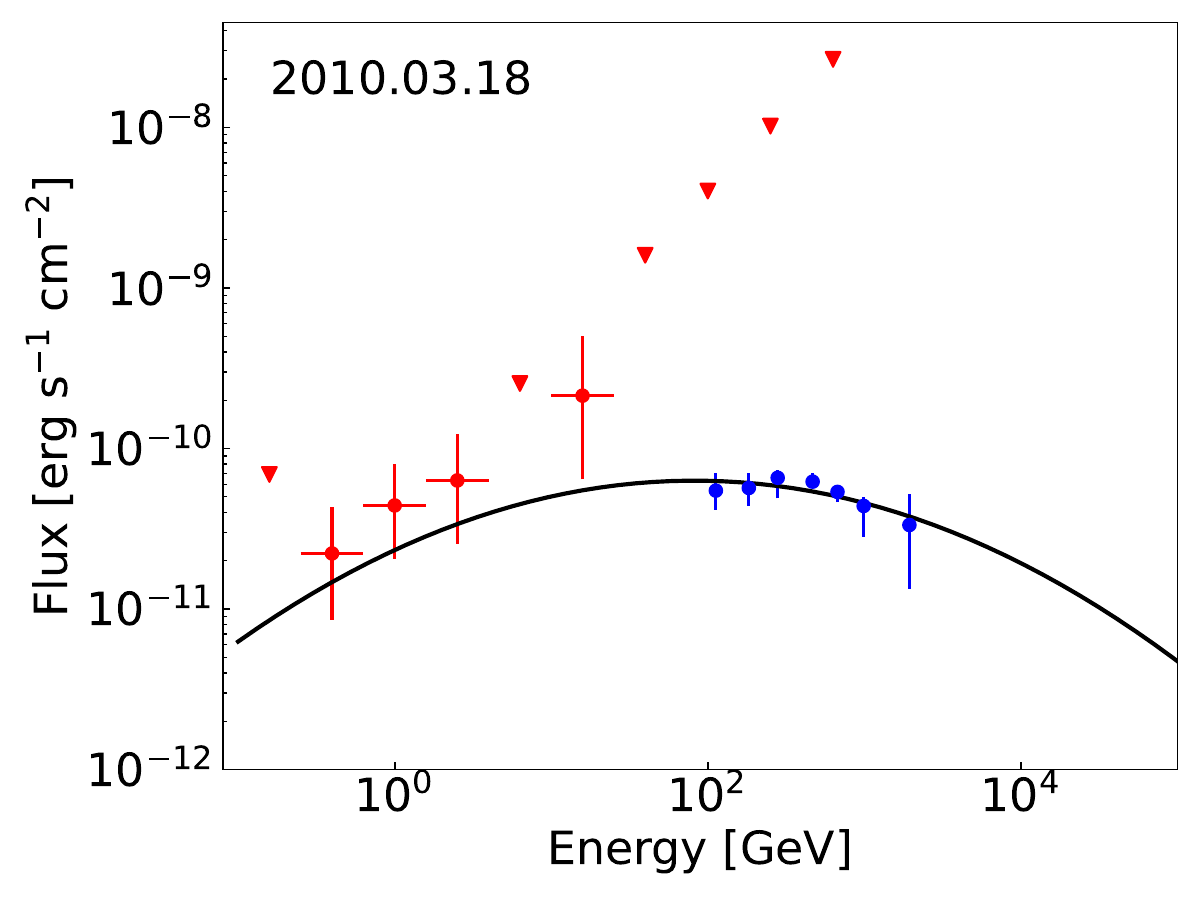} 
    \end{minipage}
    \begin{minipage}{0.3\textwidth}
        \centering
        \includegraphics[angle=0, width=\textwidth]{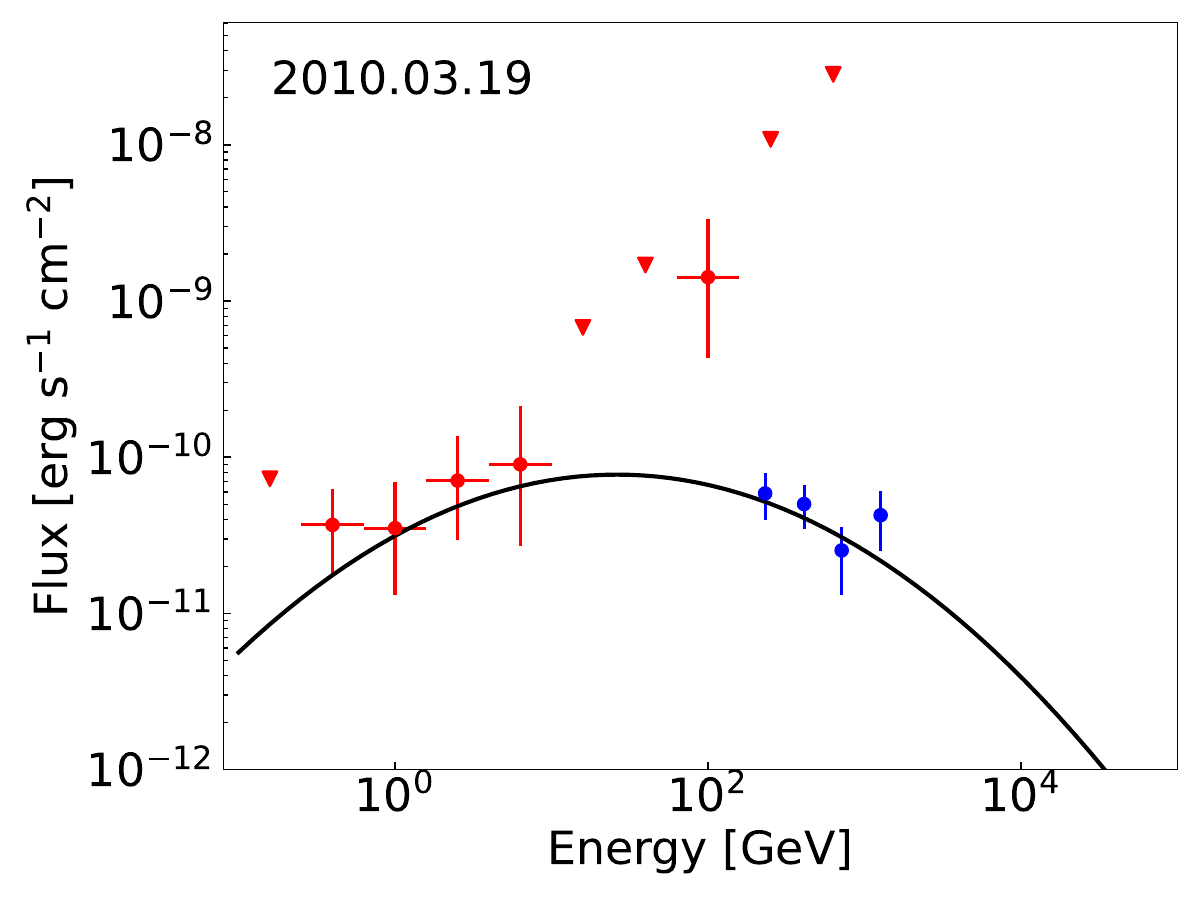} 
    \end{minipage}
    \begin{minipage}{0.3\textwidth}
        \centering
        \includegraphics[angle=0, width=\textwidth]{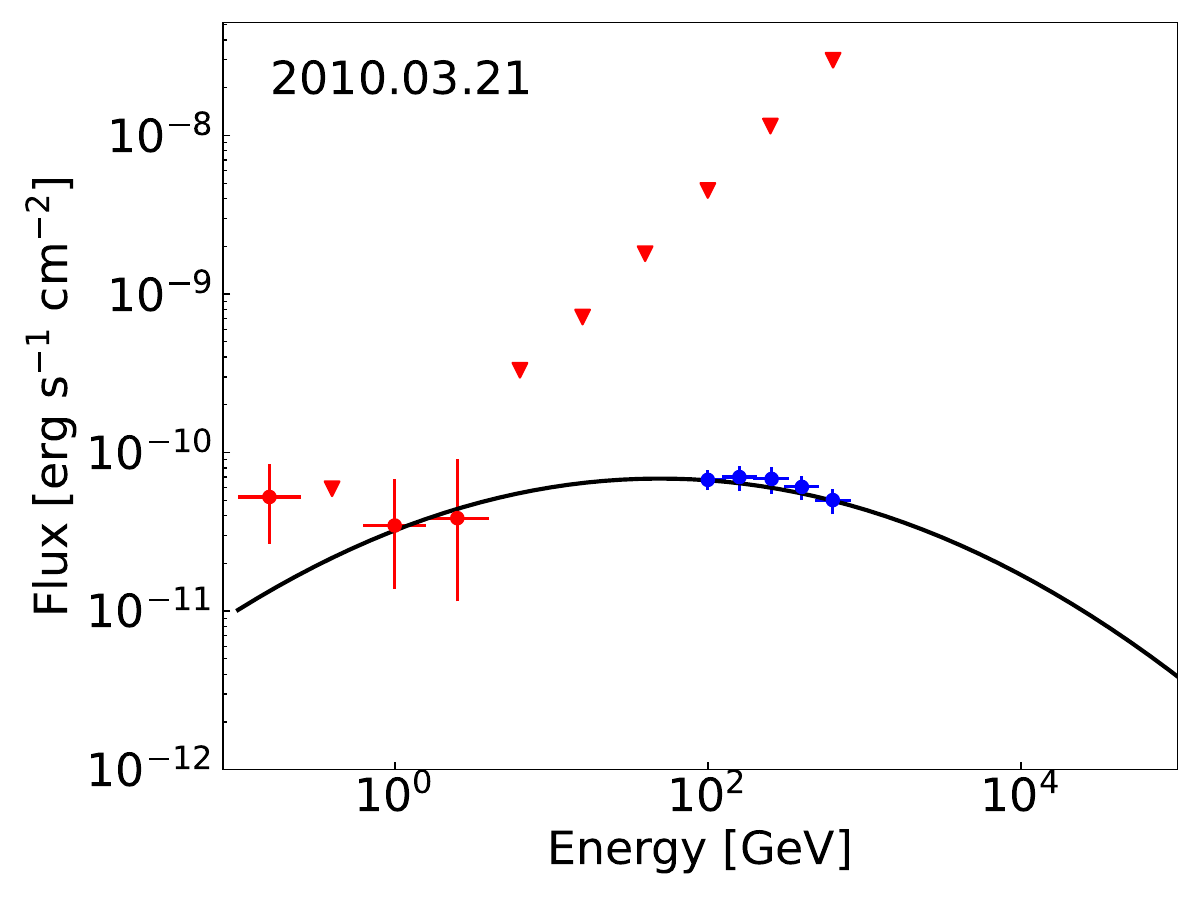} 
    \end{minipage}
     \caption{The 18 GeV--TeV spectra with a rough fitting using the LP model (black solid lines). The GeV spectra (red symbols) for each panel are same as those in Figure \ref{Spe_LAT_24}, while the TeV spectra (blue symbols) are taken from the literature. For the details of the TeV data, refer to Section \ref{sec:TeV}.}
    \label{Spe_GeV-TeV}
\end{figure*}

\begin{figure*}
    \ContinuedFloat 
    \centering
       \begin{minipage}{0.3\textwidth}
        \centering
        \includegraphics[angle=0, width=\textwidth]{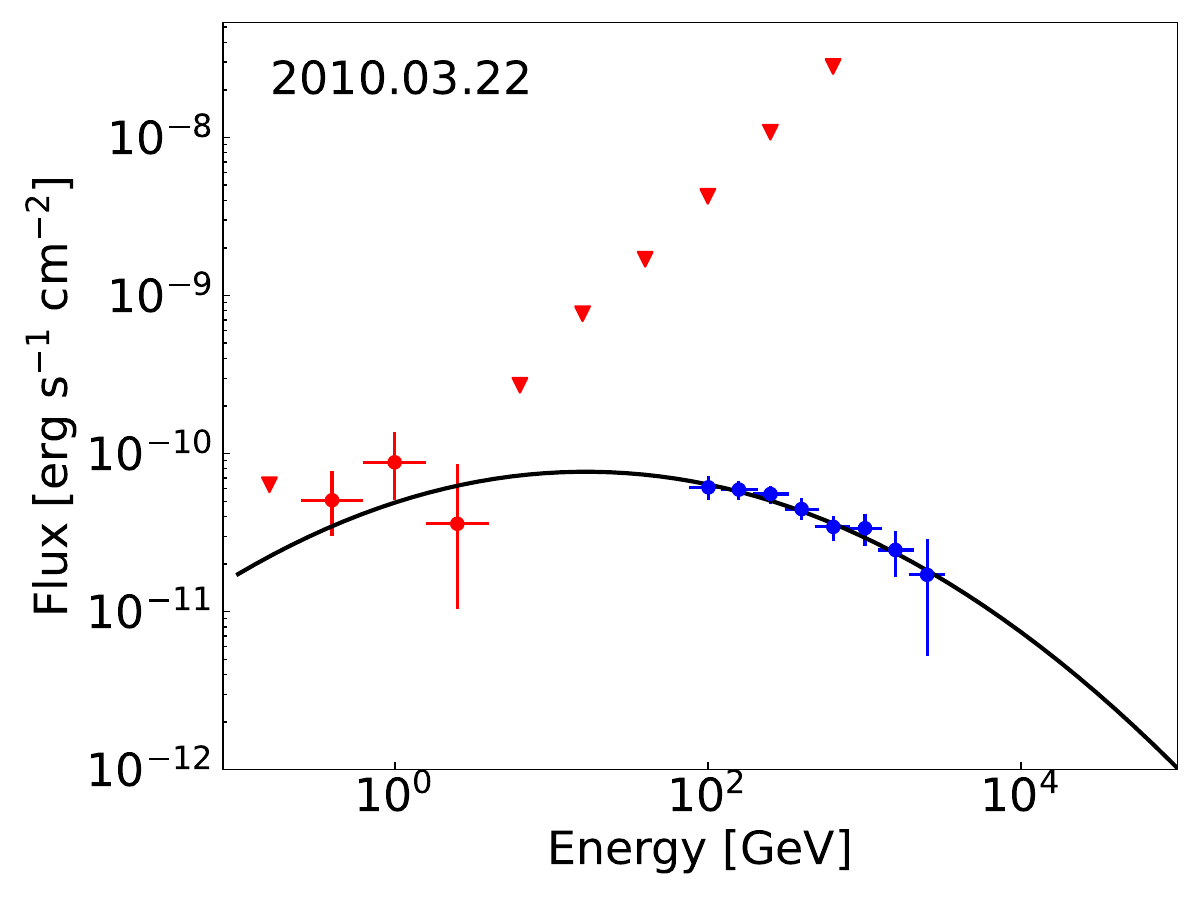} 
    \end{minipage}
    \begin{minipage}{0.3\textwidth}
        \centering
        \includegraphics[angle=0, width=\textwidth]{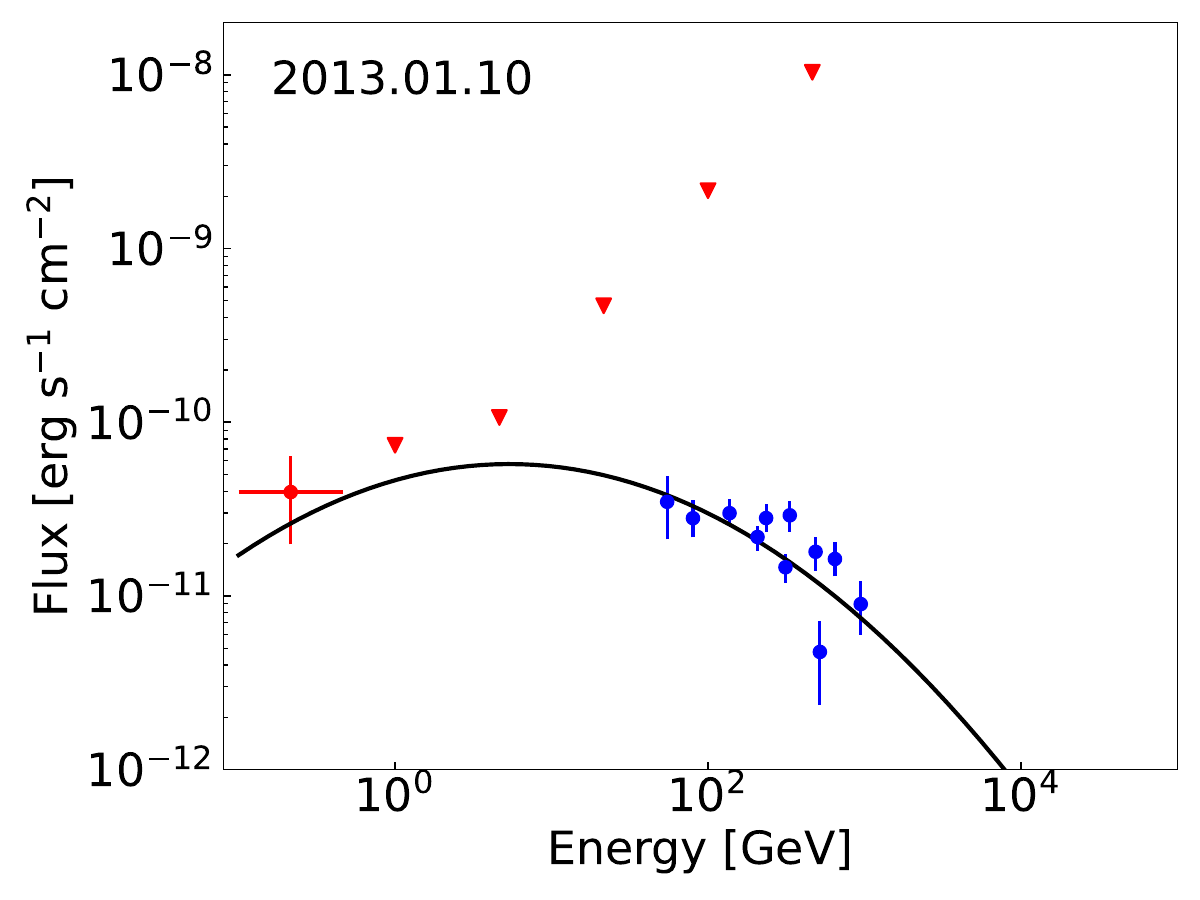} 
    \end{minipage}
    \begin{minipage}{0.3\textwidth}
        \centering
        \includegraphics[angle=0, width=\textwidth]{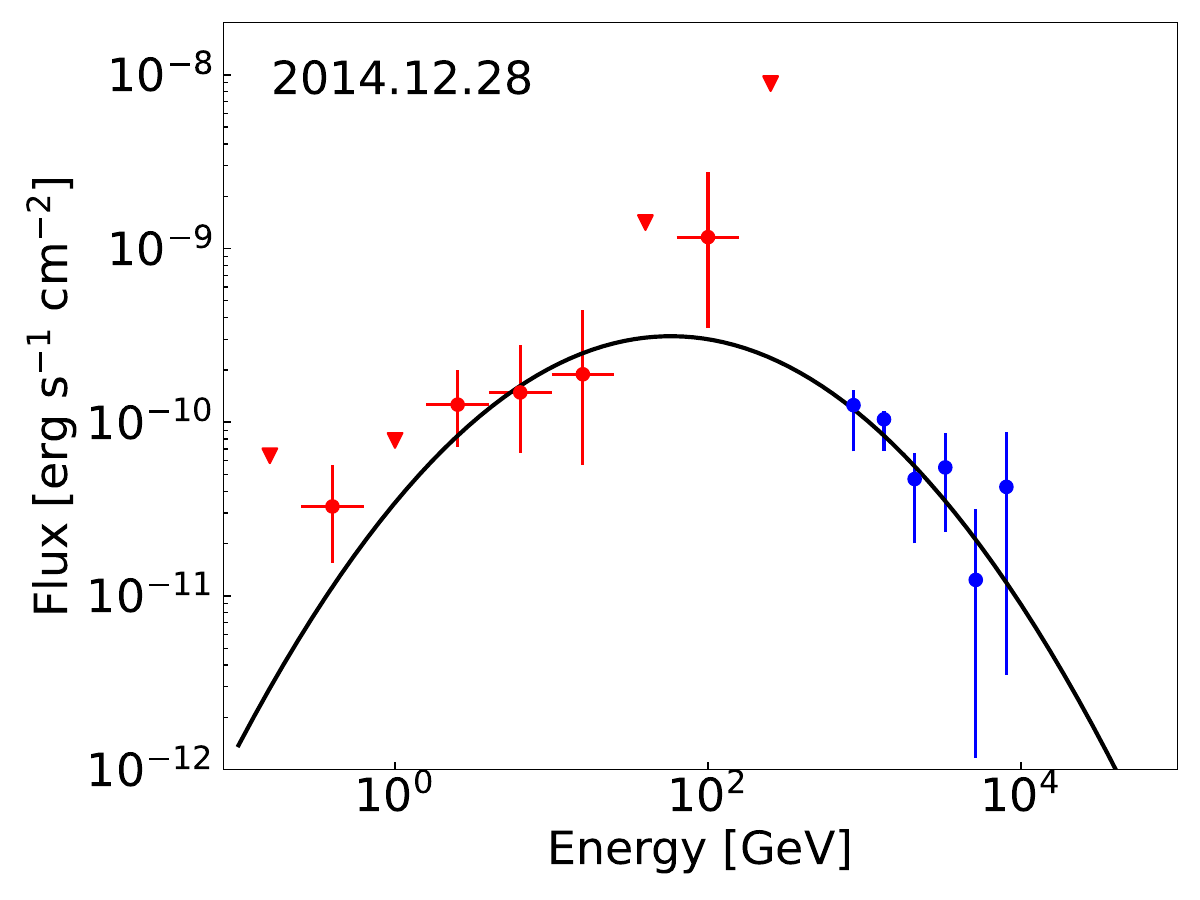} 
    \end{minipage}
    \begin{minipage}{0.3\textwidth}
        \centering
        \includegraphics[angle=0, width=\textwidth]{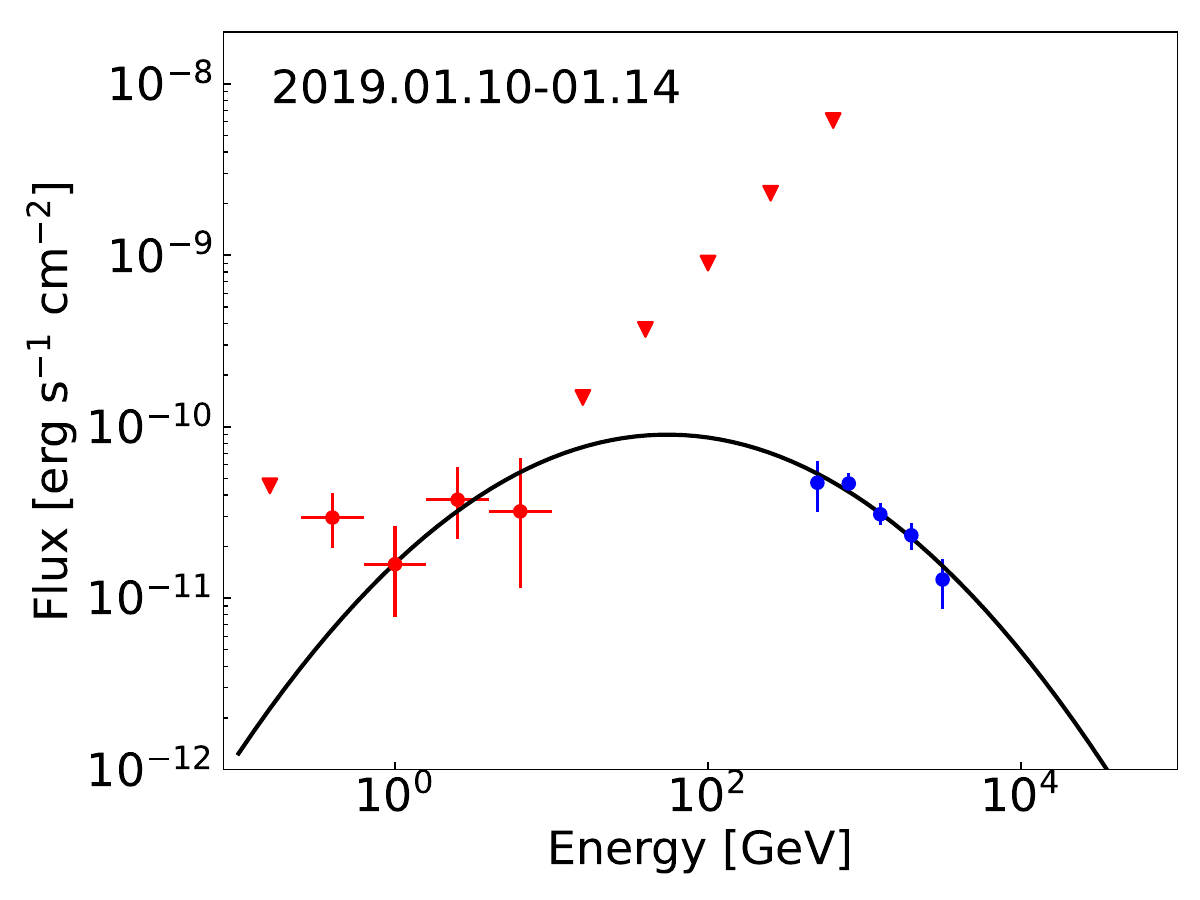} 
    \end{minipage}
    \begin{minipage}{0.3\textwidth}
        \centering
        \includegraphics[angle=0, width=\textwidth]{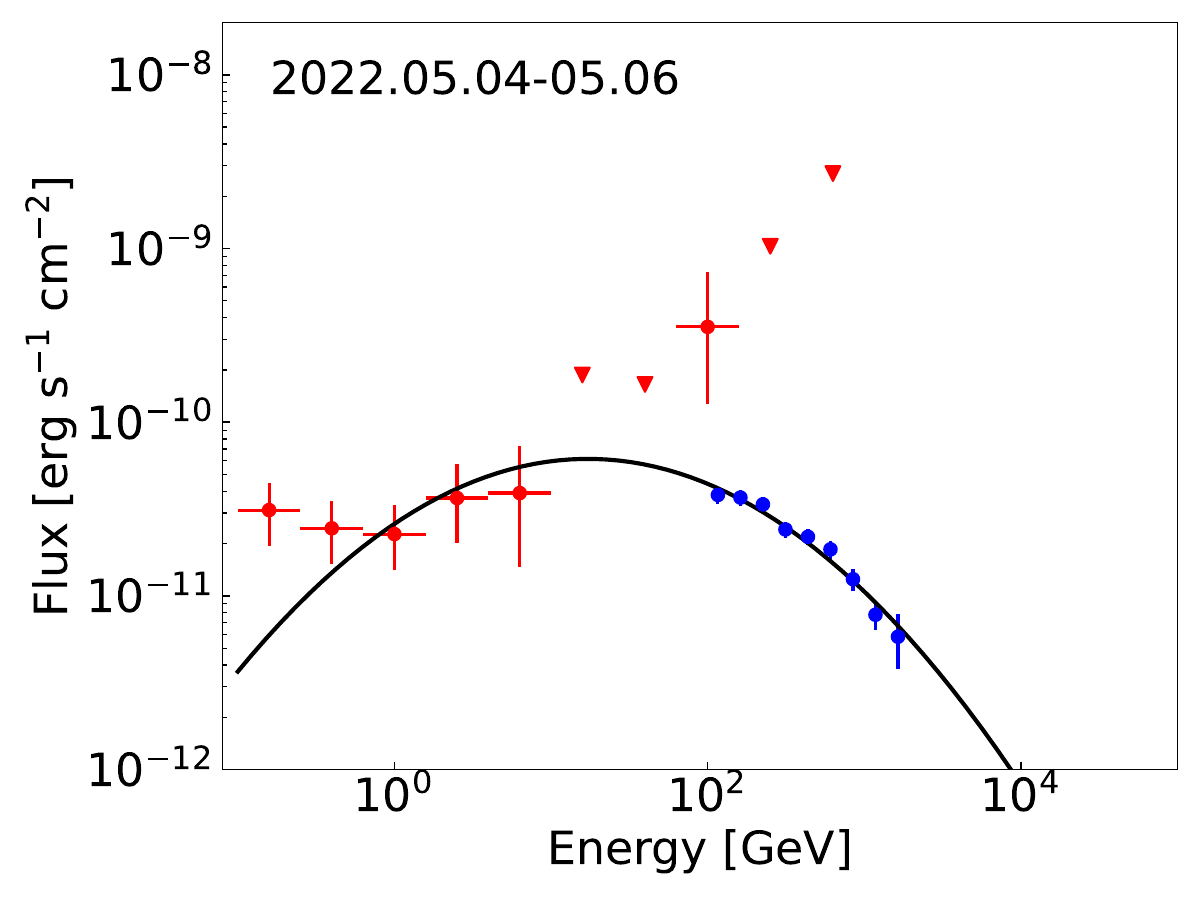} 
    \end{minipage}
    \begin{minipage}{0.3\textwidth}
        \centering
        \includegraphics[angle=0, width=\textwidth]{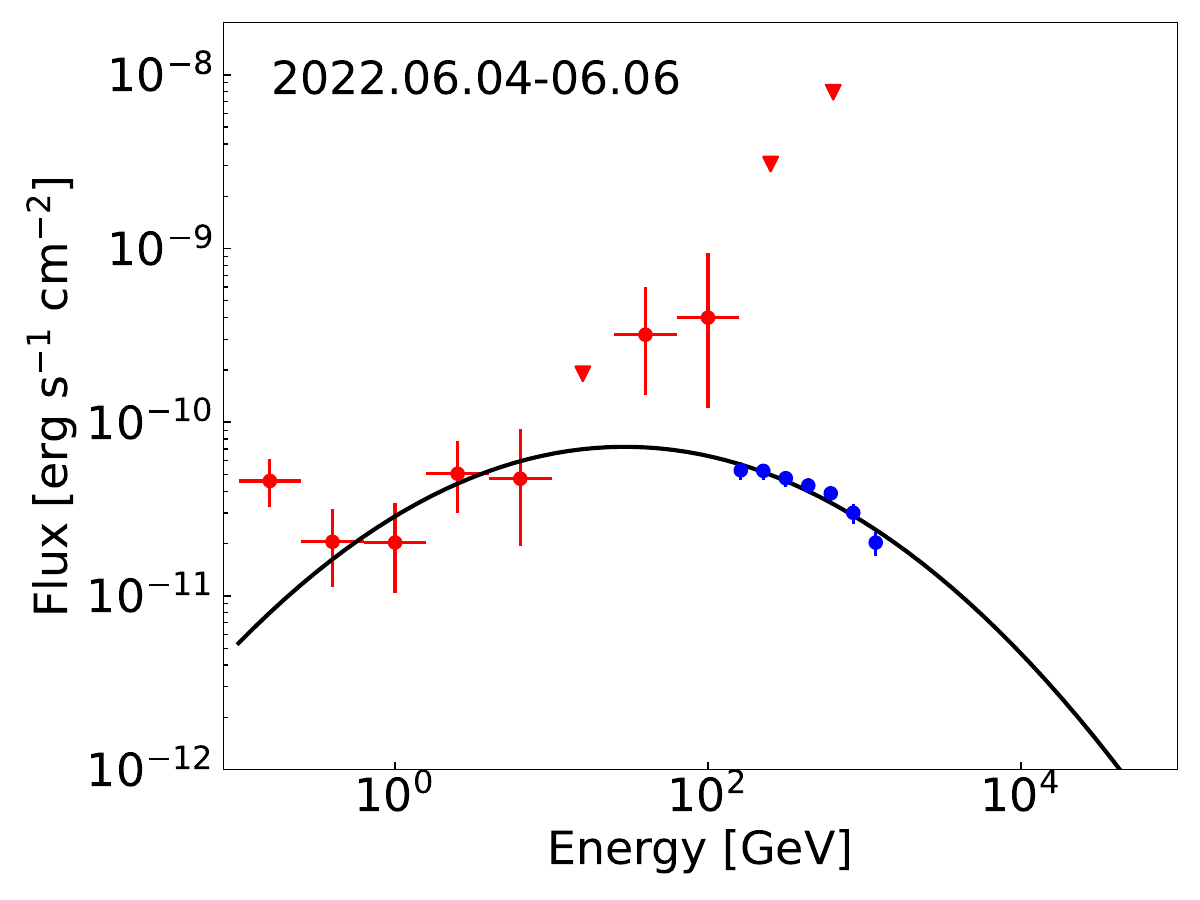} 
    \end{minipage}
      \caption{(Continued.)}
\end{figure*}

\clearpage

\begin{table*}
    \centering
    \renewcommand{\arraystretch}{1.1} 
    \setlength{\tabcolsep}{8pt}    
    \caption{Results of Fermi-LAT Data Analysis}
    \label{tab_LAT}
    \begin{tabular}{lccccc}
    \hline
    \hline
    Time & Flux$^{\clubsuit}$ & Spectral Model & $\Gamma_{\gamma}$ & $\beta$ & TS\\
    {} & {[$\times10^{-10}\,\mathrm{erg\,cm^{-2}\,s^{-1}}$]} & & & {[$\times$0.1]} & \\
    \hline
    2008.08.04 - 2025.03.26 (total) & 5.83$\pm$0.11 & PLSuperExpCutoff4 & 1.74$\pm$0.01 & 0.09$\pm$0.01 & 215187.89\\
    2014.08.11 - 2014.08.25 (low) & 0.25$\pm$0.12 & Power-law & 2.14$\pm$0.36 & & 11.79 \\
    2024.04.08 - 2024.04.22 (low) & 0.75$\pm$0.14 & Power-law & 2.13$\pm$0.13 & & 128.03 \\
    2025.02.24 - 2025.03.10 (low) & 0.55$\pm$0.28 & Power-law & 2.13$\pm$0.35 & & 17.03 \\
    2012.08.13 - 2012.08.27 (high) & 18.15$\pm$2.38 & PLSuperExpCutoff4 & 1.71$\pm$0.03 & 0.08$\pm$0.01 & 2618.49\\
    2018.01.22 - 2018.02.05 (high) & 8.53$\pm$1.92 & PLSuperExpCutoff4 & 1.90$\pm$0.07 & -0.20$\pm$0.07 & 639.59\\
    2023.01.02 - 2023.01.16 (high) & 10.97$\pm$2.11 & PLSuperExpCutoff4 & 1.59$\pm$0.06 & 0.08$\pm$0.01 & 811.78\\
    2010.02.17 & 24.14$\pm$9.55 & Power-law & 1.46$\pm$0.11 & & 165.68 \\
    2010.03.10 & 1.16$\pm$0.47 & Power-law & 2.78$\pm$0.51 & & 17.99 \\
    2010.03.11 & 2.06$\pm$0.87 & Power-law & 2.35$\pm$0.34 & & 31.83 \\
    2010.03.12 & 2.84$\pm$1.96 & Power-law & 1.91$\pm$0.27 & & 31.90 \\
    2010.03.13 & 2.12$\pm$1.59 & Power-law & 2.19$\pm$0.38 & & 21.59 \\
    2010.03.14 & 3.23$\pm$2.37 & Power-law & 1.74$\pm$0.27 & & 29.37 \\
    2010.03.15 & 3.13$\pm$2.27 & Power-law & 1.79$\pm$0.27 & & 31.37 \\
    2010.03.16 & 6.65$\pm$3.61 & Power-law & 1.56$\pm$0.29 &  & 28.63 \\
    2010.03.17 & 5.25$\pm$3.72 & Power-law & 1.89$\pm$0.27 & & 51.65 \\
    2010.03.18 & 5.92$\pm$4.30 & Power-law & 1.69$\pm$0.24 & & 38.07 \\
    2010.03.19 & 6.49$\pm$4.18 & Power-law & 1.61$\pm$0.20 & & 49.94 \\
    2010.03.21 & 1.25$\pm$0.80 & Power-law & 2.28$\pm$0.44 & & 18.50 \\
    2010.03.22 & 3.32$\pm$2.13 & Power-law & 1.95$\pm$0.25 & & 34.28 \\
    2013.01.10 & 0.88$\pm$0.34 & Power-law & 2.08$\pm$0.22 & & 6.78 \\
    2014.12.28 & 10.83$\pm$6.14 & Power-law & 1.46$\pm$0.16 & & 80.78 \\
    2019.01.10 - 2019.01.14 & 1.67$\pm$0.61 & Power-law & 2.09$\pm$0.19 & & 73.34 \\
    2022.05.04 - 2022.05.06 & 2.46$\pm$0.85 & Power-law & 1.83$\pm$0.14 &  & 113.85 \\
    2022.06.04 - 2022.06.06 & 4.12$\pm$1.60 & Power-law & 1.80$\pm$0.22 & & 115.00 \\
    \hline    
    \end{tabular}    
    \tablenotetext{\clubsuit}{Except for the 17-year average flux, which is obtained in the 0.1--1000 GeV band, the fluxes of the remaining 24 time-resolved spectra are in the 0.1--100 GeV band.}
\end{table*}

\end{document}